\documentclass[]{jfm}

\usepackage{graphicx}
\usepackage{newtxtext}
\usepackage{newtxmath}
\usepackage{natbib}
\usepackage{hyperref}
\usepackage[section]{placeins}
\usepackage{float}
\usepackage{caption}
\usepackage{subcaption}
\usepackage{xcolor}
\usepackage[normalem]{ulem}
\usepackage{cleveref}
\usepackage{color,soul}
\hypersetup{
    colorlinks = true,
    urlcolor   = blue,
    citecolor  = black,
}

\newcommand{\RomanNumeralCaps}[1]
\linenumbers

\title{Subgrid Stress Modelling with Multi-dimensional State Space Sequence Models}

\author{Andy Wu\aff{1}
  \corresp{\email{awu1018@stanford.edu}} \and
  Sanjiva K. Lele\aff{1, 2}}

\affiliation{\aff{1}Department of Aeronautics and Astronautics, Stanford University, Stanford, California, USA
\aff{2}Department of Mechanical Engineering, Stanford University, Stanford, California, USA}

\begin{document}
\maketitle

\begin{abstract}
Large Eddy Simulations (LES) are becoming increasingly viable due to the growth in computational power over the last few decades, and subgrid stress modelling plays a large role in the accuracy of LES. A new class of neural network models, S4 and S4ND models, allow for learning a continuous representation of the discrete dataset, which facilitates a principled approach to incorporating grid dependence in neural network subgrid stress modelling. A S4ND Unet neural network architecture is proposed and trained on both forced Homogeneous Isotropic Turbulence (HIT) and channel flow. \textit{A priori}, this architecture is able to extrapolate to grid spacings that are coarser than the training set grid spacings, while simpler artificial neural network (ANN) models fail. \textit{A posteriori} tests on both forced HIT and channel flow indicate that the S4ND model is more accurate than traditional models and ANN-based models on grid sizes that are in the training set. The S4ND model is also able to extrapolate robustly to grid sizes that are coarser than the training set \textit{a posteriori}, and is more accurate than many traditional subgrid stress models. Finally, the proposed model is also evaluated on flows at increasing Reynolds numbers, where it is seen that the proposed neural network remains stable even at a Reynolds number that is 500,000 times larger than what was seen in the training set. 
\end{abstract}

\begin{keywords}
Authors should not enter keywords on the manuscript, as these must be chosen by the author during the online submission process and will then be added during the typesetting process (see \href{https://www.cambridge.org/core/journals/journal-of-fluid-mechanics/information/list-of-keywords}{Keyword PDF} for the full list).  Other classifications will be added at the same time.
\end{keywords}

{\bf MSC Codes }  {\it(Optional)} Please enter your MSC Codes here

\section{Introduction}
\label{sec:headings}
Large Eddy Simulations (LES) provide an efficient paradigm to economically simulate turbulent flows and potentially flows of engineering interest by simulating the large scale, predominantly energy containing turbulent eddies while modelling the small scale turbulent structures~\citep{goc2021large}. Thus, in order to close the governing equations, subgrid-scale (SGS) modelling, the act of modelling the effect of small scale turbulent structures, is employed. SGS modelling has a significant impact on the accuracy and stability of LES~\citep{pope, sagaut}. In general, SGS models can be split into two major categories, models being driven by statistical approximations inspired by turbulence theory, or derived from data (with additional constraints that may or may not be driven by turbulence theory). In each of these two major categories, one can additionally make the characterization of the model being functional, structural, or a mix of both~\citep{evans, sagaut}. Functional models aim to approximate the effect of smaller-scale structures of turbulence on the larger scales, contributing to the stability of the LES simulation~\citep{sagaut}. Oftentimes, these models do not have high correlation with the subgrid stress tensor, $\tau_{ij}$. Meanwhile, structural models aim to capture the subgrid stress tensor more accurately, but often do not capture the true energy dissipation associated with the small scales, and can be destabilizing to LES~\citep{sagaut}. 

Traditionally, SGS models are statistical approximations inspired by turbulence theory, whether they are functional or structural. One of the first SGS models for LES, the Static Smagorinsky model~\citep{smagorinsky1963general}, assumes a perfect alignment between the resolved rate of strain tensor, $\bar{S}_{ij}$ and $\tau_{ij}$~\citep{pope}. Both the Static Smagorinsky and Dynamic Smagorinsky models~\citep{germano1991dynamic} are based on inertial range energetics, where the rate of energy transfer from large scale to small scale structures is equivalent to the viscous dissipation at small scales. Other models, such as the Sigma model~\citep{nicoud2011using}, compute the eddy-viscosity coefficient differently as compared to the Static and Dynamic Smagorinsky models to incorporate more turbulence theory (subgrid stress must vanish in laminar flows), but are still rooted in the same theory of inertial range energetics~\citep{vreman}. The Sigma model represents a significant advance in traditional subgrid stress modelling, as by using the singular values of the resolved velocity gradient tensor, the model predicts zero subgrid stress if the resolved field is two-dimensional or two-component, vanishes in laminar flow, and has cubic behavior at the wall~\citep{nicoud2011using}. Even then, all the models listed thus far have been functional models. A different subset of models, structural models, do not focus on the energy dissipation and instead are predicated on capturing higher correlation with the subgrid stress tensor. For example, Clark's Gradient model~\citep{clark1979evaluation} for the subgrid stress tensor is derived from a Taylor series expansion assuming an isotropic filter. The Scale-Similarity Model~\citep{bardina1983improved, bardina}, assumes inertial range scale-similarity, suggesting that the subgrid stress tensor can be inferred from doubly-filtered and filtered velocities. Since these models do not incorporate insights associated with turbulence energetics explicitly, often, they are destabilizing to a LES, even if the predicted subgrid stress tensor has higher correlation to the actual subgrid stress tensor~\citep{evans,beckreview}. 

Recently, with the advance of deep learning and the availability of high quality Direct Numerical Simulation (DNS) data online, learning a closure to the governing equations has become tractable instead of approaching subgrid stress modelling from a purely theoretical perspective~\citep{park, beck, stoffer, kawai, xie}. Some of the early work done using machine learning involved using the velocity gradient and stresses to predict an eddy-viscosity coefficient~\citep{sarg}. Models along this line of thought can be considered data-driven functional models. Another approach involves predicting the subgrid forcing term using neural networks, which directly approximates the term seen in the governing equations, or directly approximating $\tau_{ij}$~\citep{xie, xie2}. How $\tau_{ij}$ or the divergence of $\tau_{ij}$ is found varies, as \citet{yuan2020deconvolutional} and \citet{stolz2001approximate} adopt a deconvolution approach, while others such as \citet{xie} use a tensor basis neural network (TBNN) with associated invariants as inputs~\citep{Wu_PrF}. Others use the velocity gradient tensor as an input to a neural network to predict $\tau_{ij}$~\citep{park, Kang}, or incorporate second derivatives into the neural network input~\citep{Pawar_2020, wang2018investigations}. Another approach is to use mixed-models, where functional and structural models are mixed together to form one model, with the neural network learning the balance between the functional and structural models~\citep{beck, evans}. One can also adopt a reinforcement learning approach, where the neural network is given feedback at every ``simulation timestep" and learns according to a reward function~\citep{Kurz_2023, bae2022scientific}. In the end, all these methods are just a way to either learn an eddy viscosity coefficient (functional), some part of the subgrid stress tensor (structural), or a mix of both. 

Despite all this progress, neural network models are still not as flexible as traditional models, and the above works have all mainly focused on grid sizes that are relatively fine (only 4-8 times that of the DNS grid spacing, at most 16 times). Furthermore, it is well known that neural network models have trouble generalizing to grid spacings that are coarser than the training data~\citep{Choi_2024, Kang}. As such,~\citet{Choi_2024} adopt a recursive approach, where a neural network is continually augmented with higher Reynolds Number (\Rey) data trained on LES integrated with a neural network subgrid stress model. This approach is able to generalize to a wide range of Reynolds Numbers and grid spacing, and is one of the few investigations at a coarse LES grid spacing (when compared to the DNS grid spacing). Predominantly, neural network subgrid stress modelling only uses simple architectures such as Artificial Neural Networks (ANNs)~\citep{Kang, park, Choi_2024, stoffer, xie, xie2, yuan2020deconvolutional, yuan2021dynamic, sarg, sirignano2023deep, cheng2022deep} or simple Convolutional Neural Networks (CNNs)~\citep{Guan2022, beck}. Given the advance of deep learning, it stands to reason that using more advanced neural network architectures could gain the generalization capability necessary for neural network models to be more flexible. As seen,~\citet{Wu_PrF} use a two neural network Convolutional Unet architecture trained on both channel flow and forced Homogeneous Isotropic Turbulence (forced HIT), showing that it improves upon traditional models and machine learning (ML) models using more simple architectures at larger grid spacings (LES grid spacings that are 16 and 32 times the DNS grid spacing). However, generalization to coarser grid spacings than the training set is still lacking. 

In deep learning, a new class of models that have been gaining traction involve learning continuous convolution kernels~\citep{li2021fourierneuraloperatorparametric, nguyen2022s4nd}. Since typical convolutional neural network architectures are discrete, they are not sensitized to the grid spacing. Therefore, it is not surprising that they are unable to generalize to coarser grid spacings for subgrid stress modelling. However, continuous convolution kernels explicitly account for the grid spacing, potentially allowing for generalization to coarser grid spacing than the training data. One class of models from the computer vision community, the S4ND model~\citep{nguyen2022s4nd}, allows for continuous convolution kernels to be learned by parameterizing each spatial dimension with a State Space Model~\citep{gu2021efficiently}. By using the S4ND model in place of traditional convolutions, while enforcing the subgrid stress tensor symmetries through a Tensor Basis Neural Network (TBNN), a new neural network for subgrid stress modelling is created. 

The main contributions of this paper include introducing a new neural network architecture that can extrapolate robustly to coarser grid spacings not seen during training without any additional re-training, first evaluated in an \textit{a priori} setting. This paper also contributes \textit{a posteriori} analysis of the neural network. After the neural network is trained on both forced HIT and channel flow data, it is integrated into LES simulations of both forced HIT and channel flow. The neural network is evaluated on a coarser grid spacing unseen during training while being more accurate than traditional and other data-driven SGS models. Further analysis also shows that the neural network model can extrapolate to \Rey~up to 500 times the \Rey~in the training set, while also extrapolating in grid spacing without the need to train on any additional LES data or undergo transfer learning. This neural network subgrid stress model is also able to remain stable even when the \Rey~is over 500,000 times larger than what it was trained on.

\section{Numerical and Neural Network Details}
\label{sec:headings}
This section provides background on the LES framework used, as well as the neural network model and training details. 

\subsection {LES Governing Equations}
 The governing LES equations solved are the filtered incompressible Navier Stokes Equations. The filtering operation is defined below:

\begin{equation} \label{eqn1}
\Bar{\chi}(\mathbf{x}, t) = \int{G(\mathbf{r})\chi(\mathbf{x - r}, t) dr}
\end{equation}

\begin{equation} \label{eqn2}
\int{G(\mathbf{r}) dr} = 1
\end{equation}

where $\chi$ is the flow variable to be filtered, $G(\mathbf{r})$ is the filtering kernel, and $\Bar{\chi}$ is the filtered flow variable. In all sections, the overbar operator is used to denote filtered quantities. 

Now that the filter is defined, the governing equations are written below (in Einstein's summation notation):

 \begin{equation} \label{eqn3}
\begin{aligned}
\frac{\partial \Bar{u}_i}{\partial x_i} &= 0 \\
\frac{\partial \Bar{u}_i}{\partial t} + \frac{\partial \Bar{u}_i \Bar{u}_j }{\partial x_j} &= - \frac{1}{\rho}\frac{\partial \Bar{p}}{\partial x_i} + \nu \frac{\partial^2 \Bar{u}_i}{\partial x_j \partial x_j} - \frac{\partial \tau_{ij}}{\partial x_j} \\
\tau_{ij} &= \overline{u_i u_j} - \Bar{u}_i\Bar{u}_j
\end{aligned}
\end{equation}

where $\overline{p}$ is the filtered pressure, $\rho$ is the density, $\overline{u}$ is the filtered velocity, and $\nu$ is the kinematic viscosity. The velocity field is solenoidal as seen from the continuity equation, and filtering the momentum equation results in obtaining the unclosed term, $\tau_{ij}$. This tensor is computed by obtaining the filtered nonlinear product of the velocities before filtering, and then subtracting by the product of the filtered velocities. In LES only the filtered values of the dependent variables are available, thus, $\tau_{ij}$ can't be computed and has to be modelled. In the present work, all simulations are in the incompressible regime, and the neural network will learn to approximate $\tau_{ij}$ by learning coefficients of a predefined tensor basis expansion. 
 
\subsection {State Space Models and S4ND}
Recently, a new class of models, structured state space sequence (S4) models have emerged from the natural language processing community~\citep{gu2021efficiently}. For a 1 dimensional problem, the neural network attempts to learn a state space representation:

\begin{equation} \label{eqn3s}
\begin{aligned}
\dot{\mathbf{x}} &= \mathbf{A} \mathbf{x} + \mathbf{B} \mathbf{u} \\
\mathbf{y} &= \mathbf{C} \mathbf{x} + \mathbf{D} \mathbf{u}
\end{aligned}
\end{equation}

where $\mathbf{A}$, $\mathbf{B}$, $\mathbf{C}$, $\mathbf{D}$, are learnable matrices, $\mathbf{u}$ represents a 1D input (function of time), $\mathbf{x}$ represents the state vector (dimension of this state vector is specified by the user), while $\mathbf{y}$ represents the output vector. In general, $\mathbf{D}$ is omitted as it can be viewed as a skip connection, which can be added in the neural network architecture~\citep{gu2021efficiently}. As seen, the neural network learns an underlying continuous representation from the discrete data. However, since the data is discrete, there needs to be a connection between the continuous and discrete setting. This connection is derived for the bilinear method~\citep{gu2021efficiently}. One can write the discretized state space as (the tilde operation signifies the discretized matrices):

\begin{equation} \label{eqn3s1}
\begin{aligned}
\mathbf{x}_{k} &= \Tilde{\mathbf{A}} \mathbf{x}_{k-1} + \Tilde{\mathbf{B}} \mathbf{u}_{k} \\
\mathbf{y}_{k} &= \Tilde{\mathbf{C}} \mathbf{x}_{k} \\
\Tilde{\mathbf{A}} &= (\mathbf{I}-\frac{\Delta}{2}\cdot\mathbf{A})^{-1}(\mathbf{I}+\frac{\Delta}{2}\cdot\mathbf{A}) \\
\Tilde{\mathbf{B}} &= (\mathbf{I}-\frac{\Delta}{2}\cdot\mathbf{A})^{-1}\Delta\mathbf{B} \\
\Tilde{\mathbf{C}} &= \mathbf{C}
\end{aligned}
\end{equation}

As seen, when discrete values (such as during training) are needed since the training data is discrete, the state space can be discretized with a spacing parameter $\Delta$ and a discretization method such as the bilinear method or zero order hold~\citep{gu2021efficiently}. This ensures that the neural network, while learning from discrete data, has an underlying continuous representation, which is enabled by the additional knowledge of the spacing parameter $\Delta$. The underlying continuous representation it learns can be found by writing the non-tilde matrices as a function of the tilde matrices. By learning this underlying continuous representation and then being able to discretize using a specified grid spacing, the S4 model has been found to generalize very well to different $\Delta$ not in the training set~\citep{gu2021efficiently}.

\begin{figure}
\centering
\includegraphics[width=.75\textwidth]{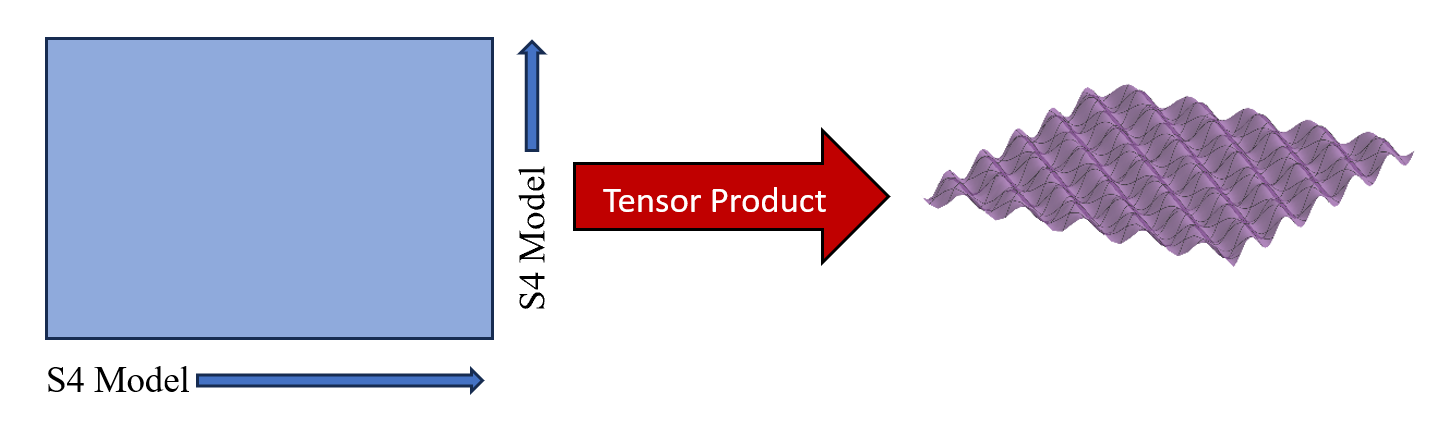}
\caption{S4ND extension of the S4 model. A S4 model is instantiated in each direction, and then an outer product is taken to produce a continuous convolution kernel. When training and during inference, the continuous convolution kernel is discretized.}
\label{fig:S4ND}
\end{figure}

This approach was extended to problems involving two dimensions in the computer vision community with the introduction of the S4ND model~\citep{nguyen2022s4nd}. The core idea behind the S4ND model, illustrated in figure~\ref{fig:S4ND}, is to learn global, continuous convolution kernels by initializing a S4 model in each spatial dimension, and then taking the tensor outer product in order to obtain a continuous convolution kernel. 
Now, the $\Delta$ spacing parameter is the grid spacing. In other words, $\Delta = \Delta_{g}$ where $\Delta_g$ denotes the spatial spacing. While the grid spacing in each cardinal direction must be the same, this formulation allows for the grid spacing in various directions to be different (for example different constant grid spacing in x, y, z directions). When convolving the continuous convolution kernel with the discrete data, the spatial spacing is used to discretize the convolution kernel into a discrete convolution kernel before applying the convolution operation. Thus, the S4ND model can be used to replace traditional convolutions as a grid spacing aware operator, while traditional convolutions are agnostic to grid spacing. Overall, the S4ND model allows the neural network to learn continuous multi-dimensional representations of the data, which is more in line with fluid dynamics (continuum approximation), as the governing equations are continuous and then discretized onto a grid for LES.

In addition, since the state space is now discretized with a specific spatial spacing $\Delta_g$, S4ND introduces a novel idea called bandlimiting~\citep{nguyen2022s4nd}, which seeks to remove spurious frequencies in the convolution kernel that can't be supported by the grid spacing implied by $\Delta_g$. If $\mathbf{A}$ is diagonal, then the state space convolution kernel ``basis" can be represented by $e^{a_nt}B_n$, where $a_n$ denotes the nth diagonal element of $\mathbf{A}$. Therefore, $a_n$ is a parameter that controls the frequency of the basis function associated with the convolution kernel. In order to mask out spurious frequencies in the convolution kernel, S4ND introduces a bandlimiting parameter, $\alpha$, which is a hyperparameter. If $a_n \Delta_g > \frac{1}{2}\alpha$ , then this convolution kernel basis function is masked out, as this means that the frequency of the basis function is larger than what the grid can support (it is spurious). When $\alpha = 1$, bandlimiting corresponds to masking out everything above the Nyquist wavenumber. However, due to the finite state representation and also the decay associated with the real part of $a_n$, $\alpha$ is set lower empirically (between 0.1 and 0.6), which creates smoother convolution kernels~\citep{nguyen2022s4nd}. In this work, $\alpha = 0.1$ is used. Appendix A provides a justification for using this specific bandlimit parameter.

These mechanisms make S4ND more suitable for subgrid stress modelling in LES. First, the idea of bandlimiting functions as a rigorous filter for aliasing. In LES, the grid spacing $\Delta_g$ functions as a low-pass filter, where beyond the Nyquist wavenumber, $k_c = \pi/ \Delta_g$, the grid can no longer support higher wavenumber oscillations. Standard discrete convolutional neural networks learn fixed weights that ignore this limitation set by the grid. Therefore, after training on a set of resolutions, when applied to coarser grids, they may predict high-frequency oscillations that alias back into the resolved scales, which is seen as a failure to extrapolate to larger grid spacings. In contrast, S4ND's bandlimiting framework ensures that the continuous convolution kernel (due to the state space representation) suppresses the frequencies that exceed the spatial grid's spectral resolution, enforcing the consistency between the data-driven subgrid stress model and the resolution of the mesh. 

Specifically, this is directly connected to the notion of LES filter consistency. In LES, the subgrid stress tensor must be mathematically consistent with the implied filter that is a consequence of both the grid width and the numerical method. Since standard discrete convolutions learn fixed, grid-dependent weights that treat the LES simulation as an array of pixels, they have no inductive biases to enforce this consistency outside of the training set grid resolutions. Thus, the learned closure is oftentimes only valid for filter width ranges seen during training. Meanwhile, S4ND's continuous formulation with bandlimiting provides an inductive bias, ensuring that the subgrid stress model adapts its spectral content as the filter width changes, even at untrained grid resolutions. This ensures that the S4ND model can robustly produce a more filter-consistent closure even on grid resolutions outside of the training set, instead of injecting unphysical frequency content at untrained resolutions. For S4ND, this comes as part of the architectural framework instead of only as a learned consequence of the data.

Furthermore, S4ND allows for a continuous representation instead of a discrete representation, which allows it to better learn scale similarity. By learning the underlying subgrid stress field as a continuous quantity, it allows the neural network model to extrapolate physically consistent predictions to coarser grid resolutions where scale-invariance is important but where discrete correlations learned by a convolutional neural network would break down.

\subsection {Neural Network Architecture}
The neural network involves replacing discrete convolutions with the S4ND block in a U-net architecture, followed by a backend tensor basis neural network (TBNN). 
 
\subsubsection {Tensor Basis Neural Network}
The TBNN, originally introduced by~\citet{ling}, involves computing the invariants and the tensor basis assuming that the subgrid stress tensor is a function of the resolved rate of strain tensor, $\mathbf{\bar{S}_{ij}}$, and the resolved rotation rate tensor, $\mathbf{\bar{R}_{ij}}$. A detailed derivation of the tensor basis and invariants for the subgrid stress tensor can be found in works such as~\citet{stallcup} and~\citet{Wu_PrF}. Below, the invariants and tensor basis are listed for reference. 

If one assumes that:
\begin{equation} \label{eqn4}
\tau_{ij} = F(\mathbf{\Bar{S}_{ij}},\mathbf{\Bar{R}_{ij}})
\end{equation}
\begin{equation} \label{eqn5}
\mathbf{\Bar{S}} \equiv \mathbf{\Bar{S}}_{ij} = \frac{1}{2}(\frac{\partial \Bar{u}_i}{\partial x_j}+\frac{\partial \Bar{u}_j}{\partial x_i})
\end{equation}
\begin{equation} \label{eqn6}
\mathbf{\Bar{R}} \equiv \mathbf{\Bar{R}}_{ij} = \frac{1}{2}(\frac{\partial \Bar{u}_i}{\partial x_j}-\frac{\partial \Bar{u}_j}{\partial x_i})
\end{equation}

then the invariants can be written as:
\begin{equation} \label{eqn7}
\{tr(\mathbf{\Bar{S}}), tr(\mathbf{\Bar{S}^2}), tr(\mathbf{\Bar{S}^3}), tr(\mathbf{\Bar{R}^2}), tr(\mathbf{\Bar{S}\Bar{R}^2}), tr(\mathbf{\Bar{S}^2\Bar{R}^2}), tr(\mathbf{\Bar{S}^2\Bar{R}^2\Bar{S}\Bar{R}}) \}
\end{equation}

and the tensor basis can be written as:
\begin{equation} \label{eqn8}
\{\mathbf{I}, \mathbf{\Bar{S}}, \mathbf{\Bar{S}^2}, \mathbf{\Bar{R}^2}, \mathbf{\Bar{S}\Bar{R} - \Bar{R}\Bar{S}}, \mathbf{\Bar{R}\Bar{S}\Bar{R}}, \mathbf{\Bar{S}^2\Bar{R} - \Bar{R}\Bar{S}^2}, \mathbf{\Bar{R}\Bar{S}\Bar{R}^2 - \Bar{R}^2\Bar{S}\Bar{R}} \}
\end{equation}

As seen, the invariants involve extremely high powers of the resolved strain rate and rotation rate tensors, which are not ideal due to aliasing errors, which increase with increasing power~\citep{roberts1987digital}. Thus, instead of using the invariants, which are mathematically complete, one can simply use the velocity gradient tensor in place of the invariants, which is an input commonly used for neural network subgrid stress models~\citep{park, Kang, Choi_2024}. It is noted that there is no performance degradation using the velocity gradient tensor instead of the invariants as inputs to the neural network \textit{a priori}. Even if the input to the neural network is changed, the backend of the neural network remains the same. The neural network will predict coefficients of the tensor basis expansion, listed here as $c_1-c_8$ (8 scalar values are predicted given the filtered velocity gradient tensor as an input):
\begin{equation}\label{eqn9}
\tau_{ij} = c_1 \mathbf{I} + c_2 \mathbf{\Bar{S}} + c_3 \mathbf{\Bar{S}^2} + c_4 \mathbf{\Bar{R}^2} + c_5 (\mathbf{\Bar{S}\Bar{R} - \Bar{R}\Bar{S}}) 
\\ + c_6 \mathbf{\Bar{R}\Bar{S}\Bar{R}} + c_7 (\mathbf{\Bar{S}^2\Bar{R} - \Bar{R}\Bar{S}^2}) + c_8 (\mathbf{\Bar{R}\Bar{S}\Bar{R}^2 - \Bar{R}^2\Bar{S}\Bar{R}}) 
\end{equation}

\subsubsection {Overall Architecture}
\begin{figure}
  \centerline{\includegraphics[scale=0.65]{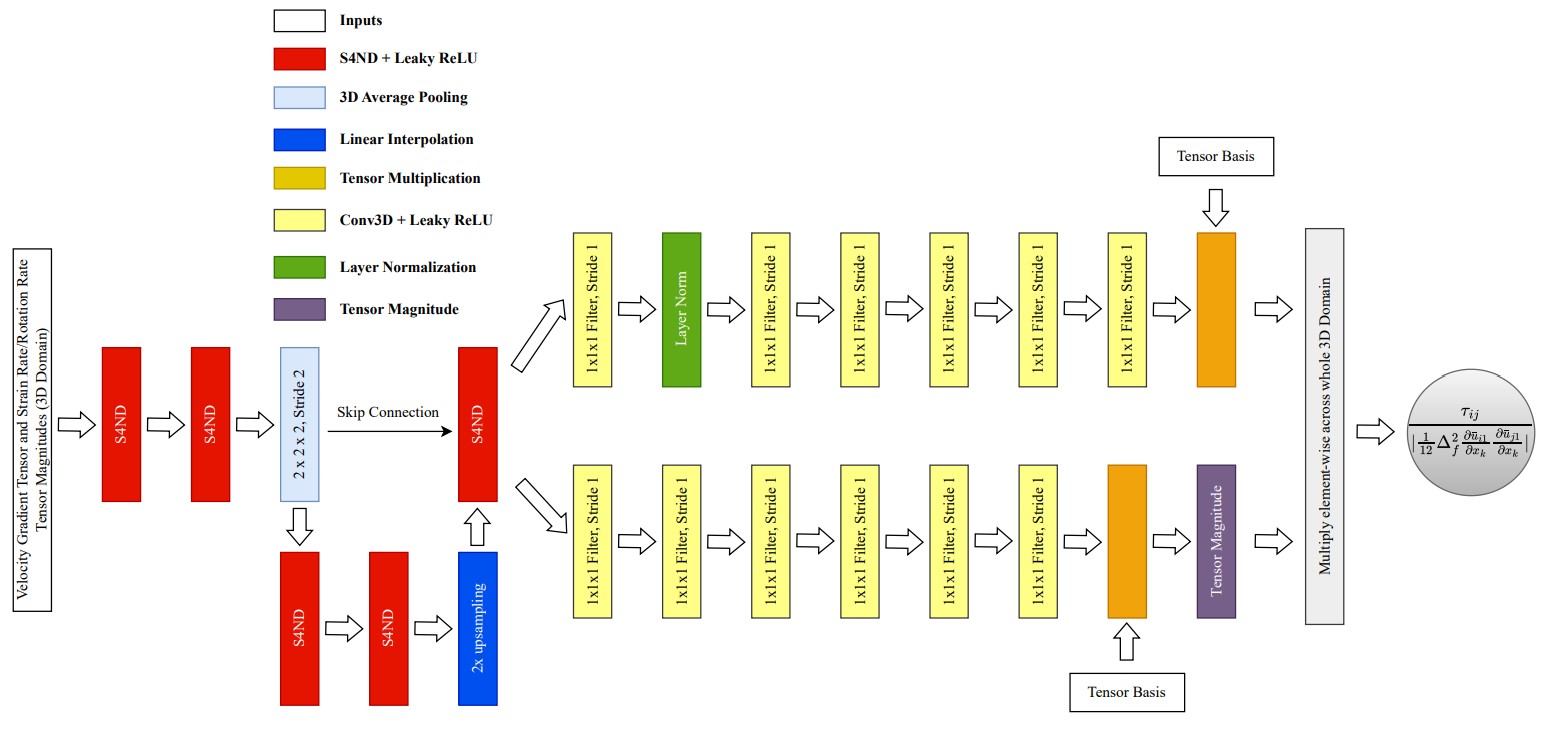}}
  \caption{The overall neural network architecture. The whole 3D domain is inputted into the neural network, and the output is the corresponding subgrid stress at each point. For each S4ND layer, a state space dimension of 8 was chosen to keep the trainable parameters comparable to existing models in literature. The average pooling operation will downsample all spatial dimensions by a factor of 2. As seen, after the final S4ND layer, the neural network architecture splits into two ``prediction heads" to predict the ``magnitude" and ``structure" of the subgrid stress tensor, and both prediction heads use the tensor basis (each prediction head predicts its own set of coefficients to multiply the tensor basis components). See~\cite{Wu2025_Git} for downloadable version of the model and an example input/output.}
\label{fig:nnarch}
\end{figure}

Now that the backend TBNN is introduced, the overall neural network architecture can be seen in figure~\ref{fig:nnarch}. 
The inputs to the neural network are the three dimensional, spatially varying velocity gradient tensor, as well as the strain rate and rotation rate tensor magnitudes. These inputs are passed into a U-net S4ND architecture that allows for multiscale fusion of different spatial scales of turbulence. After the U-net architecture, the neural network splits into two streams.~\citet{Wu_PrF} found that using two neural networks, one to predict the magnitude of $\tau_{ij}$ while the other predicts the structure (normalized) $\tau_{ij}$ results in better performance. This neural network architecture retains the idea of predicting a ``structure" component and a ``magnitude" component, but is able to do it in one single neural network instead of splitting the task of predicting a ``structure" or ``magnitude" component to separate neural networks. As seen, the neural network splits into two paths after the U-net architecture, with both paths using the tensor basis neural network (predicting 8 coefficients that multiply the tensor basis). However, instead of using fully connected layers, which are unable to handle different domain sizes, 1x1x1 convolutions are used, which can emulate fully connected layers while still being able to generalize to various domain sizes.

Non-dimensionalizing the model and scaling the inputs and outputs are crucial to training a neural network model. Here, the inputs (velocity gradient tensor, strain rate and rotation rate tensor magnitudes, so a total of 11 quantities at each point in the domain) are normalized with a max-min scaling:
\begin{equation} \label{eqn10}
x_{norm} = \frac{x-x_{min}}{x_{max}-x_{min}}*2-1
\end{equation}
where $x$ corresponds to every component of the velocity gradient tensor (or the strain rate/rotation rate tensor magnitudes). The maximum and minimum correspond to the 3D domain's maximum and minimum. This input normalization ensures that the neural network model inputs are always scaled between [-1, 1]. The tensor basis are also scaled, although with the Frobenius Norm instead:
\begin{equation} \label{eqn11}
W^{norm}_{ij} = \frac{W_{ij}}{|W_{ij}|}
\end{equation}
where $W_{ij}$ is a tensor corresponding to a tensor in the tensor basis expansion. Note that in Equation~\ref{eqn11}, repeated indices do not imply summation. Furthermore, the output is also normalized accordingly to ensure that the model is non-dimensional. Following observations made by~\citet{Kang}, the Clark's gradient model term is taken as a normalization for $\tau_{ij}$, as seen in figure~\ref{fig:nnarch}. However, one small modification is made. Work done by~\citet{GenSmag} introduced the anisotropic mesh length factor, $f$, which multiplies the effective grid width $\Delta_e$ as grid anisotropy may play a role in determining $\tau_{ij}$. Therefore, the normalization used for $\tau_{ij}$ becomes:
\begin{equation} \label{eqn12}
\frac{\tau_{ij}}{|\frac{1}{12}\Delta_f^2\frac{\partial   \bar{u}_{l}}{\partial x_k}\frac{\partial   \bar{u}_{m}}{\partial x_k}|}
\end{equation}
where $\Delta_f = f \Delta_e$, $f = 1 + \frac{2}{27}[ln(a_1)^2-ln(a_1)ln(a_2)+ln(a_2)^2]$ and $\Delta_e = (\Delta_1 \Delta_2 \Delta_3)^{\frac{1}{3}}$. $\Delta_i$ denotes the grid spacing in the ith dimension, while $a_1$ and $a_2$ denote the grid spacing ratios (both less than 1). 

\subsection {Neural Network Training Details}
This section goes over the process of querying the training data, as well as introducing the loss function and some metrics that can be used to quantify the performance of the neural networks in a LES \textit{a posteriori}. 
\subsubsection {Training Data}
Both filtered HIT and channel flow data are queried from the Johns Hopkins Turbulence Database (JHTDB), a service that provides high-quality DNS data for research purposes~\citep{JHTB1, JHTB2, graham2016web}. From~\citet{park}, training on two different filter widths provides the neural network with more generalizability to different grid spacings. Thus, since this paper is focused on applications to aggressive, coarse grid spacing settings, after querying the DNS data, the DNS data is filtered to a filter width corresponding to $\Delta_g = 16 \Delta_{DNS}$ and $\Delta_g = 32 \Delta_{DNS}$, where $\Delta_{DNS}$ denotes the DNS grid spacing. The filtering operation involves using a box filter in physical space. Following~\citet{Wu_PrF}, two different domain sizes are also queried in order to help the neural network generalize to various domain sizes. The querying procedure used here is identical to that used in~\citet{Wu_PrF}, and is also listed here for completeness. 

The training data is split into a 80-10-10 split, where 80\% of the total training data is used for training the neural network, 10\% of the data is used as a validation set during training for early stopping, and 10\% of the data is used as the hold-out test set in the \textit{a priori} testing section. To produce the training data for forced HIT, 5 snapshots are queried from the JHTDB database. To assemble the channel flow training data, 4 snapshots are queried. Each snapshot is queried at equal time intervals that are statically de-correlated from each other (time intervals different for HIT and channel flow). The DNS HIT resolution queried is $1024^3$, with a Taylor Reynolds number ($Re_{\lambda}$) of 433 and a domain size of $[2\pi, 2\pi, 2\pi]$. The channel flow resolution queried is $2048 \times 512 \times 1536$, with a friction Reynolds number ($Re_{\tau}$) of 1000 and a domain size of $[8\pi, 2, 3\pi]$ when non-dimensionalized by the channel half height. After the training data is queried, it is filtered. For example, filtering at a $16 \Delta_{DNS}$ filter width would result in the filtered HIT resolution being $64^3$ and the filtered channel flow resolution being $128\times32\times96$. Then, the full domains are split into ``sub-domains" of sizes $16^3$ and $32^3$. This results in the total training data content in Table~\ref{tab:tabletrain}.

\begin{table}
\caption{\label{tab:tabletrain} Training Data}
\centering
\begin{tabular}{lccc}
\hline
Number of Realizations & Filter Width & Domain Size & Flow Type\\\hline
320 & $16\Delta_{DNS}$ & $16^3$ & HIT \\
320 & $32\Delta_{DNS}$ & $8^3$ & HIT \\
40 & $16\Delta_{DNS}$ & $32^3$ & HIT \\
40 & $32\Delta_{DNS}$ & $16^3$ & HIT \\
320 & $16\Delta_{DNS}$ & $16^3$ & Channel \\
320 & $32\Delta_{DNS}$ & $8^3$ & Channel \\
40 & $16\Delta_{DNS}$ & $32^3$ & Channel \\
40 & $32\Delta_{DNS}$ & $16^3$ & Channel \\
\hline
\end{tabular}
\end{table}

As stated before, the training data is divided into 80 percent for training, 10 percent for validation, and 10 percent as a held-out test set. Each of the categories of training data is split according to the percentages above and then recombined into the total training, total validation, and total test set. In addition, filter width $\Delta_g = 24 \Delta_{DNS}$ and filter width $\Delta_g = 64 \Delta_{DNS}$ data are also queried, but are not included in the training set and are instead held out as part of the test set to test the neural network generalization capabilities. 

\subsubsection {Loss Function and Metrics}
The loss function used for training follows from the loss function introduced in~\cite{Wu_PrF}, involving a two-term loss function that enforces both the root mean squared error between the predicted and actual subgrid stress tensor, as well as the root mean squared error between the predicted and actual interscale energy transfer field, $\epsilon = - \tau_{ij} \Bar{S}_{ij}$. As such, the loss function involves:
\begin{equation} \label{eqn12}
L = \mathbf{RMSE}(\tau_{ij,pred}-\tau_{ij})/\mathbf{RMS}(\tau_{ij})+ \lambda \mathbf{RMSE}(\epsilon_{pred}-\epsilon)/\mathbf{RMS}(\epsilon)
\end{equation}

where $\lambda$ is a constant that can bias the neural network towards emphasizing the energy dissipation or the predicted values of $\tau_{ij}$. In the present work, following~\cite{Wu_PrF}, $\lambda = 1$ is used, which indicates an equal bias between both loss function terms. The first term of this loss function is termed ``RMSE" (unbolded) and the second term of this loss function is termed ``DRMSE" (unbolded). Meanwhile, the bolded terms signify the operation of taking the RMSE or RMS of the particular quantity in the parenthesis. During training, the neural networks are trained with a batch size of 8, and the denominators of the loss function are computed per batch. As seen, this loss function directly enforces that the predicted and actual subgrid tensors are close, while simultaneously ensuring that the predicted and actual interscale energy transfer fields are also close. This ensures that the overall energy dissipation is accurate, meaning that the subgrid stress model is dissipating the correct energy from the resolved field.

For \textit{a posteriori} tests, both qualitative and quantitative evaluations are important. However, quantitative metrics are often case dependent. Therefore, a few metrics are introduced that can be used to compare the various subgrid stress models with the filtered DNS solution that work for both forced HIT and channel flow. Note that these are necessary but not sufficient metrics to evaluating a subgrid stress model's quality. The first metric involves computing a sum-absolute error (SAE) between the filtered DNS spectra and the LES spectra:
\begin{equation} \label{eqn13}
SAE_{E} = \int_k |(log(E_{DNS}(k))-log(E_{LES}(k)))| dk \approx \sum_{k} |log(E_{DNS_k})-log(E_{LES_k})|
\end{equation}
where the energy spectra is summed over a specific wavenumber band, with $k$ denoting the wavenumber. The logarithm ensures that differences in the spectra at large wavenumbers can still meaningfully increase this metric. Another metric involves computing the SAE between the filtered DNS and LES correlations (spatial or temporal):
\begin{equation} \label{eqn14}
SAE_{corr} = \int_{\delta} |C_{DNS}(\delta)-C_{LES}(\delta)| d\delta \approx \sum_{\delta} |C_{DNS_{\delta}}-C_{LES_{\delta}}|
\end{equation}
where $C$ denotes the temporal or spatial correlation and $\delta$ represents the spacing (temporal or spatial) used to calculate the value of the correlation function. For channel flow, additional metrics involving the Reynolds Stress profiles can also be developed. One involves computing the SAE over the various Reynolds Stress profiles summed in the wall-normal direction:
\begin{equation} \label{eqn15}
SAE_{R_{xx}} = \int_y |R_{xx_{DNS}}(y)-R_{xx_{LES}}(y)| dy \approx \sum_{y} |R_{xx_{DNS_{y}}}-R_{xx_{LES_{y}}}|
\end{equation}
where $R_{xx}$ represents the Reynolds Stress over the ``xx" direction (ie 11, 22, 33), while $y$ denotes the wall-normal direction. Another quantitative metric for channel flow involves computing the SAE for the mean velocity profiles, summed over the wall-normal direction:
\begin{equation} \label{eqn16}
SAE_{u} = \int_{y} |\overline{u(y)}_{DNS}-\overline{u(y)}_{LES}| dy \approx \sum_{y} |\overline{u(y)}_{DNS}-\overline{u(y)}_{LES}|
\end{equation}
where $y$ once again is the wall-normal direction and $\overline{u(y)}$ is the LES or filtered DNS mean velocity profile (averaged over the homogenous directions, in this case x and z). The above 4 quantitative metrics do not represent a complete characterization of turbulence, but serve as ``first-order" metrics in order to compare various subgrid stress models quantitatively in an \textit{a posteriori} setting. 

\section{\textit{A priori} Results}
Here, neural network training hyperparameters are covered, along with information on various other neural network architectures before discussing \textit{a priori} results. 

\subsection{Neural Network Training Details}
All neural networks are trained using the Adam Optimizer using three learning rates of 0.01, 0.001 and 0.0001, and the best performing model is selected. We note that in training all the neural network models, each neural network performed the best using a learning rate of 0.001. In order to facilitate comparisons between the neural network proposed and other neural network models in literature, a TBNN using an artificial neural network frontend was also trained (labeled as ``TBNN"), as well as a two neural network convolutional Unet architecture introduced by~\citet{Wu_PrF}, which is labeled as ``NN". The ``TBNN", which uses an artificial neural network as a frontend, has around 13200 trainable parameters and uses normalized invariants and a normalized tensor basis. Meanwhile, the ``NN" is a CNN architecture following~\citet{Wu_PrF}, and also uses normalized invariants and a normalized tensor basis as inputs. The S4ND model was trained with the bandlimit parameter set to 0.1. This neural network has inputs that are simplified from the invariants, and is seen in table~\ref{tab:table1}. Appendix A provides a more detailed hyperparameter sweep of the bandlimit parameter, which provides justification for setting the bandlimit parameter to 0.1. Table~\ref{tab:table1} also shows the total number of trainable parameters for each neural network model. As seen, all neural networks are roughly the same size (orders of magnitude for the parameters are similar), allowing for a fair comparison between the various neural network models. To isolate the architectural effects of the S4ND model, a second convolutional neural network, mirroring the exact same architecture as the S4ND model with only the S4ND layers replaced with convolutions is also trained, which is labeled as NN-V. This neural network is also trained on the exact same inputs and outputs and trained the same way as the S4ND model. NN-V is evaluated \textit{a posteriori} and will be compared to all the other models to isolate the architectural effects of S4ND \textit{a posteriori}. For all neural networks, early stopping was employed to ensure that the neural networks do not overfit, with the criteria being that if the validation loss did not decrease for 300 epochs, the neural network weights would be restored to the instance of the lowest validation loss and training would terminate. 

\begin{table}
\caption{\label{tab:table1} Various Neural Network Model Inputs and Trainable Parameters}
\centering
\begin{tabular}{lcc}
\hline
NN Model& Inputs & Trainable Parameters \\\hline
S4ND& $\bf{V}_{norm}$, $\bf{|S|}_{norm}$, $\bf{|R|}_{norm}$ & 9256\\
NN-V& $\bf{V}_{norm}$, $\bf{|S|}_{norm}$, $\bf{|R|}_{norm}$ & 18526\\
NN& Normalized Invariants and Normalized Tensor Basis& 12523\\
TBNN& Normalized Invariants and Normalized Tensor Basis& 13210\\
\hline
\end{tabular}
\end{table}

\subsection{HIT Filtering}
The \textit{a priori} results will be demonstrated on forced HIT. To gain intuition on the amount of filtering introduced by different filter widths, Table~\ref{tab:filter} illustrates the amount of energy filtered as a function of filter width.

\begin{table}
\caption{\label{tab:filter} Filter Widths and Energy Filtered for HIT}
\centering
\begin{tabular}{l|c|c}
$\Delta_g$& Average Energy Filtered & Maximum Observed Energy Filtered \\
1$\Delta_{DNS}$& 0\%& 0\%\\
2$\Delta_{DNS}$& 1\%& 2.1\%\\
4$\Delta_{DNS}$& 2\%& 5.3\%\\
8$\Delta_{DNS}$& 4\%& 12.2\%\\
16$\Delta_{DNS}$& 9.6\%& 23.4\%\\
32$\Delta_{DNS}$& 14.3\%& 40.4\%\\
64$\Delta_{DNS}$& 22.6\%& 56.4\%\\
\end{tabular}
\end{table}

The values are computed by evaluating the average turbulent kinetic energy (TKE) filtered out as compared to the reference DNS simulation over multiple ensembles, as well as the maximum TKE filtered out that was observed over the ensembles. As seen, for small filter widths, only a small amount of energy has been filtered out as compared to the DNS simulation. Therefore, the subgrid stress models do not need to model much of the physics, as most of the flow has been resolved. However, from a filter width of 16 $\Delta_{DNS}$ onward, the energy filtered out becomes increasingly significant. In these situations, the subgrid stress models contribute directly to the accuracy of the simulation, and inaccuracies in the subgrid stress model have a larger effect on the resolved flow. Therefore, the \textit{a priori} results are evaluated on filter widths 16 $\Delta_{DNS}$ or larger.

\subsection{\textit{A Priori} Results}
\begin{figure}
  \centerline{\includegraphics[scale=0.6]{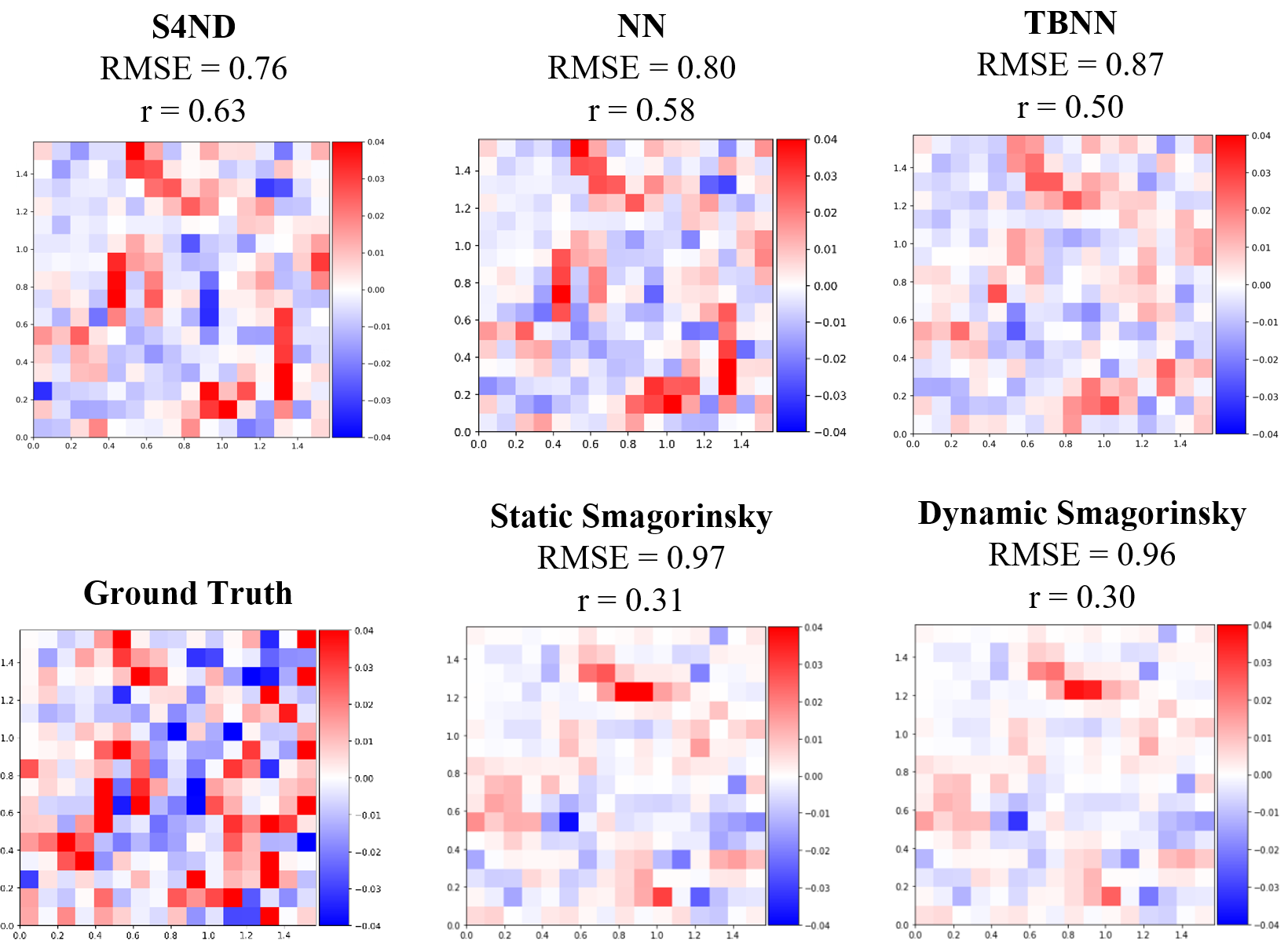}}
  \caption{Spatial distribution of $\tau_{11}$ for a 2D slice of the 3D domain for a filter width of 16 $\Delta_{DNS}$ for the forced HIT case. All numerical values are ensemble averaged over 10 ensembles, while the spatial fields are chosen randomly for visual reference. The letter $r$ denotes the correlation coefficient. The correlation coefficients are highest for S4ND, while the other, less complex neural networks have lower correlation coefficients. The x and y axis correspond to the coordinates in the physical domain.}
\label{fig:16RMSE}
\end{figure}

For \textit{a priori} tests, the S4ND model is compared with the TBNN, NN model, Static Smagorinsky, and Dynamic Smagorinsky models. As defined in the previous sections, the TBNN model refers to a tensor basis neural network with an aritificial neural network frontend, while the NN model refers to the convolutional neural network based model. First, the spatial distribution of the subgrid stress can be compared for a filter width of 16 $\Delta_{DNS}$, which is shown in figure~\ref{fig:16RMSE}. 

From figure~\ref{fig:16RMSE}, all the neural network models are far better than the Smagorinsky models in capturing the overall correlation with the filtered DNS subgrid stress, as evidenced by both the higher correlation coefficient and the lower RMSE. The trend among neural networks is also of note, as the simplest neural network (TBNN) has the worse RMSE and correlation coefficient out of all the neural network models, yet it has the most trainable parameters. Meanwhile, the more complex neural network models (such as the NN model and the S4ND model) perform far better. Due to the coarse-grid regime that the models are evaluated in, the traditional models, as well as the TBNN model perform poorly, and the more complex models show improvement on top of these models. However, even the more advanced models are not perfect when compared to the ground truth, which is expected in the coarse-grid regime, where a significant portion of the energy has been filtered out~\citep{Wu_PrF}. In addition, the interscale energy transfer field, $-\tau_{ij}\Bar{S}_{ij}$ is also shown.  

\begin{figure}
  \centerline{\includegraphics[scale=0.6]{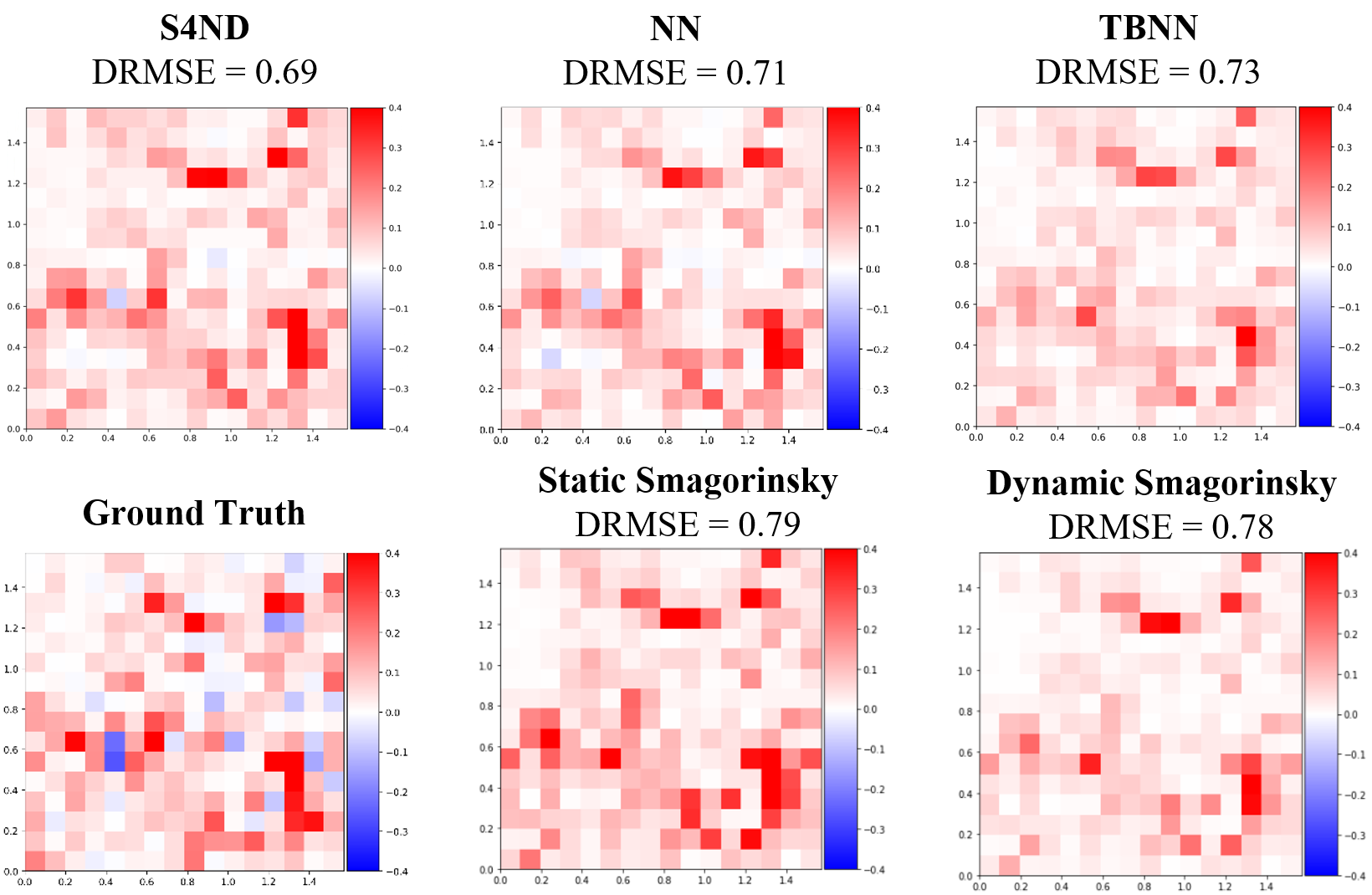}}
  \caption{Spatial distribution of $-\tau_{ij}\Bar{S}_{ij}$ for a 2D slice of the 3D domain for a filter width of 16 $\Delta_{DNS}$. All numerical values are ensemble averaged over 10 ensembles, while the spatial fields are chosen randomly for visual reference. The x and y axis correspond to the coordinates in the physical domain.}
\label{fig:16DRMSE}
\end{figure}

Overall, as seen from figure~\ref{fig:16RMSE} and figure~\ref{fig:16DRMSE}, visually, the TBNN field is more diffuse as compared to the S4ND and NN fields, but they are quite similar. In fact, all the interscale energy transfer fields look relatively similar visually, and the DRMSE are not too far off from one another. This suggests that the neural network models are able to represent the energetics of turbulence faithfully \textit{a priori}, and therefore have a good chance of being stable \textit{a posteriori}, as the interscale energy transfer field is similar to the Static and Dynamic Smagorinsky models, which are known to be stabilizing towards a LES simulation. Despite the spatial distribution of $-\tau_{ij}\Bar{S}_{ij}$ looking visually similar to the Smagorinsky models, the neural network models are still able to predict actual subgrid stress tensor spatial distributions that are more in line with the ground truth seen in Figure~\ref{fig:16RMSE}. This suggests that the neural network model performance \textit{a priori}, when coupled with the two-term loss function, are not behaving like ``structural models", which oftentimes sacrifice stability (prediction of energy dissipation) for higher correlation with the subgrid stress tensor.

\begin{figure}
\centering
     \begin{subfigure}{0.375\textwidth}
         \centering
         \includegraphics[width=\textwidth]{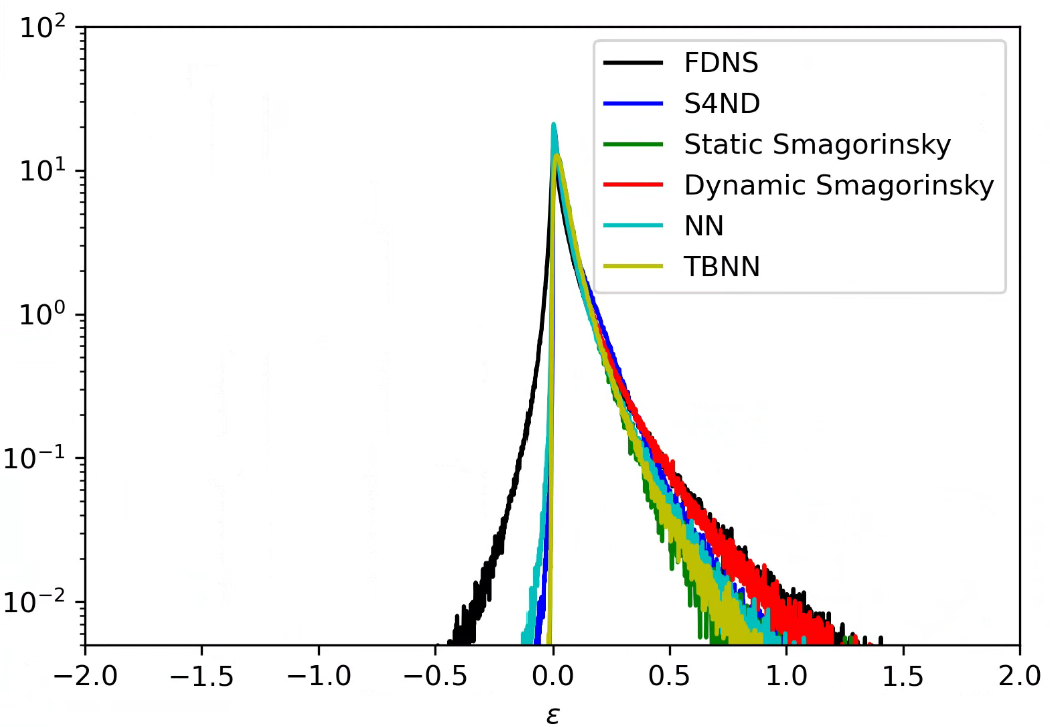}
         \caption{PDF of $\epsilon$.}
         \label{fig:PDF_eps}
     \end{subfigure}
     \hspace{1em}
     \begin{subfigure}{0.375\textwidth}
         \centering
         \includegraphics[width=\textwidth]{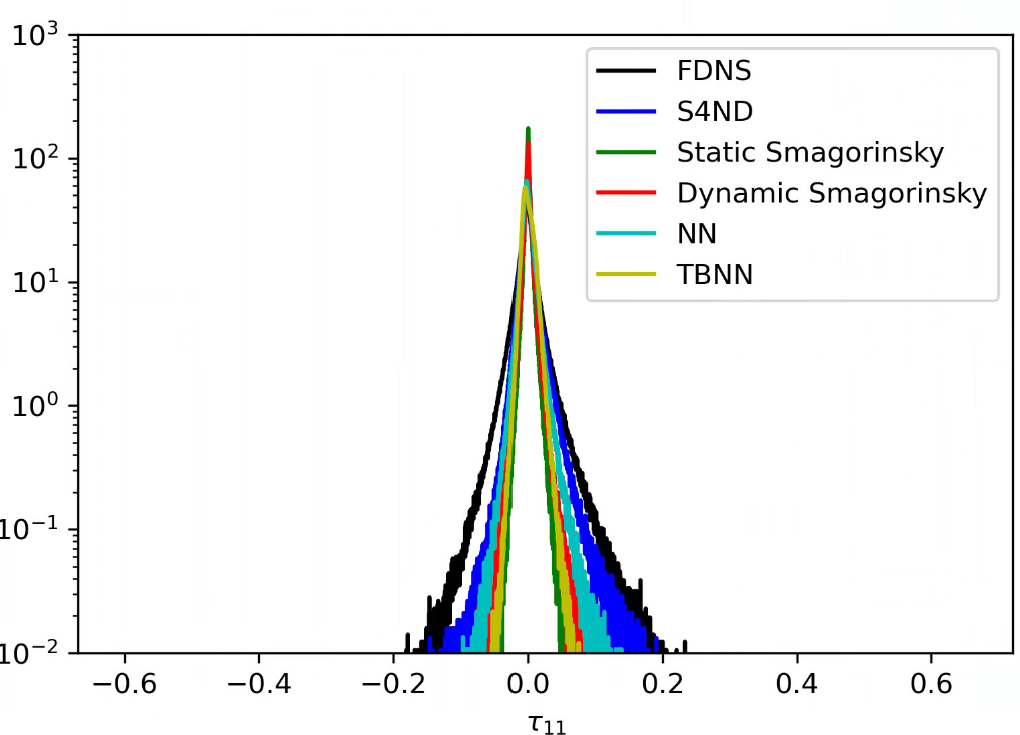}
         \caption{PDF of $\tau_{11}$}
         \label{fig:PFD_T11}
     \end{subfigure}
     \caption{Probability Density Functions for 16 $\Delta_{DNS}$ forced HIT. As seen, the S4ND model is more able to accurately capture the distribution of $\tau_{11}$ as compared to other models. The S4ND model and NN model (convolution-based) both capture more backscatter than traditional models, although it is still less than FDNS.}
\end{figure}

In addition, probability density functions (PDFs) showing the overall distribution associated with various models are shown in figure~\ref{fig:PDF_eps} and figure~\ref{fig:PFD_T11}. From figure~\ref{fig:16DRMSE}, the overall spatial distribution of $-\tau_{ij}\Bar{S}_{ij}$ is similar across most models, and this is corroborated when evaluating the PDF of $\epsilon$. From figure~\ref{fig:PDF_eps}, one can see that all the models do a good job representing the forward scatter (energy transfer to smaller scales) while they do worse on the back-scatter (energy transfer from small to large scales). However, both the NN and S4ND models have more back-scatter than the traditional models, which barely even have any values smaller than zero. From figure~\ref{fig:PFD_T11}, the overall conclusions drawn from evaluating the spatial distribution are confirmed. Overall,the S4ND model is the closest to the filtered DNS PDF, followed by the CNN-based neural network model (NN), and then followed by the TBNN and the traditional models. Note that in the aggressive filtering regime, the TBNN behaves quite poorly, oftentimes being no better than the Dynamic Smagorinsky model when comparing PDFs. As seen from others in literature, in general, simple ANN-based neural network models can still perform well when compared to traditional models when the filtering is less aggressive~\citep{Choi_2024, xie2}. 

From all these results, while adding complexity to the neural network model can help increase performance, as seen since the S4ND and NN models have better performance as compared to the TBNN model, in a setting where the filter width is the same as the training data, it is hard to distinguish between whether the S4ND or NN model is better. This is to be expected, as the continuous convolution kernels that S4ND uses should perform better in extrapolative settings, not necessarily in situations that are part of the training data. 

\subsection{Interpolative and Extrapolative \textit{A Priori} Results}

While the performance when the filter width is the same as the training set is good, one also needs to evaluate the neural network on different filter widths. First, a less stringent test is applied, where the neural network models, trained on $16 \Delta_{DNS}$ and $32 \Delta_{DNS}$ data, is evaluated on a filter width of $24 \Delta_{DNS}$. 

 \begin{figure}
  \centerline{\includegraphics[scale=0.6]{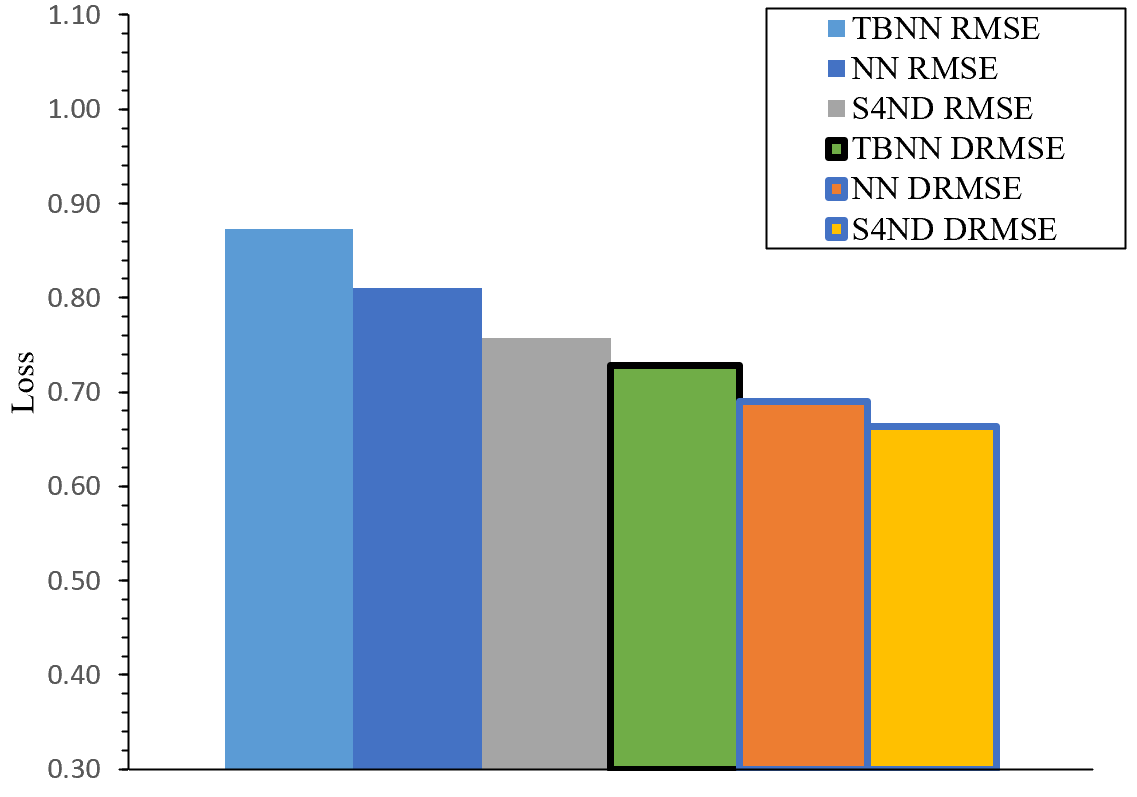}}
  \caption{RMSE and DRMSE for various neural network models evaluated on $24 \Delta_{DNS}$ HIT data. As seen, the S4ND is able to handle this case the best (no increase in the RMSE and DRMSE as compared to the test set at other filter widths), followed by the NN model and then the traditional TBNN, which shows a noticeable increase in both the RMSE and DRMSE when compared to the S4ND model.}
\label{fig:24DeltaBars}
\end{figure}

\begin{figure}
  \centerline{\includegraphics[scale=0.6]{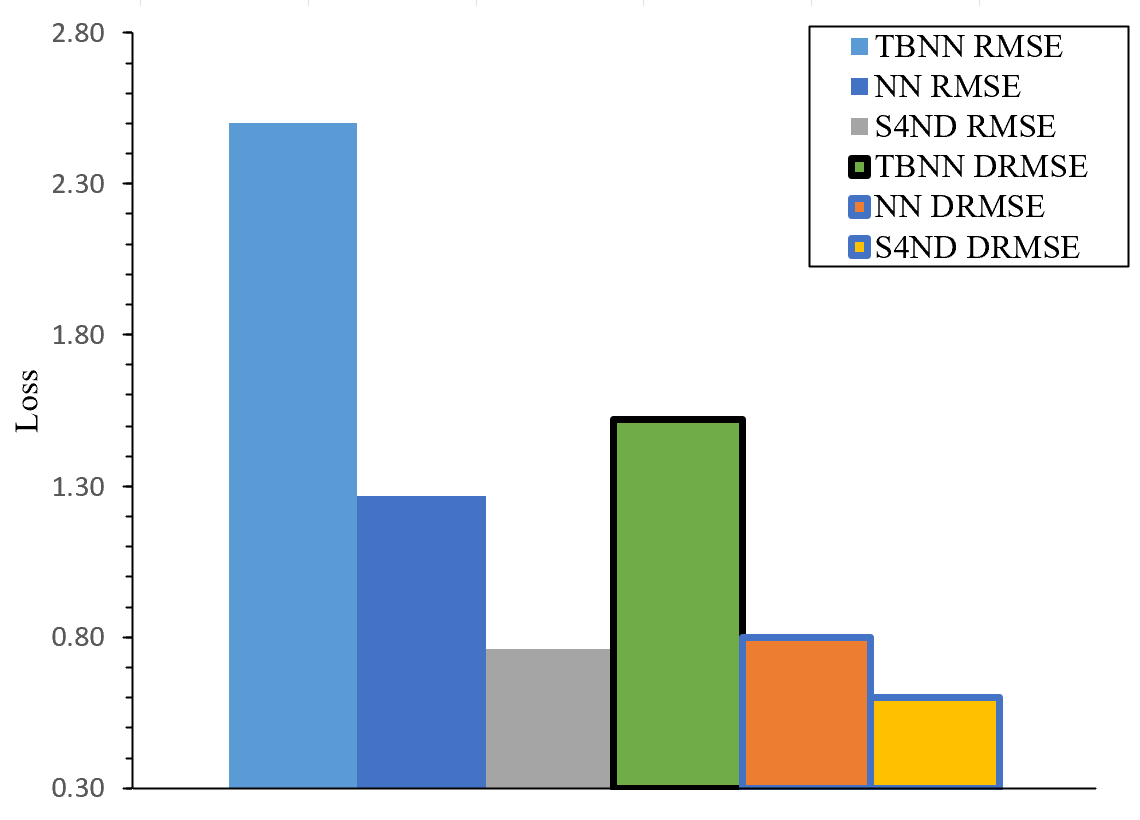}}
  \caption{RMSE and DRMSE for various neural network models evaluated on $64 \Delta_{DNS}$ HIT data. As seen, the S4ND model is able to extrapolate to this more aggressive filter width without any significant increase in the RMSE and DRMSE, while the NN model has around a factor of 2 increase in the RMSE and a factor of 1.3 in DRMSE. Meanwhile, the traditional TBNN has a loss increase of almost a factor of 4 for RMSE and a loss increase of a factor of 2.5 for DRMSE.}
\label{fig:64DeltaBars}
\end{figure}

From figure~\ref{fig:24DeltaBars}, the overall trends hold, where the ANN-based model (TBNN) is the worst-performing, followed then by the NN and then the S4ND models for both RMSE and DRMSE. Overall, all the neural network models are able to interpolate to this ``in-between" filter width, since the training data consists of $16 \Delta_{DNS}$ and $32 \Delta_{DNS}$ data. A more challenging case is evaluating the neural network models on $64 \Delta_{DNS}$ data, as even more energy is filtered out at this aggressive filter width, meaning that the neural network models must learn scale-similarity in order to extrapolate to this more aggressive filter width. These results are shown in figure~\ref{fig:64DeltaBars}. 

Figure~\ref{fig:64DeltaBars} shows that the only neural network model that can successfully extrapolate to this filter width is the S4ND model. The TBNN model does extremely poorly, as the RMSE loss is now multiple times higher than the test set loss at an in-training set filter width, completely failing to generalize. Even the NN model is unable to generalize, as the loss is around twice as high as compared to the in-training set filter width loss. Meanwhile, the S4ND model shows no increase in the loss as compared to filter widths that are within the training set, being the only model that can extrapolate successfully to a setting where the filter width is more aggressive than the training set it was trained on. Referring to table~\ref{tab:filter}, this extrapolative case corresponds to a setting where on average, 22.6 percent of the energy has been filtered out as compared to DNS. In particular, the DRMSE is an important quantity. While the RMSE is important, as it is an indirect way to ensure that the momentum equation is closed properly (since the divergence of the subgrid stress tensor is taken in the momentum equation), if the energy equation for the flow were written, the term $\tau_{ij}S_{ij}$ arises directly in the equation. The DRMSE is a direct measure of how well the neural network represents this as compared to the ground truth data. Therefore, when the S4ND model is able to produce DRMSE values on a coarser grid that has no performance degradation as compared to the training set, this suggests that the model is correctly representing the interscale energy transfer for various grid resolutions (filter cutoffs). Capturing this rate of energy transfer correctly is a necessary condition for the model to remain stable and accurate \textit{a posteriori}, since incorrect energy transfer rates can lead to energy pile-up at large wavenumbers or over-dissipation of resolved structures.

\section{\textit{A Posteriori} Results}
All neural network models are integrated into an in-house code, PadeOps~\citep{ghate}, to be evaluated in an \textit{a posteriori} setting. In these sections, two traditional models (the Static Smagorinsky and Sigma models) are compared with four neural network models (S4ND, TBNN, NN, and NN-V as defined in the earlier section). These two traditional models were chosen as it represents one of the earliest subgrid stress models and a modern subgrid stress model that has many desirable properties such as correct behavior near-wall. The Static Smagorinsky model is used with a coefficient set to $C = 0.18$, while the Sigma model is also run with a coefficient set to $C=1.3$. These are all within the ranges commonly used for these two models~\citep{nicoud2011using, smagorinsky1963general, meyers2006model}. The Sigma model was chosen as the coefficient range works for many different flows and does not require clipping for channel flow, whereas models such as the Dynamic Smagorinsky model may need clipping~\citep{meneveau1996lagrangian}. Four neural network models are compared to highlight the benefits gained from increasing the complexity of neural network models, as well as to highlight the architectural effects of the S4ND model. For all cases, using no subgrid stress model results in the simulations producing NaNs.

\subsection{In-training Set Simulation Results}
First, all models are evaluated on LES simulations with grid resolutions corresponding to in-training set filter width cases. 

\subsubsection{Forced HIT at $16 \Delta_{DNS}$}
\begin{figure}
  \centerline{\includegraphics[scale=0.65]{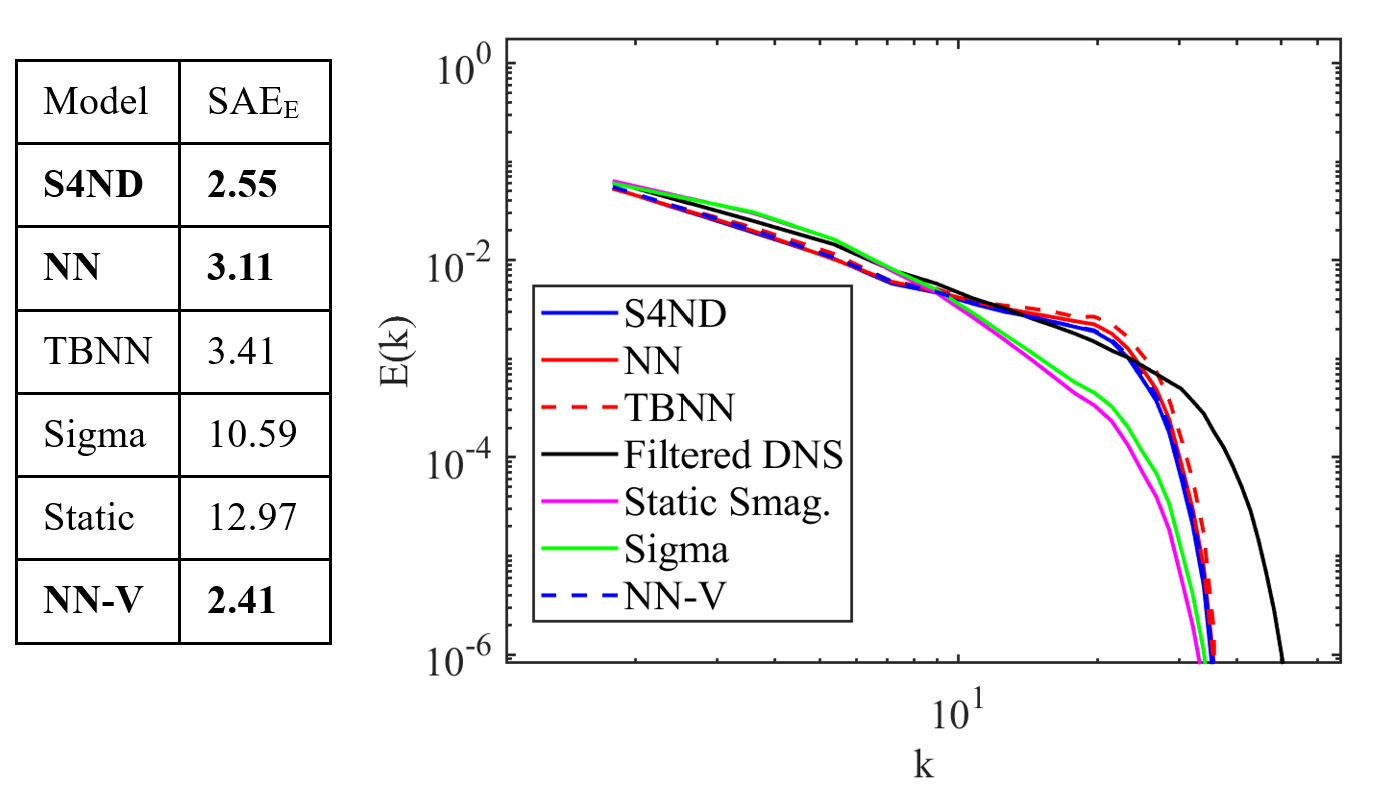}}
  \caption{Forced HIT spectra for various SGS models \textit{a posteriori} on a grid size corresponding to a filter width of 16 $\Delta_{DNS}$.}
\label{fig:16HITSpectra}
\end{figure}

The forced HIT simulation is run with a RK4-5 time stepping scheme~\citep{Bogacki} on a $64^3$ grid, which corresponds to a $16\Delta_{DNS}$ filter width. The spatial discretization scheme is Fourier collocation, possible since the boundary conditions are triply periodic, with a domain length of $2\pi$ in each direction. The forcing scheme forces the energy in the wavenumber band below 2 to be constant, and the non-dimensional dissipation, $\epsilon$ is set to 0.103 for all simulations. Due to Fourier collocation, $\frac{2}{3}$ dealiasing is applied. For this simulation, all cases have a Taylor Reynolds number ($Re_{\lambda}$) of approximately 433. As seen in figure~\ref{fig:16HITSpectra}, all the LES simulations with various models have reasonable spectra. 
The eddy viscosity models, as expected, are more dissipative as compared to filtered DNS, while the neural network models better approximate the overall shape of the spectra but are slightly under-dissipative. From the ``bump" in the spectra, the TBNN subgrid stress model is the most under-dissipative, followed by the NN model, then the S4ND and NN-V model. The $SAE_{E}$ table validates this, as this metric suggests that the S4ND model and the NN-V model are the closest to the filtered DNS spectra, as lower $SAE_{E}$ values are closer to filtered DNS statistical quantities. Overall, the NN-V and S4ND models have spectra that are extremely similar, suggesting that S4ND, even with bandlimiting, is not significantly reducing the performance of the neural network in forced HIT.

\begin{figure}
  \centerline{\includegraphics[scale=0.55]{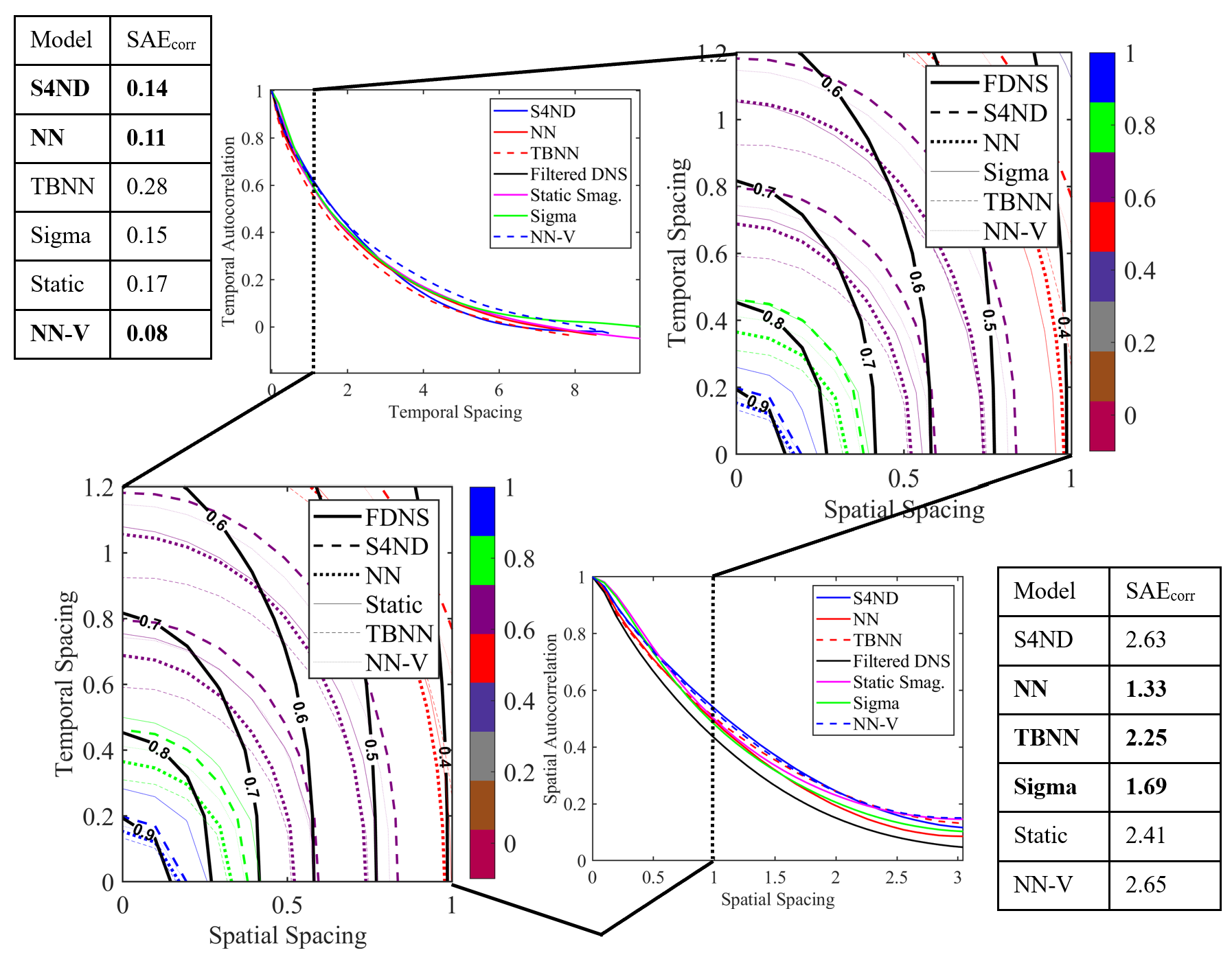}}
  \caption{Forced HIT spatial, temporal, and space-time correlations for various SGS models \textit{a posteriori} on a grid size corresponding to a filter width of 16 $\Delta_{DNS}$.}
\label{fig:16HITCorrelations}
\end{figure}

From figure~\ref{fig:16HITCorrelations}, all neural network models are fairly good at predicting the temporal decorrelation, as they all follow the overall expected decay shape with the S4ND, NN-V, and NN neural network models being slightly more accurate. When evaluating the spatial correlations, all models are fairly accurate for small spacings, but the S4ND and NN-V models are less accurate towards the larger spacings compared to both the Sigma and NN models. While the overall spatial correlation shapes are preserved for all models, as seen from the $SAE_{corr}$ for the spatial decorrelation, the S4ND and NN-V models have the largest error compared to the filtered DNS decorrelation. This suggests that the performance degradation for S4ND is not necessarily due to the state space model, but is likely a product of the normalization and inputs. From the space-time correlations, one can conclude that all the neural network models are far more accurate than either of the traditional models, especially for small spatial and temporal spacings. It is hard to determine which neural network model is more superior when it comes to space-time correlations. Overall, the benefits of neural network complexity is not clear in this set of simulations, as sometimes the S4ND model is better, while other times (such as when evaluating spatial correlations) it is worse. 
However, this simulation is run with an in-training set filter width of $16\Delta_{DNS}$, where it has already been established that one does not expect much difference between the S4ND and the CNN-based models. This is reflected in this set of simulations as the S4ND model is sometimes better, and sometimes worse than the NN model (although it is quite consistent with the NN-V model due to the same input/output normalizations). The S4ND model's strength, as seen \textit{a priori}, is in extrapolative scenarios. 

\subsubsection{Channel Flow at $16 \Delta_{DNS}$}
Similarly to the HIT cases, the channel flow simulation is also discretized with Fourier Collocation in the homogeneous directions, while a 6th order compact scheme is used in the wall-normal direction. The wall model used is not data-driven, and instead follows from the wall model used by~\citet{bou2005scale}. The simulation grid size is 128x96x32, while the domain size is [$8\pi$, $3\pi$, 2], with the streamwise velocity pointing in the ``first" direction and the wall-parallel direction being the ``second" direction. All simulations run in this section correspond to a filter width of $16 \Delta_{DNS}$ with a friction Reynolds number of $Re_{\tau} = 1000$, with $Re_{\tau} = \frac{u_{\tau}H}{\nu}$, where $u_{\tau}$ is the friction velocity, $H$ is the channel half-height, and $\nu$ is the kinematic viscosity. 
\begin{figure}
  \centerline{\includegraphics[scale=0.65]{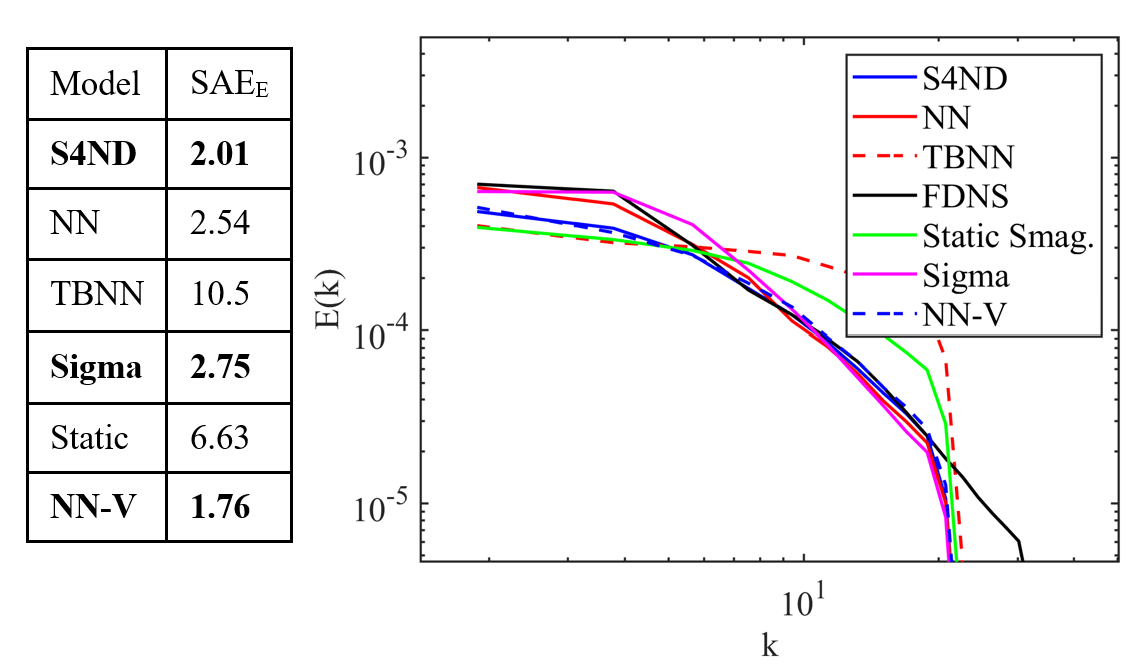}}
  \caption{Channel flow streamwise spectra at $z/H = 0.46875$ for various SGS models compared to filtered DNS. Both the Static Smagorinsky model and TBNN fail at predicting the overall spectra shape.}
\label{fig:16ChannelSpectra}
\end{figure}
As seen in figure~\ref{fig:16ChannelSpectra}, in a more complex flow (as compared to HIT), the utility of more complicated neural network models is starting to show. The TBNN is unable to predict even the spectra properly, and is even worse than Static Smagorinsky. Meanwhile the overall shapes associated with the NN and S4ND model is significantly closer to that of the filtered DNS profile. While the S4ND model is slightly over-dissipative at small scales, it follows the inertial range of the spectra better than the NN model when comparing these two profiles with the filtered DNS profile. However, as seen with NN-V, the S4ND model being over-dissipative is not due to the state space model, as the NN-V model also follows a similar spectra, suggesting that the differences in the spectra are more due to normalization and different inputs into the neural network. Overall, from the $SAE_{E}$ table, one can see that the S4ND model, NN-V, and the Sigma model are more similar to filtered DNS as compared to the other models. 
\begin{figure}
\centering
     \begin{subfigure}{0.375\textwidth}
         \centering
         \includegraphics[width=\textwidth]{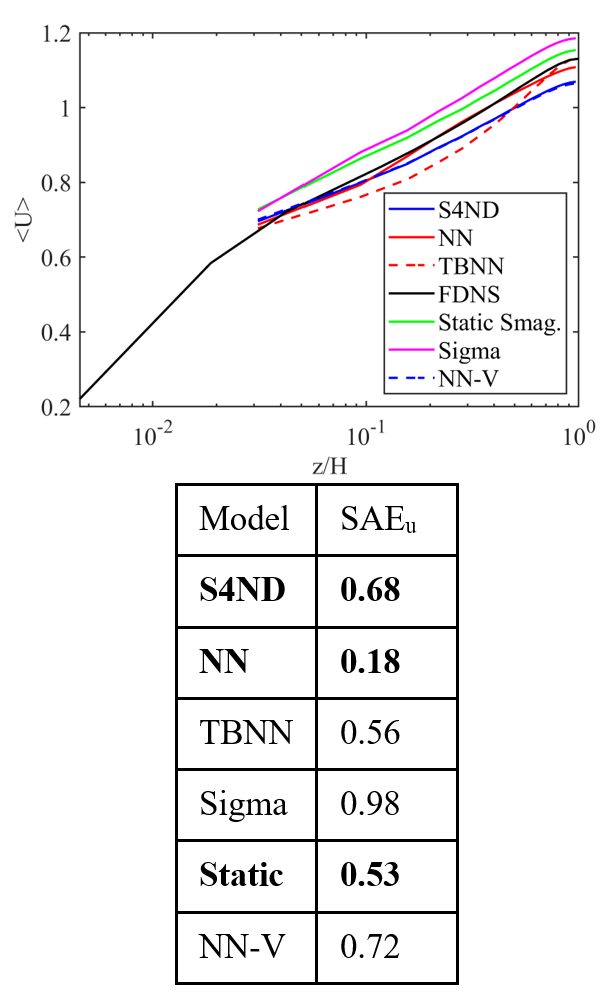}
         \caption{Mean streamwise velocity profiles.}
         \label{fig:16ChannelMean}
     \end{subfigure}
     \hspace{1em}
     \begin{subfigure}{0.38\textwidth}
         \centering
         \includegraphics[width=\textwidth]{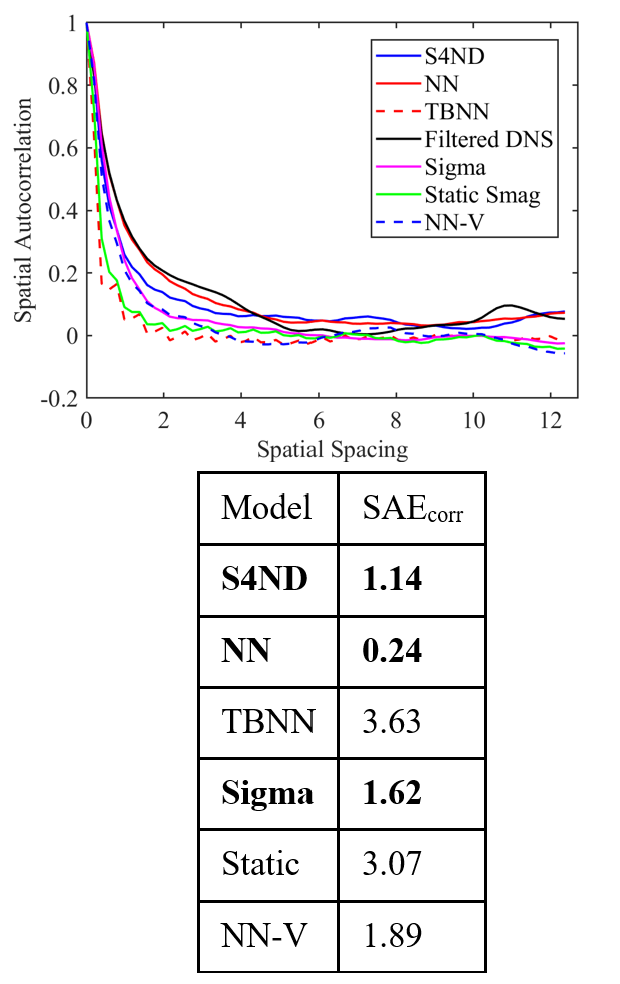}
         \caption{Spatial correlations at $z/H = 0.46875$.}
         \label{fig:16ChannelSpatial}
     \end{subfigure}
     \caption{Channel flow macro-scale quantities analyzed for various SGS models. The mean velocity profiles and spatial correlations for the S4ND and NN model look the closest compared to filtered DNS. As seen, the TBNN completely fails in this setting, predicting jagged spatial correlations.}
\end{figure}

From~\crefrange{fig:16ChannelMean}{fig:16ChannelSpatial}, one can see a similar story when comparing the TBNN model with its more complex counterparts. The mean profiles associated with the TBNN model are inaccurate as compared to filtered DNS, not even predicting the shape properly, while the other three neural network models perform far better. In fact, the spatial correlations associated with the TBNN model at aggressive grid spacings highlights that it is completely inaccurate, as the spatial correlations become jagged. Meanwhile, the mean profiles predicted by all the other models that are not the TBNN are reasonable, with the NN model being more accurate as compared to the S4ND model. The NN-V and S4ND models have very similar mean profiles, which highlights that the S4ND architecture and the CNN architectures have relatively similar performance when the training set is still in filter-width. For the spatial correlations, both the S4ND and the NN model are far better than the traditional models, with the NN model being more accurate. When comparing NN and NN-V, one can isolate the effects of the normalization scheme and inputs. The improvements associated with the NN model is due to the inputs and normalization scheme used, not the architecture as NN-V has the same exact architecture as the NN model. From figure~\ref{fig:16ChannelSpatial}, it is seen that the NN-V model has worse spatial correlations as compared to the S4ND model, which highlights that even for an in training set filter width, when comparing across the same normalizations, the convolutional models can start to lose performance. This is because channel flow operates on anisotropic grids (different $dx, dy, dz$ spacings, but constant spacing in each direction), while convolutional based models implicitly assume isotropic grids (HIT grids). This means that for anisotropic grids, the convolutional models will not perform as well when all things are considered equal (such as same normalization and same inputs).

\begin{figure}
\centering
     \begin{subfigure}{0.27\textwidth}
         \centering
         \includegraphics[width=\textwidth]{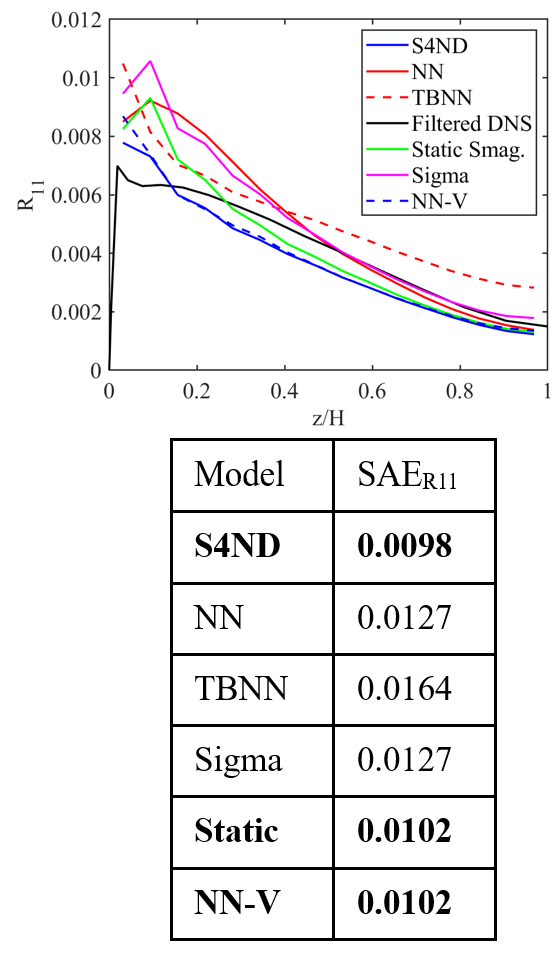}
         \caption{$R_{11}$ varying as a function of wall normal distance.}
         \label{fig:16ChannelR11}
     \end{subfigure}
     \hspace{1em}
     \begin{subfigure}{0.27\textwidth}
         \centering
         \includegraphics[width=\textwidth]{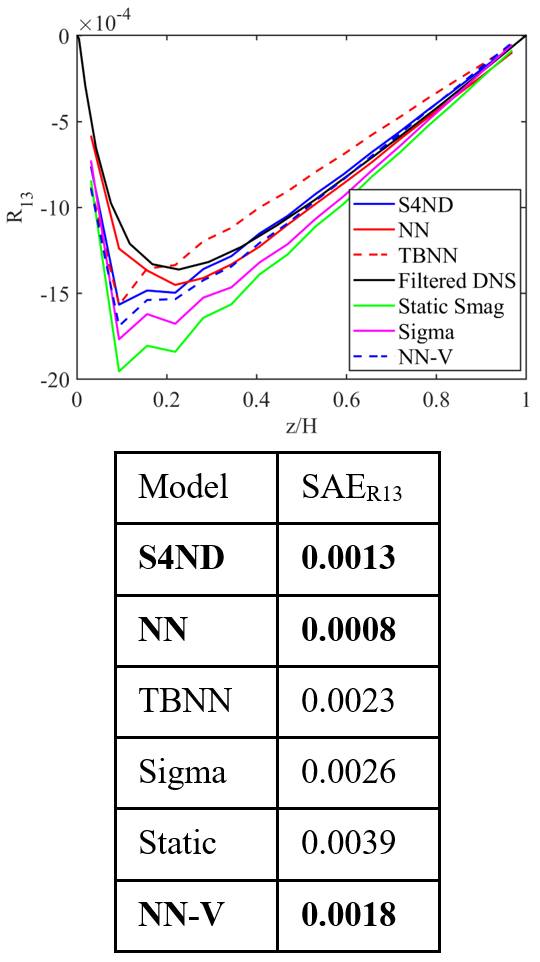}
         \caption{$R_{13}$ varying as a function of wall normal distance.}
         \label{fig:16ChannelR13}
     \end{subfigure}
     \hspace{1em}
     \begin{subfigure}{0.28\textwidth}
         \centering
         \includegraphics[width=\textwidth]{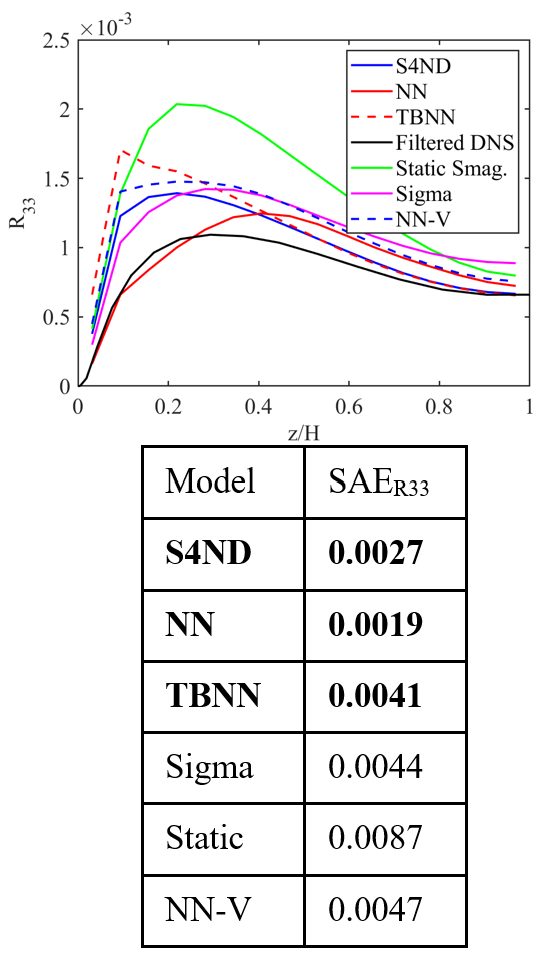}
         \caption{$R_{33}$ varying as a function of wall normal distance.}
         \label{fig:16ChannelR33}
     \end{subfigure}
     \caption{Channel flow Reynolds stress profiles for various SGS models at a filter width corresponding to 16 $\Delta_{DNS}$.}
\end{figure}

Reynolds stress profiles are also plotted, which are seen in~\crefrange{fig:16ChannelR11}{fig:16ChannelR33}. Overall, both the S4ND model and NN model are more accurate compared to the other models, with the S4ND model and NN model trading off in accuracy. When comparing the NN model to the S4ND model, it would seem like for channel flow at this resolution, the NN model is slightly better than the S4ND model. This is seen in the $R_{13}$ profile, as the overall shape of the Reynolds Stress profile is better captured by this model (although S4ND is better at capturing the $R_{11}$ profile). A different trend emerges when comparing the S4ND and NN-V model. In all cases, the NN-V model is either slightly worse ($R_{11}$) or worse than the S4ND model ($R_{13}, R_{33}$). This suggests that when accounting only for architectural effects, the S4ND model provides more accuracy on anisotropic grids as the NN-V model, which uses discrete convolutions. Since discrete convolutions assume isotropic grids, they will suffer a decrease in accuracy in flows with anisotropic grids such as channel flow when compared to the S4ND model.

While the NN-based model (with different inputs and normalization) is indeed slightly better than the S4ND model at this resolution, LES grid spacing is ultimately determined by the user, and therefore the models must also be evaluated on an out-of-training set filter width. As seen in~\cite{nguyen2022s4nd}, the S4ND model, through the use of bandlimiting, explicitly learns smoother kernels which would generalize better. However, this is an implicit form of regularization that will slightly decrease the performance of the S4ND model on filter widths that it was trained on. In the in-training set filter width \textit{a posteriori} results for HIT and channel flow, the bandlimiting used actually doesn't decrease the performance of the S4ND model. For HIT, the S4ND model is on par with NN-V and for channel flow, since it accepts the different grid spacings in each dimension, the S4ND model is actually better than NN-V. This suggests that even though the S4ND model is biasing more towards extrapolation by regularizing the continuous convolution kernel to be more smooth, the larger scales of the flow that are captured with the regularized convolution kernel are still enough to infer the subgrid scales \textit{a posteriori}.

\subsection{Filter Width Extrapolation Results}
Now, all the models are evaluated on an extrapolative filter width that is out of the training set. This will test the generality of the neural network models to see if true ``scale-similarity" has been learned. For true scale-similarity to be learned, the neural network must produce results on coarser grids that are reasonable for both HIT and channel flow. Given that the TBNN model was completely inaccurate for channel flow already (shown in figure~\ref{fig:16ChannelSpatial}), this section focuses on evaluating the more complex neural network models and comparing them to the traditional models. 

\subsubsection{Forced HIT at $64 \Delta_{DNS}$}
\begin{figure}
  \centerline{\includegraphics[scale=0.6]{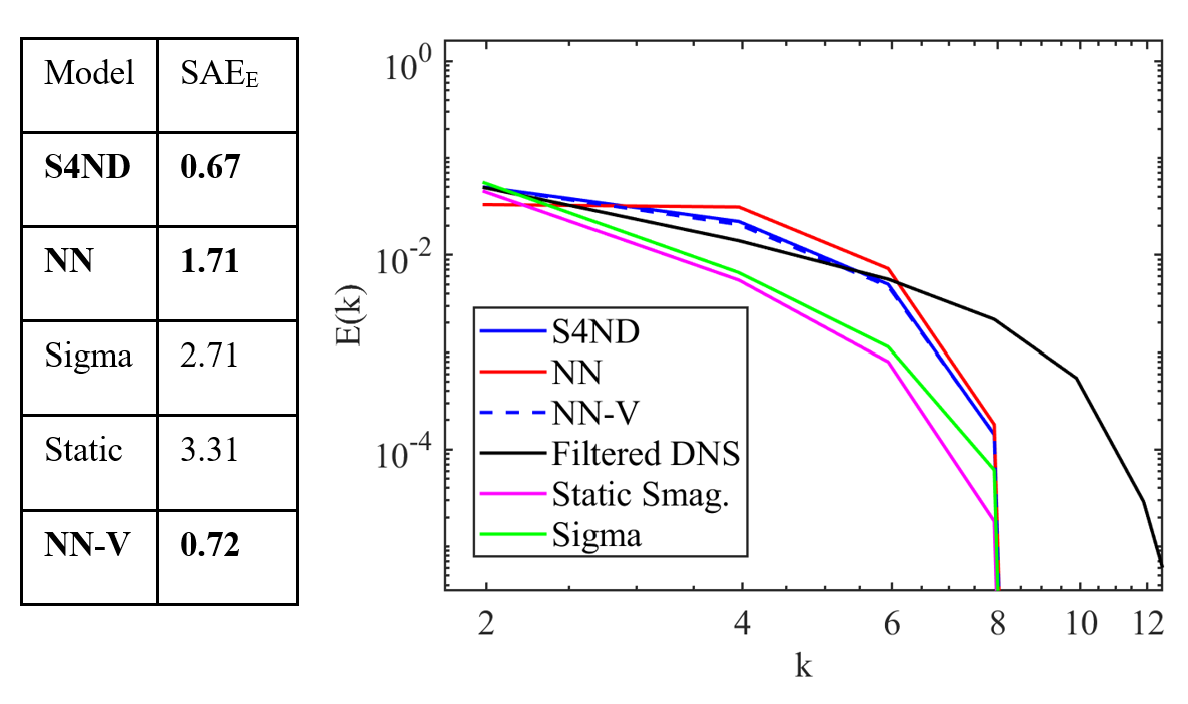}}
  \caption{Spectra for various SGS models on a grid corresponding to an extrapolative filter width of $64 \Delta_{DNS}$.}
\label{fig:64HITSpectra}
\end{figure}

Now, fixing all the previous simulation parameters to be identical to the $16 \Delta_{DNS}$ HIT simulation except for grid spacing, all models are run on a significantly coarser LES where the number of points is $16^3$, corresponding to a filter width of $64 \Delta_{DNS}$. The spectra is shown in figure~\ref{fig:64HITSpectra}. 

\begin{figure}
  \centerline{\includegraphics[scale=0.6]{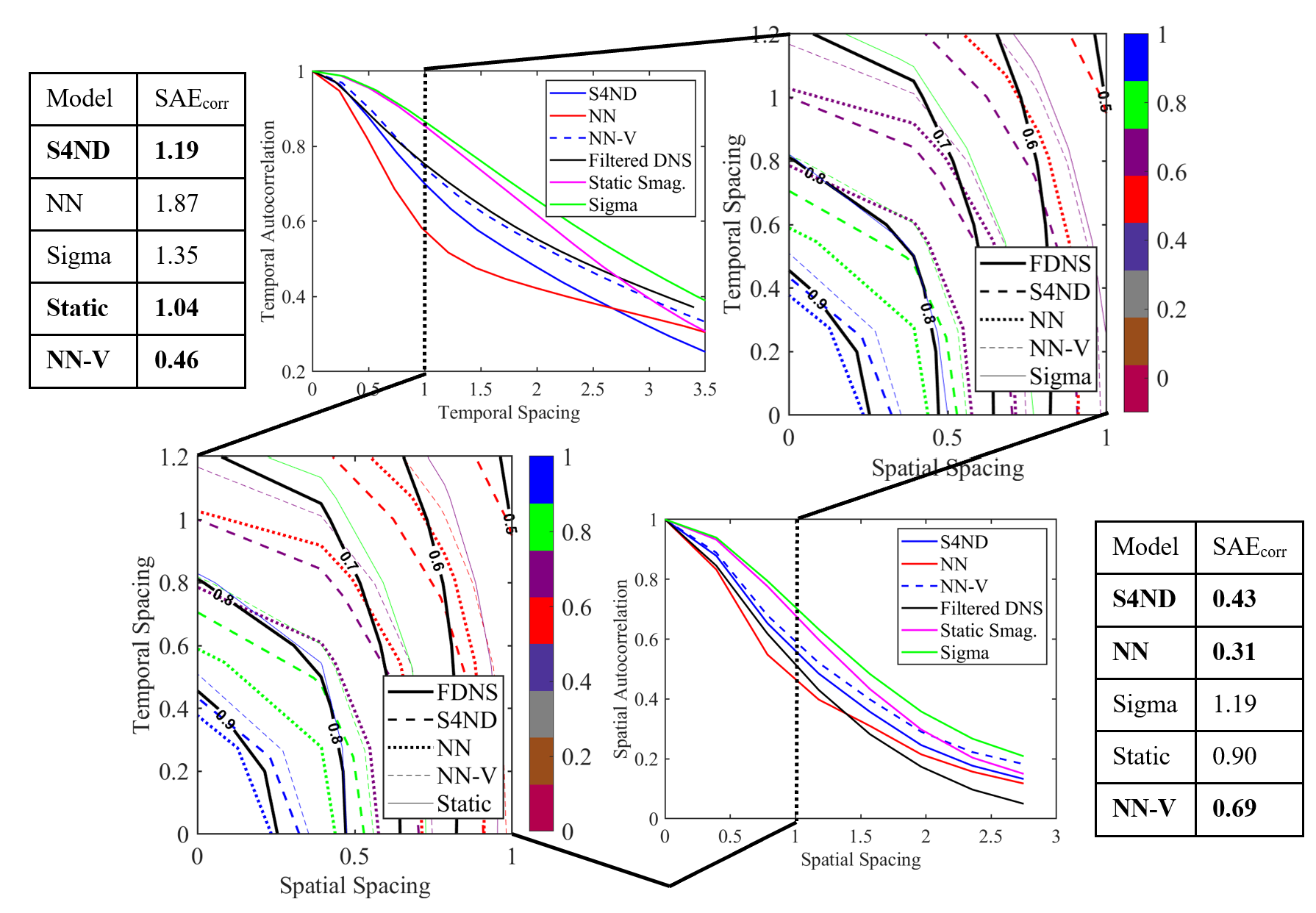}}
  \caption{Forced HIT spatial, temporal, and space-time correlations for various SGS models on a grid corresponding to an extrapolative filter width of $64 \Delta_{DNS}$.}
\label{fig:64HITCorrelations}
\end{figure}

From the spectra, one can conclude that the Sigma and Static Smagorinsky models, consistent with the in-training set filter widths, are still over-dissipative. In addition, the S4ND model produces more accurate spectra as compared to the NN model. Furthermore, the S4ND and NN-V models produce similar spectra. Overall, the spectra suggest that the data-driven models are better at predicting the filtered DNS spectra, even when extrapolating, as compared to the traditional models. Furthermore, additional quantities such as correlations are shown in figure~\ref{fig:64HITCorrelations}.

At this more aggressive grid spacing, the temporal autocorrelations suggest that the S4ND model is far better at extrapolating than the NN-based model, and outperforms both traditional models. First, the traditional models are unable to predict the initial temporal decay properly, significantly overpredicting the temporal autocorrelation. Meanwhile, the NN model decorrelates far too quickly, and then decorrelates far too slowly as the temporal spacing increases. When comparing NN-V with S4ND, the temporal correlations for NN-V are surprisingly better than S4ND. This suggests that with this output normalization, some sense of scale-similarity is ``baked-in". In addition, since the HIT grid is isotropic, and the models are trained on two grid resolutions, the discrete convolutions in NN-V are able to learn a sense of scale similarity. This also highlights the importance of output normalizations and inputs, as the NN model, with different inputs and output normalization, was unable to extrapolate while the NN-V model was able to, despite both using convolutional neural networks. In general, the top three lowest $SAE_{corr}$ include the S4ND, NN-V, and Static Smagorinsky models. However, both quantitative and qualitative analysis must be used, as $SAE_{corr}$ for the Static Smagorinsky model is lower than the S4ND model, but this can be attributed to the Static Smagorinsky model crossing the filtered DNS line and thus the integrated distance between the two curves is smaller due to this crossover. As seen from the temporal decorrelation graph, the overall shape of the Static Smagorinsky correlation is quite different from filtered DNS. 

For the spatial correlations, when comparing the S4ND model with the NN model and traditional models, one can see that the S4ND model produces the most similar decorrelation shape to the filtered DNS profile. Meanwhile, the $SAE_{corr}$ for the NN model is artificially lowered due to the line crossing the filtered DNS line. When comparing the S4ND model and the NN-V model, one can see that the S4ND model is more accurate than the NN-V model. For the spatial correlations, the S4ND model is the most accurate out of all the models. From the space-time correlations, one can conclude that the traditional models completely fail in predicting the correlations, as the black lines denote the predicted contour for filtered DNS, and the colored lines denote the contours associated with a particular SGS model. For example, the blue contours denote the 0.9 contours for all the SGS models. Meanwhile, the traditional model ``thin blue contours" in the plots are extremely off, showing a significant overprediction of the correlation as compared to both the neural network models. While both the NN and S4ND model do well, the NN-V model performs a little worse, especially for small decorrelation scales. This is seen as it is slower to decorrelate than the S4ND model, as the blue contour is further away from the 0.9 filtered DNS contour. Thus, one can see that the S4ND model, with the continuous kernel and bandlimiting, overall produces more accurate space-time correlations and spatial decorrelations as compared to all the other SGS models evaluated. Meanwhile, while the NN-V model produces good temporal decorrelations, it still suffers in accuracy when comparing the small scale space-time decorrelations and the overall spatial autocorrelations, suggesting that S4ND's architectural effects are directly responsible for this increase in accuracy. By comparing NN and NN-V, one can also conclude that correct inputs and output normalization is essential for scaling to coarser grid spacings.

\subsubsection{Channel Flow at $64 \Delta_{DNS}$}
\begin{figure}
  \centerline{\includegraphics[scale=0.8]{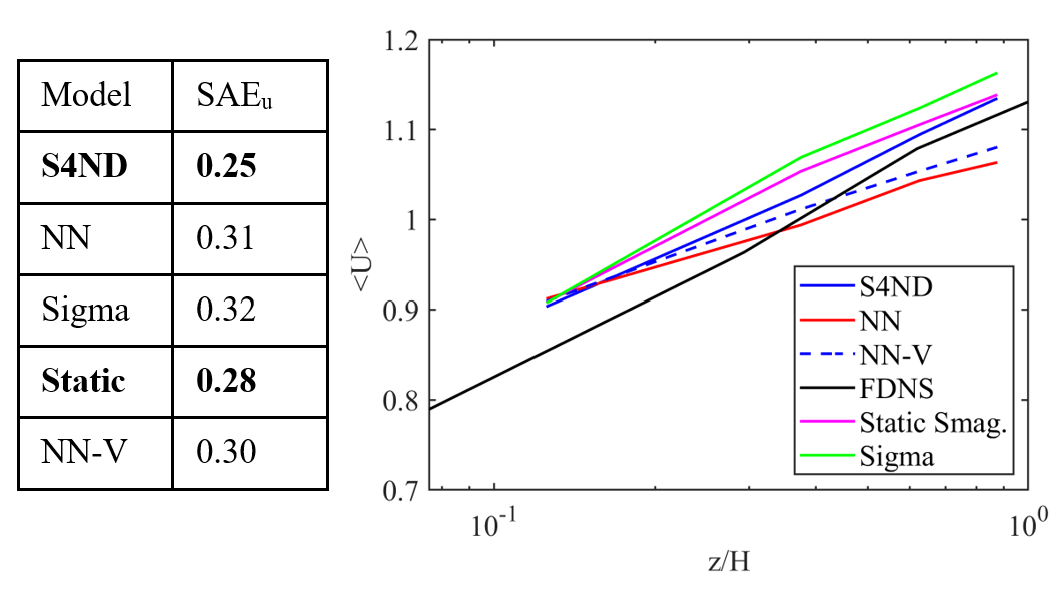}}
  \caption{Channel flow mean profiles for various SGS models on a grid corresponding to an extrapolative filter width of $64 \Delta_{DNS}$.}
\label{fig:64ChannelMean}
\end{figure}

Channel flow is also evaluated at a more aggressive, extrapolative grid spacing. Keeping all other simulation parameters fixed, the grid spacing is changed to 32x24x8, corresponding to a filter width of $64 \Delta_{DNS}$. Otherwise, the simulation parameters are kept the same as the finer grid simulation, corresponding to a $Re_{\tau}$ of 1000. From the mean profiles, shown in figure~\ref{fig:64ChannelMean}, the S4ND model is able to preserve the slope of the mean streamwise velocity profile as compared to filtered DNS, while the NN model fails. Meanwhile, the two traditional models are more inaccurate as they are further from the filtered DNS profile. When comparing the S4ND model with the NN-V model, one can see that while the NN-V model is able to improve upon the NN model form by using different inputs and output normalization, the overall slope is still incorrect when compared with the S4ND model. This is also seen in the $SAE_u$ tables, with the S4ND being the lowest (and the NN model being not far behind only because it crosses the filtered DNS profile). 

The Reynolds Stress profiles are also plotted. From these profiles, the accuracy of the S4ND model in an extrapolative setting is highlighted. Comparing the S4ND model to the NN model, the S4ND model is far more accurate as the NN model has significant trouble extrapolating. The NN model profile shapes are extremely off and the magnitudes are incorrect even when compared to the traditional models. As seen from every single profile, the S4ND is far better at preserving the overall shape and magnitude as compared to both the NN model and all the traditional models. However, this could also potentially be because the NN model's output scaling, which was different from the S4ND's output scaling, was just not as optimized for extrapolation to coarse grids. This hypothesis can be ruled out by looking at the NN-V model, which has the same exact inputs and output scaling as the S4ND model. Here, the NN-V model also is significantly worse than the S4ND model. While it doesn't overpredict as much as the NN model, suggesting that the output scaling used for the NN-V model and S4ND model are better at scaling, the overall profile shapes are oftentimes incorrect, especially $R_{33}$, where it overpredicts the Reynolds stress peak such that it is over 5 times that of filtered DNS. This suggests that it is not the output scaling that is causing the performance improvement of S4ND for channel flow extrapolation, but solely due to the S4ND architecture. This highlights the extrapolatory capability of the S4ND model as it outperforms the NN and NN-V model while also being significantly better than the traditional models. In all situations for coarse-grid channel flow, the S4ND-based model, when integrated into a LES simulation, gives results closest to the filtered DNS profiles. 

Overall, combined with the HIT results from the previous section, the S4ND model is a more flexible model since it is more accurate in extrapolating than any of the other neural network models tested (especially for channel flow), and is more accurate than the traditional SGS models even when extrapolating in grid resolution. Furthermore, by comparing with NN-V, one can highlight that while the output scaling does play a role in helping to improve extrapolation, the S4ND architecture is still key in extrapolating, especially in channel flow. This also suggests that the in-training set filter width results, where oftentimes the NN model was better, highlight that the normalization used for the NN model helped contribute to the NN model ``overfitting" to the in-training set filter width cases, producing better results \textit{a posteriori} on these cases and then completely failing to extrapolate to HIT or channel flow on coarser grids. This can be contrasted with NN-V and S4ND, where the inputs and output normalization allowed for less \textit{a posteriori} performance for the in-training set filter width cases, but then allowed for better extrapolation to coarser grids. In the simpler HIT case with isotropic grids, the normalization allowed the NN-V model, which uses convolutional neural networks, to extrapolate well to HIT, but it still failed in the more complex channel flow case with anisotropic grids. The S4ND meanwhile, was able to extrapolate to both coarser grid cases reasonably well, as a consequence of its architecture. This validates what was brought up in the previous section, where the use of bandlimiting is effectively a regularizer that smooths out the convolution kernel, allowing the S4ND model to be more accurate in out of training set filter widths, while the models with convolutional neural networks are more likely to ``overfit" to the specific grid resolutions that it has been trained on, particularly for channel flow, where due to anisotropic grids, the primary assumption of convolutional neural networks is already violated (isotropic grids). Physically, the S4ND coarse grid extrapolation cases highlight that the continuous convolution kernels with bandlimiting allow the model to respect the scale-dependent structure of turbulence across filter widths. When the filter width increases (grid gets coarser), more energy is filtered out, and the subgrid stress model must account for a larger fraction of the unresolved energy. Particularly for channel flow, since the S4ND model produces more accurate Reynolds stress profiles and mean profiles on coarser grids when compared to the NN-V (which has identical inputs, outputs, and normalization), this suggests that the continuous formulation adjusts to the balance between resolved and modelled scales more naturally than discrete convolutions. The improved performance of the S4ND model on coarser grid resolutions is therefore attributed to the continuous formulation with bandlimiting, which ensures that the learned closure adapts to changing filter width, maintaining filter-consistent spectral content while capturing the interscale energy transfer (which can be seen directly in the \textit{a priori} results).

\begin{figure}
\centering
     \begin{subfigure}{0.27\textwidth}
         \centering
         \includegraphics[width=\textwidth]{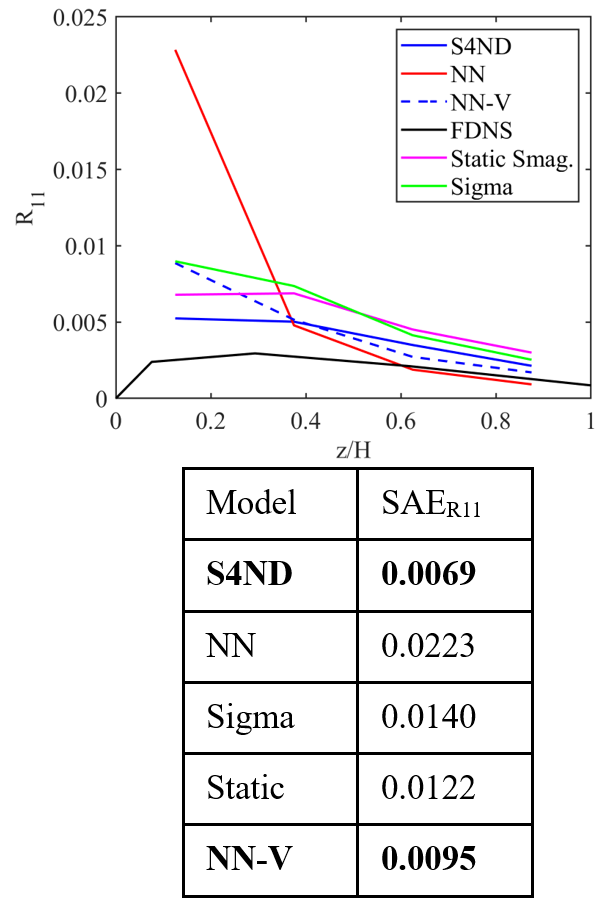}
         \caption{$R_{11}$ varying as a function of wall normal distance.}
         \label{fig:64ChannelR11}
     \end{subfigure}
     \hspace{1em}
     \begin{subfigure}{0.27\textwidth}
         \centering
         \includegraphics[width=\textwidth]{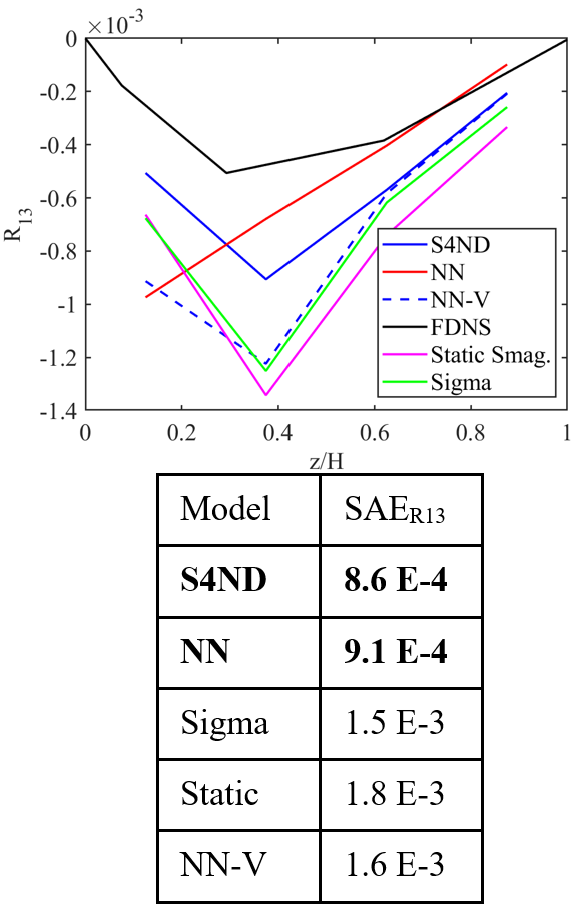}
         \caption{$R_{13}$ varying as a function of wall normal distance.}
         \label{fig:64ChannelR13}
     \end{subfigure}
     \hspace{1em}
     \begin{subfigure}{0.27\textwidth}
         \centering
         \includegraphics[width=\textwidth]{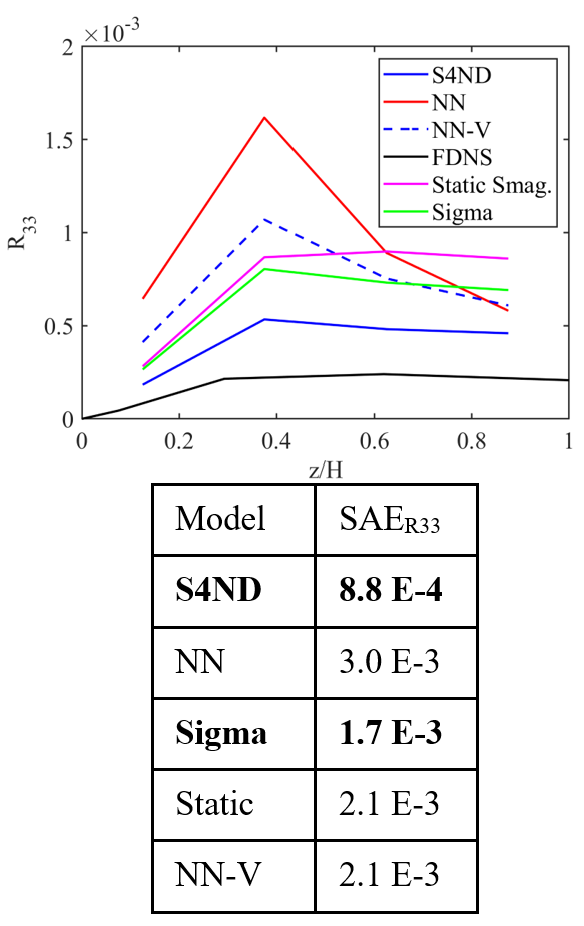}
         \caption{$R_{33}$ varying as a function of wall normal distance.}
         \label{fig:64ChannelR33}
     \end{subfigure}
     \caption{Channel flow Reynolds stress profiles for various SGS models on a grid corresponding to an extrapolative filter width of $64 \Delta_{DNS}$.}
\end{figure}

\section{Reynolds Number Scaling}
The Reynolds number scaling capabilities of the S4ND model are now investigated. First, the S4ND model, which is trained on forced HIT with $Re_{\lambda} = 433$ and channel flow with $Re_{\tau} = 1000$ data, is tested \textit{a posteriori} on both forced HIT with $Re_{\lambda} = 2400$ and channel flow with $Re_{\tau} = 5200$. The reference data are also queried from the JHTDB, with the DNS HIT resolution being $4096^3$ on the same $[2\pi, 2\pi, 2\pi]$ domain, while the DNS channel flow resolution is $10240 \times 1536 \times 7680$ on the same $[8 \pi, 2, 3\pi]$ domain when normalized by channel half-height. Next, the S4ND model is then tested on cases where not only the grid width, but the Reynolds number is also extrapolated. 

\subsection{Reynolds Number Scaling with DNS Data}
Both HIT and channel flow Reynolds number scaling is investigated. The HIT simulation is conducted on a $64^3$ grid on a triply periodic domain of length $2 \pi$ in each direction, with the dimensional dissipation, $\epsilon$, set to 1.4144. The channel flow simulation is conducted on a grid of size $128$ x $32$ x $96$ on a domain of size $8 \pi$ x $2$ x $3 \pi$. Due to the increase in Reynolds number, while the $\Delta_{LES} = 64 \Delta_{DNS}$ for HIT (reference DNS is now 4x finer than the previous DNS) and $\Delta_{LES} = 80 \Delta_{DNS}$ for the channel flow simulation (reference DNS is now 5x finer than the previous DNS), the overall grid spacing is still considered in training set. This is because the grid spacing $dx$, $dy$, $dz$ are grid spacings that the neural network has been trained on in the training data. As seen from~\cite{Choi_2024}, neural networks extrapolate well to higher Reynolds number as long as $dx$, $dy$, $dz$ is the same as the grid spacing of the training data, even if $\frac{\Delta_{LES}}{\eta}$ is larger than the training data, where $\eta$ is the Kolmogorov length scale. As such, we will still consider this a ``in training set" filter width. 

\subsubsection{HIT}
\begin{figure}
\centering
     \begin{subfigure}{0.375\textwidth}
         \centering
         \includegraphics[width=\textwidth]{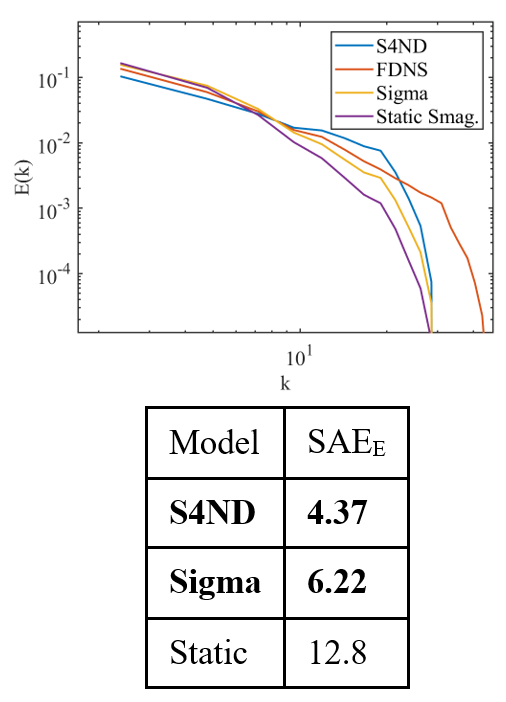}
         \caption{Forced HIT spectra.}
         \label{fig:HITReSpectra}
     \end{subfigure}
     \hspace{1em}
     \begin{subfigure}{0.35\textwidth}
         \centering
         \includegraphics[width=\textwidth]{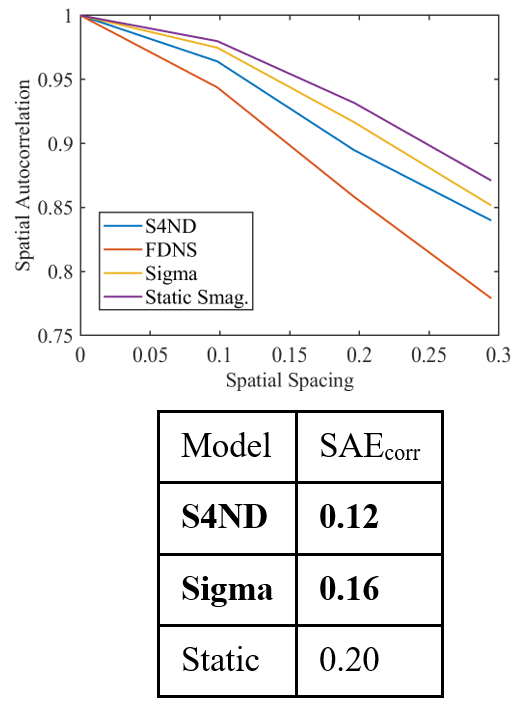}
         \caption{Forced HIT spatial correlations.}
         \label{fig:HITReSpatial}
     \end{subfigure}
     \caption{Macro-scale quantities analyzed for forced HIT at a higher Reynolds number not seen in the training set.}
\end{figure}

The Reynolds number scaling results for forced HIT are shown in figures~\ref{fig:HITReSpectra} and~\ref{fig:HITReSpatial}. In these figures, the S4ND model is compared to both the Static Smagorinsky and Sigma models, as well as filtered DNS to evaluate the extrapolatory capability of the S4ND model. 

\begin{figure}
  \centerline{\includegraphics[scale=0.2]{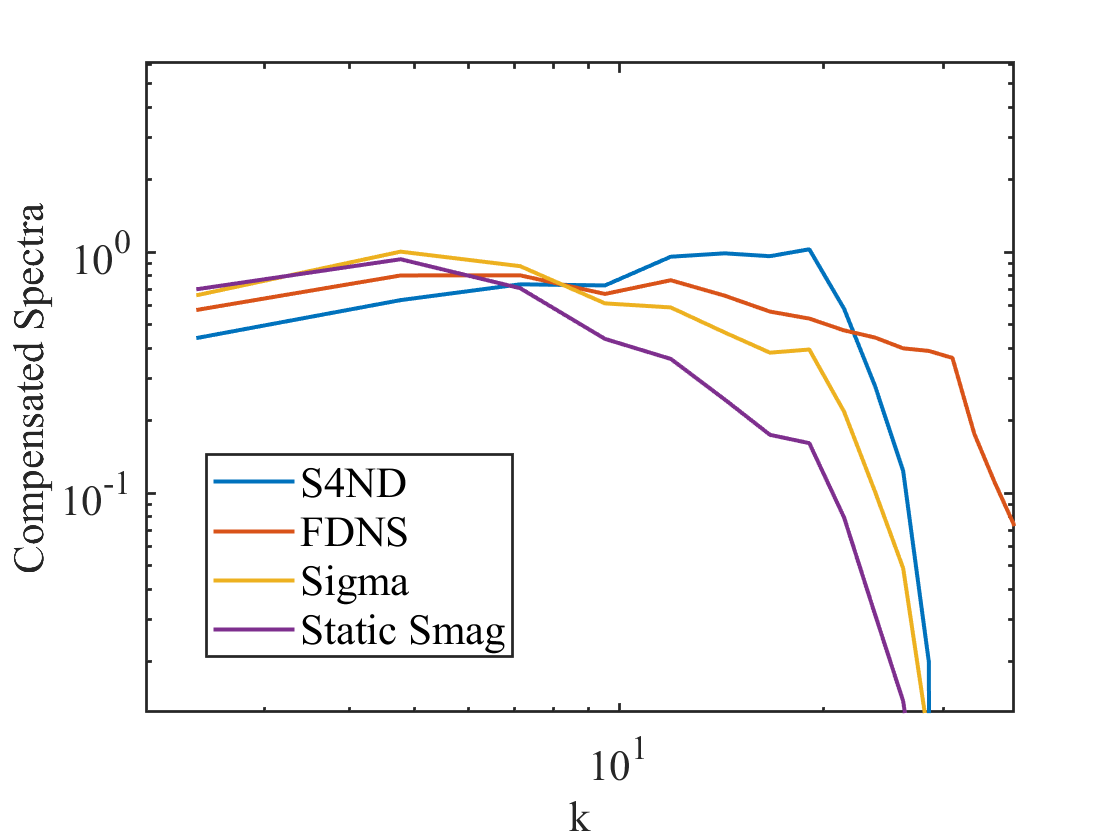}}
  \caption{Compensated Spectra for forced HIT at a higher Reynolds number not seen in the training set.}
\label{fig:HITCompensated}
\end{figure}

As seen, when the Reynolds number is larger than the training set Reynolds number, the S4ND model still produces reasonable results for both spectra and spatial correlations. Overall, the same trends hold when comparing the S4ND simulations that were run with an in-training set Reynolds number versus an extrapolatory Reynolds number. For example, the energy spectra for S4ND is still slightly underdissipative, while the spatial correlations are more accurate for the S4ND model compared to the spatial correlations produced by the traditional subgrid stress models. In this case, the spatial correlations are truncated since there is only one ensemble at this Reynolds number for the filtered DNS profile (JHTDB provides one snapshot), and thus the spatial correlation is truncated at a point where it is still reasonably converged. When taking a look at the SAE for both the energy spectra and the correlations, the S4ND model has the lowest error out of all the models compared. The compensated spectra, which pre-multiplies the energy spectra with a $k^{\frac{5}{3}}$ factor, is shown in figure~\ref{fig:HITCompensated}. As seen, the filtered DNS reference has an inertial range, but the inertial range varies from being slightly less than the $k^{\frac{5}{3}}$ scaling to slightly more than the $k^{\frac{5}{3}}$ scaling. This is also seen in the SGS models, with the S4ND model being consistent with previous results (slightly under-dissipative) while the traditional SGS models are also consistent with previous results (slight over-dissipative). Overall, the S4ND model is able to successfully extrapolate to a higher Reynolds number for forced HIT. 

\subsubsection{Channel Flow}
\begin{figure}
\centering
     \begin{subfigure}{0.45\textwidth}
         \centering
         \includegraphics[width=\textwidth]{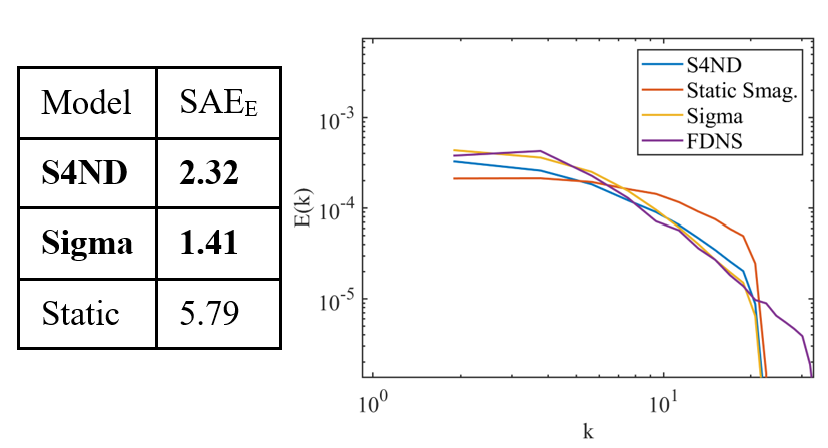}
         \caption{Streamwise spectra at z/H = 0.46875}
         \label{fig:ChannelReSpectra}
     \end{subfigure}
     \hspace{1em}
     \begin{subfigure}{0.45\textwidth}
         \centering
         \includegraphics[width=\textwidth]{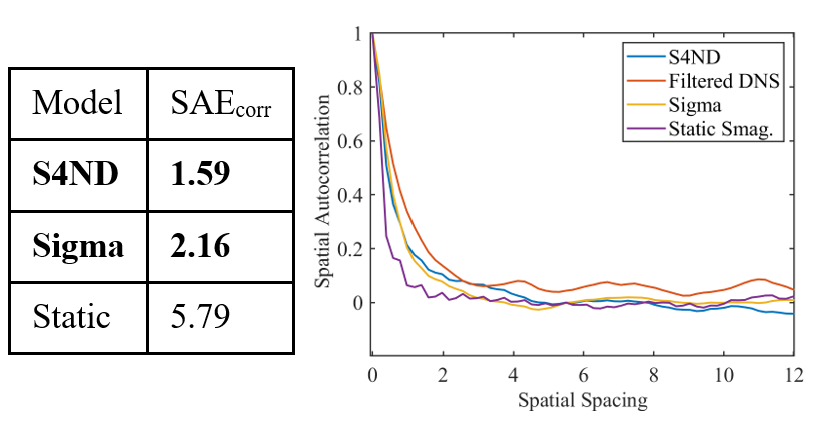}
         \caption{Spatial correlations at $z/H = 0.46875$.}
         \label{fig:ChannelReSpatial}
     \end{subfigure}
     \caption{Macro-scale quantities analyzed for channel flow at a higher Reynolds number not seen in the training set.}
\end{figure}

\begin{figure}
\centering
     \begin{subfigure}{0.27\textwidth}
         \centering
         \includegraphics[width=\textwidth]{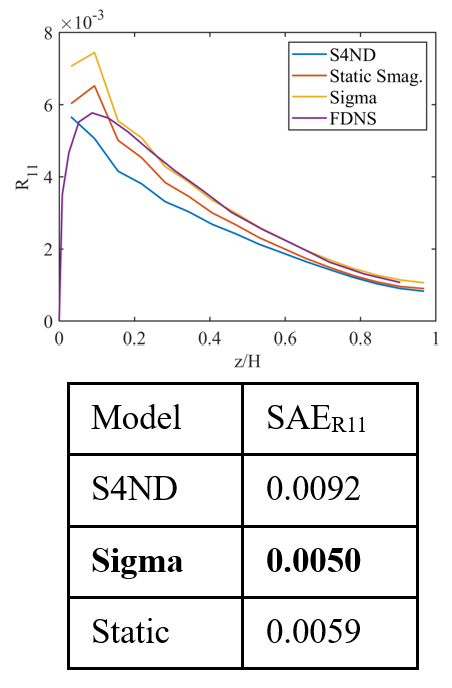}
         \caption{$R_{11}$ varying as a function of wall normal distance.}
         \label{fig:ChannelReR11}
     \end{subfigure}
     \hspace{1em}
     \begin{subfigure}{0.27\textwidth}
         \centering
         \includegraphics[width=\textwidth]{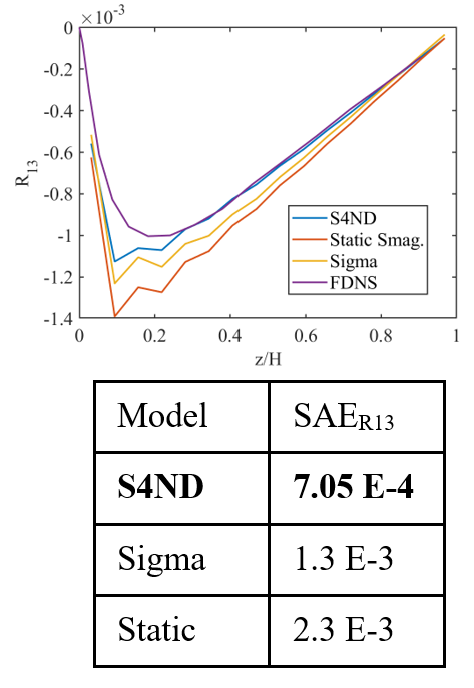}
         \caption{$R_{13}$ varying as a function of wall normal distance.}
         \label{fig:ChannelReR13}
     \end{subfigure}
     \hspace{1em}
     \begin{subfigure}{0.27\textwidth}
         \centering
         \includegraphics[width=\textwidth]{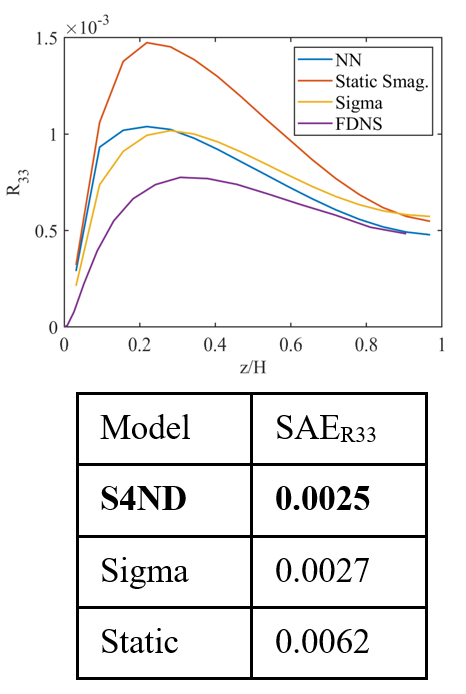}
         \caption{$R_{33}$ varying as a function of wall normal distance.}
         \label{fig:ChannelReR33}
     \end{subfigure}
     \caption{Channel flow Reynolds stress profiles for various SGS models at a higher Reynolds number not seen in the training set.}
\end{figure}

For channel flow Reynolds number scaling, first the spectra at a specific wall normal location and the spatial correlations are shown in figures~\ref{fig:ChannelReSpectra} and~\ref{fig:ChannelReSpatial}.
As seen in the plots, the S4ND model produces spectra and spatial correlations that match filtered DNS profiles when the Reynolds number is out of the training set. There is no obvious performance degradation when compared to the in-training set filter width, and the overall profiles match the filtered DNS profiles. Furthermore, the S4ND Reynolds stress profiles are seen in figures~\ref{fig:ChannelReR11},~\ref{fig:ChannelReR13} and~\ref{fig:ChannelReR33}. 
Overall, the S4ND model is able to capture the overall shape and magnitude of the Reynolds stress profiles associated with channel flow. From all these plots, the S4ND model is in general, consistent with the in-training set Reynolds number, with no obvious performance degradation when the Reynolds number is out of the training set. This is corroborated by looking at the SAE, where the S4ND model produces reasonable SAE for all quantities. Combining both the channel flow and forced HIT results, one can conclude that the S4ND model is able to extrapolate to higher Reynolds number flow cases and still match filtered DNS reasonably well. 

\subsection{Extreme Reynolds Number Scaling Without DNS Data}
Now, true ``out of training set" filter width and Reynolds number extrapolation is run, where the grid spacings $dx$, $dy$, $dz$ are larger than the training set, and where  $\frac{\Delta_{LES}}{\eta}$ is also far larger than the training set. While there are no DNS results at these larger Reynolds numbers, one can evaluate the stability of the subgrid stress models in extreme Reynolds number scaling scenarios. 

\subsubsection{HIT}
\begin{figure}
\centering
     \begin{subfigure}{0.375\textwidth}
         \centering
         \includegraphics[width=\textwidth]{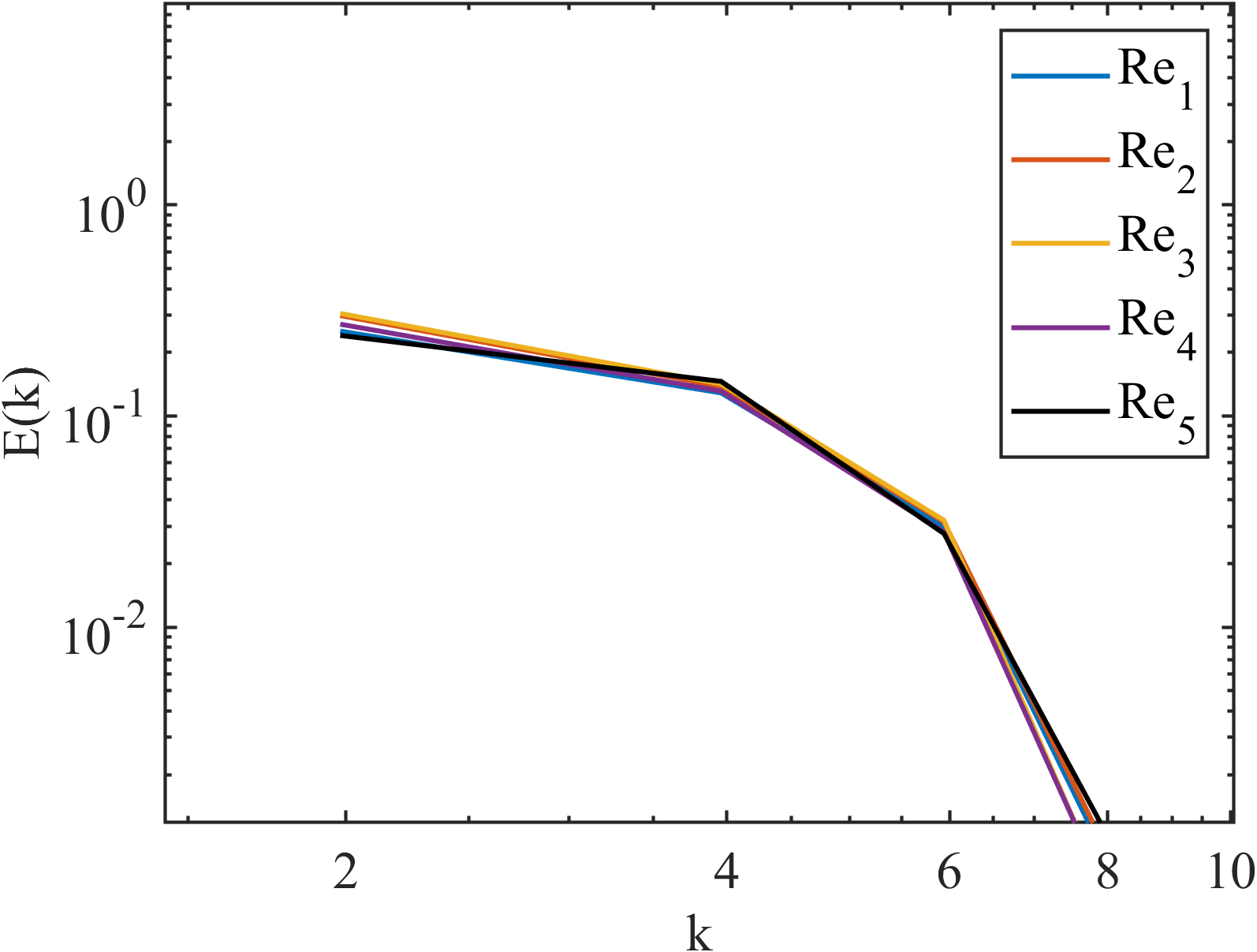}
         \caption{Energy spectra at various Reynolds numbers.}
         \label{fig:HITExtreme1}
     \end{subfigure}
     \hspace{1em}
     \begin{subfigure}{0.375\textwidth}
         \centering
         \includegraphics[width=\textwidth]{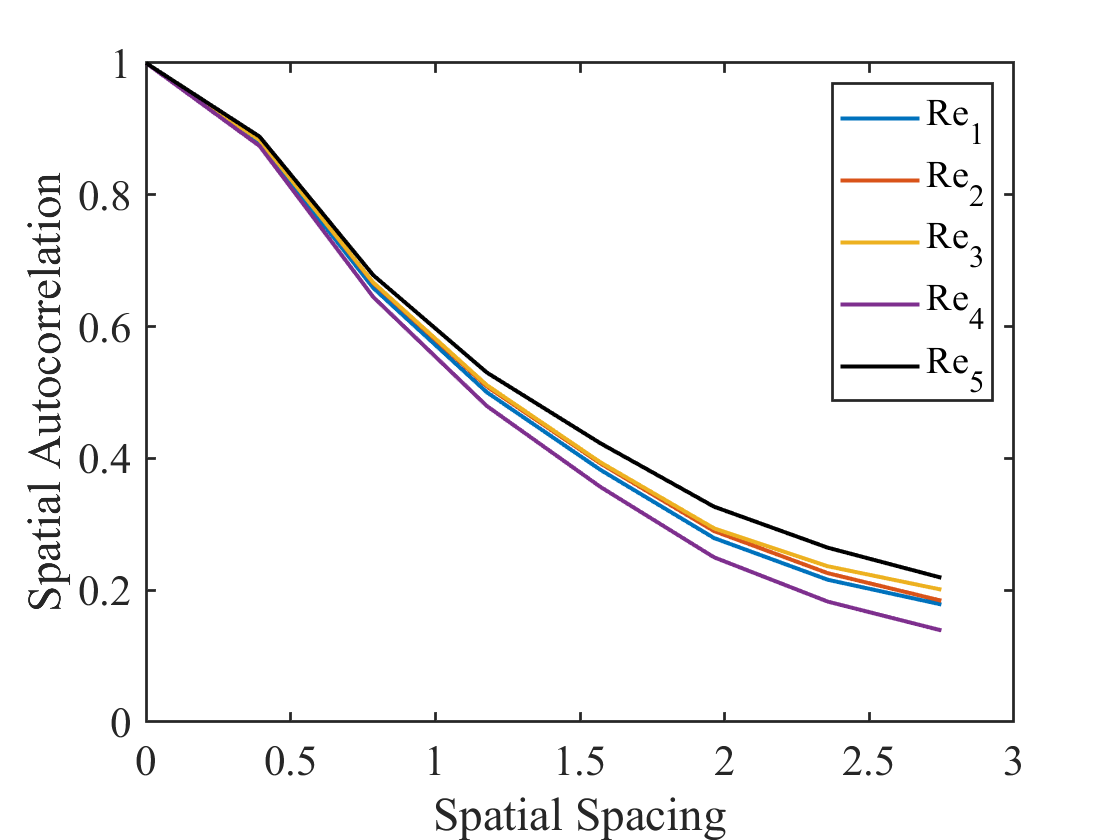}
         \caption{Spatial correlations at various Reynolds numbers.}
         \label{fig:HITExtreme2}
     \end{subfigure}
     \caption{Extreme Reynolds number scaling for HIT}
\end{figure}

The HIT simulation is evaluated on a $16^3$ grid, corresponding to a grid spacing in each direction that is twice as large as the training set. Furthermore, 5 Reynolds numbers based on the Taylor microscale are tested: $\Rey_{1} = 4800$, $\Rey_{2} = 19500$, $\Rey_{3} = 56900$, $\Rey_{4} = 515000$, $\Rey_{5} = 508900000$. The S4ND model is only trained on $\Rey = 433$. As seen in figures~\ref{fig:HITExtreme1} and~\ref{fig:HITExtreme2}, the neural network is able to ensure that the LES simulation is stable, even when extrapolating in both grid spacing and $\Rey$. Overall, the spectra and spatial correlations look reasonable for all $\Rey$, showing that the S4ND model produces reasonable results even if the Reynolds number is over 1000000 times that of the training set. Given the coarseness of the grid, the results should not change drastically with Reynolds number, which is demonstrated in these plots. In addition, up to $\Rey_3$, the trends are monotonic, suggesting that the S4ND model can capture trends up to around 70 times the Reynolds number in the training set. From $Re_1$ to $Re_5$, at no point is a ``tail-up" in the spectra seen, which is expected for ANN-based subgrid stress models that are trained on only one Reynolds number, as seen in~\citet{Choi_2024}. 

\subsubsection{Channel Flow}
The same neural network is also integrated into a channel flow simulation with varying Reynolds numbers to test the extrapolatory capabilities of the neural network, on a grid of size $32$ x $8$ x $24$ on a domain of size $8 \pi$ x $2$ x $3 \pi$ (grid spacing is coarser than the training set). The Reynolds number based on friction velocity tested involve $\Rey_1 = 5,200$, $\Rey_2 = 20,000$, $\Rey_3 = 50,000$, $\Rey_4 = 500,000$, and $\Rey_5 = 500,000,000$. The neural network is trained only on Reynolds number based on friction velocity of $\Rey = 1000$. 

\begin{figure}
\centering
     \begin{subfigure}{0.45\textwidth}
         \centering
         \includegraphics[width=\textwidth]{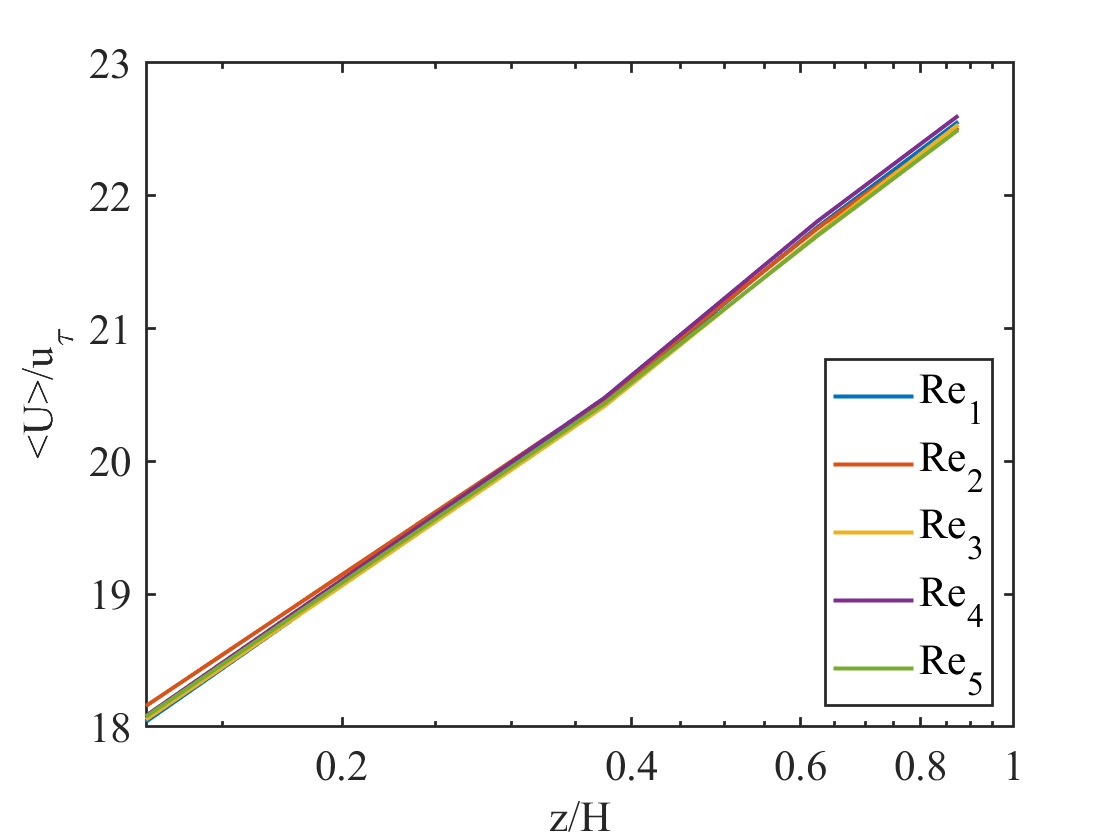}
         \caption{Channel flow mean profiles for various Reynolds numbers.}
         \label{fig:ExtremeChannelMean}
     \end{subfigure}
     \hspace{1em}
     \begin{subfigure}{0.45\textwidth}
         \centering
         \includegraphics[width=\textwidth]{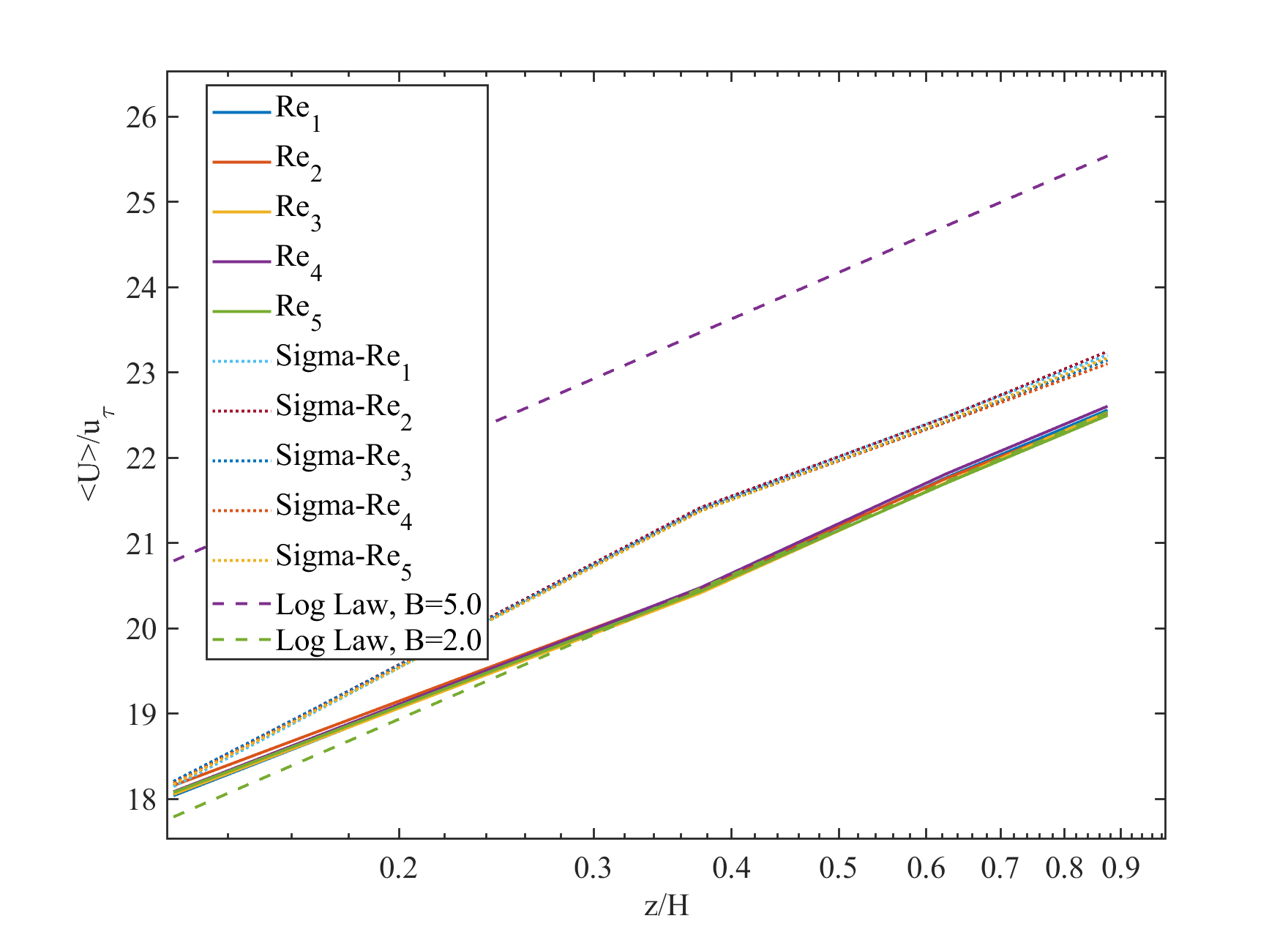}
         \caption{Channel flow mean profiles for various Reynolds numbers with Sigma model and fits.}
         \label{fig:ExtremeChannelMean2}
     \end{subfigure}
     \caption{Various mean profiles for channel flow extrapolation to extreme Reynolds numbers.}
\end{figure}

\begin{figure}
\centering
     \begin{subfigure}{0.3\textwidth}
         \centering
         \includegraphics[width=\textwidth]{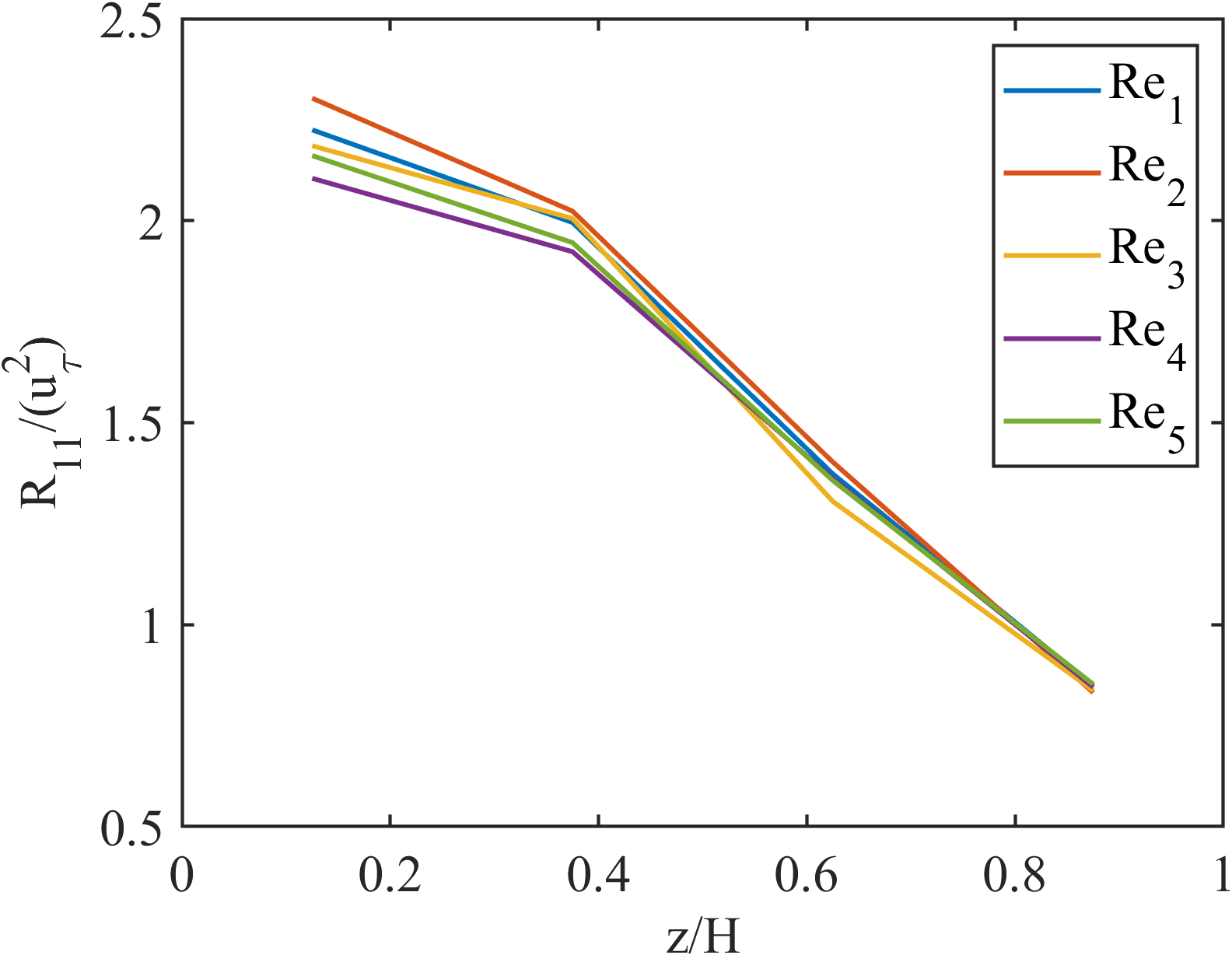}
         \caption{$R_{11}$ varying as a function of wall normal distance.}
         \label{fig:ExtremeChannelR11}
     \end{subfigure}
     \hspace{1em}
     \begin{subfigure}{0.32\textwidth}
         \centering
         \includegraphics[width=\textwidth]{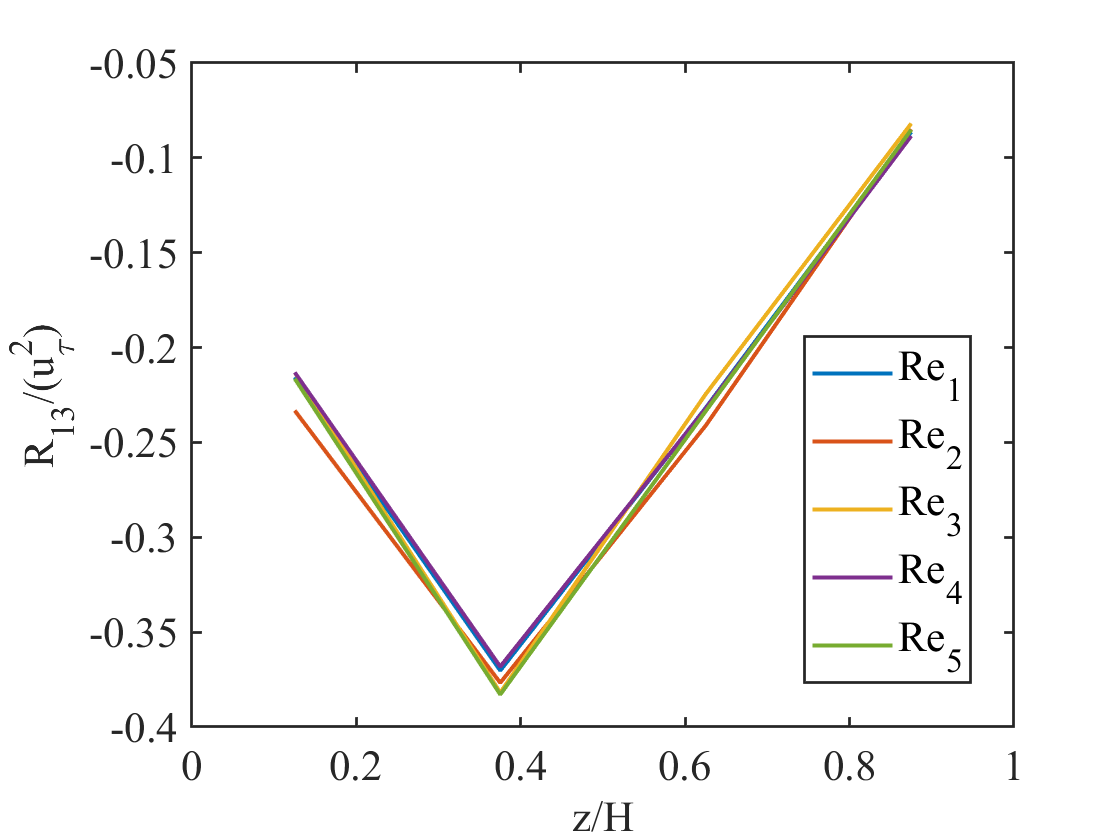}
         \caption{$R_{13}$ varying as a function of wall normal distance.}
         \label{fig:ExtremeChannelR13}
     \end{subfigure}
     \hspace{1em}
     \begin{subfigure}{0.31\textwidth}
         \centering
         \includegraphics[width=\textwidth]{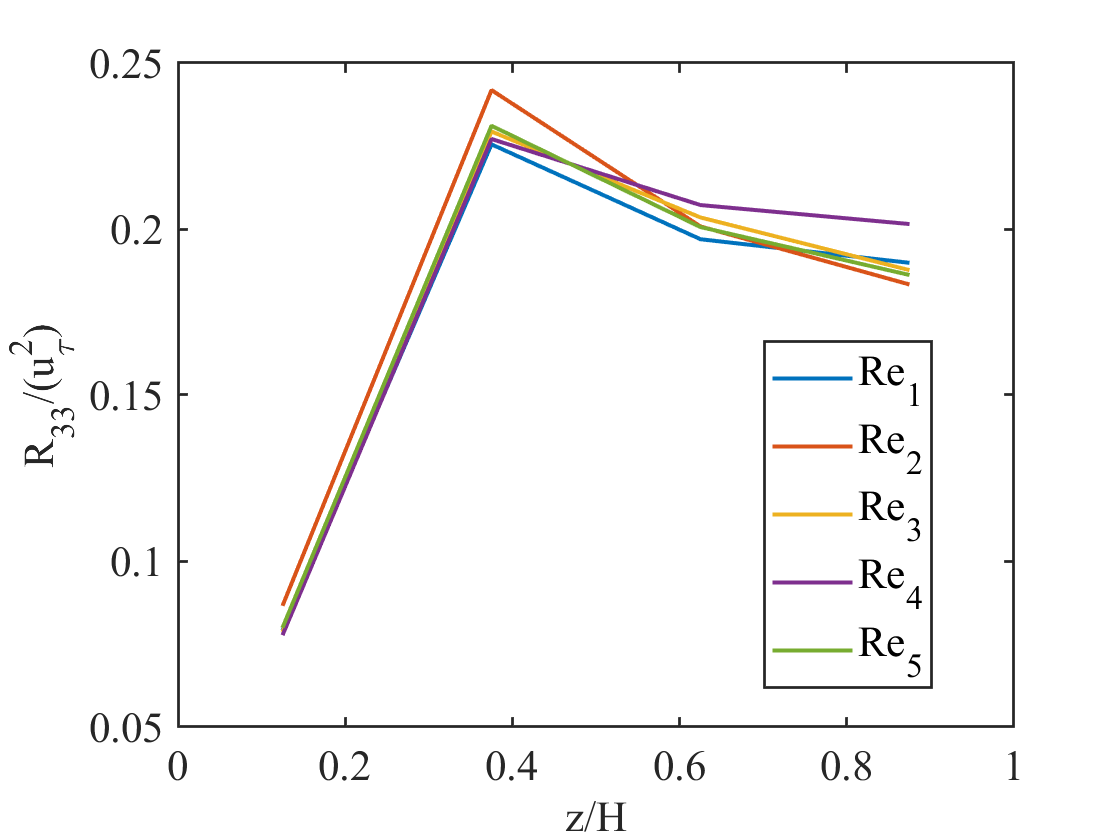}
         \caption{$R_{33}$ varying as a function of wall normal distance.}
         \label{fig:ExtremeChannelR33}
     \end{subfigure}
     \caption{Reynolds stress profiles for various extreme Reynolds numbers for the S4ND model.}
\end{figure}

From figure~\ref{fig:ExtremeChannelMean}, when scaled with friction velocity, the channel mean streamwise velocity profiles collapse nicely, suggesting that these profiles are reasonable for all Reynolds numbers. To fully verify this, the log-law profiles are plotted, along with the Sigma model mean profile results in figure~\ref{fig:ExtremeChannelMean2}. Due to the large offset from the S4ND model, the Sigma model was also run as verification to isolate if this offset was associated with the subgrid stress model or the wall model. As seen, using the classic offset of $B=5.0$ is insufficient, likely due to the error in the wall model. This error can be attributed to the wall model since two different SGS models (Sigma model and data-driven S4ND model) produce extremely similar results at the first off-wall grid point, which is dominated by the wall model error. In this extremely coarse and extremely aggressive Reynolds number extrapolation case, the S4ND model is able to produce a profile that is closer to the log-law profile than the Sigma model. From all this, the neural network is seen to provide reasonable results even when the Reynolds number is extrapolated by over 500000 times the Reynolds number of the training set when considering mean profiles. This is further corroborated in figures~\ref{fig:ExtremeChannelR11}, ~\ref{fig:ExtremeChannelR13}, and ~\ref{fig:ExtremeChannelR33}, where the Reynolds stresses also collapse reasonably well when scaled with friction velocity. There is no significant performance degradation visible at first glance as the Reynolds number increases, suggesting that the S4ND model remains stable even at Reynolds numbers significantly larger than the training set Reynolds number. While exact comparisons can not be made as there is no DNS at these higher Reynolds number cases ($Re_2-Re_5$), if the neural network is unable to extrapolate, as seen in~\citet{Choi_2024}, the change in shape of the profiles of turbulent statistical quantities is quite obvious to see.

\begin{figure}
  \centerline{\includegraphics[scale=0.2]{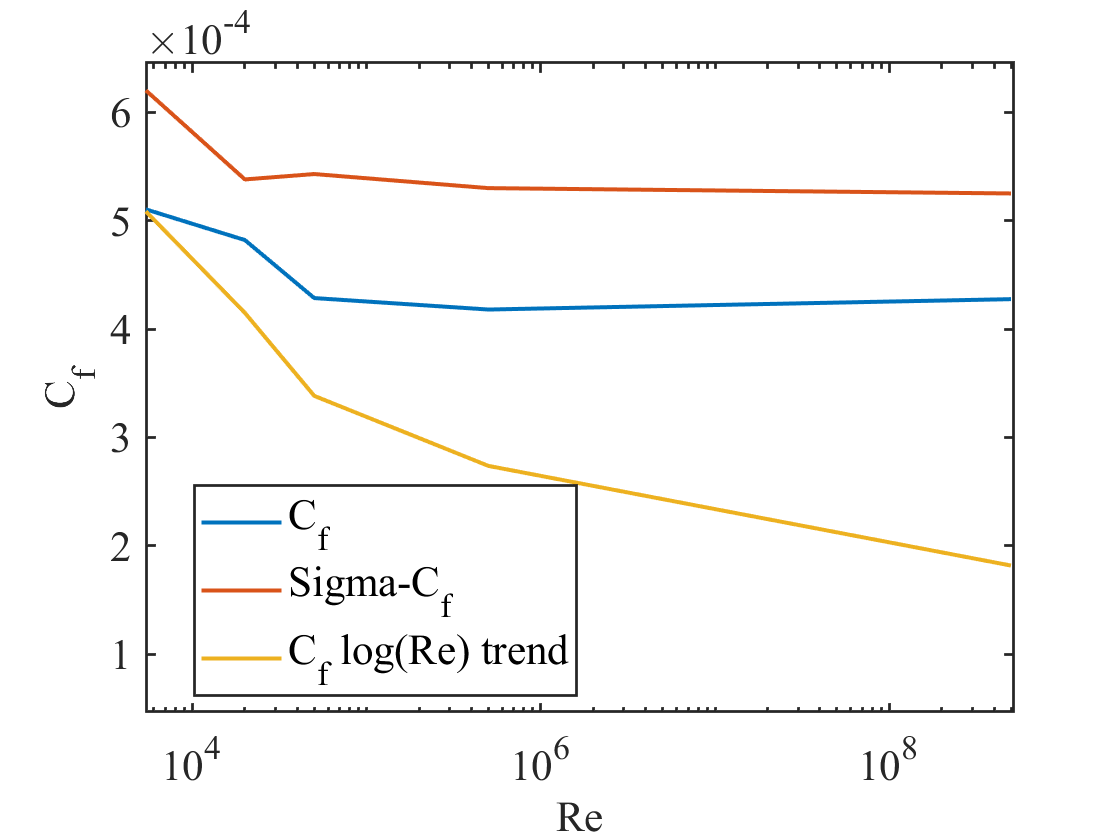}}
  \caption{Implied wall skin friction coefficient versus a first-order trend, with the x-axis being friction Reynolds number. The trends show that the error is likely due to the wall model, as both the S4ND model (plotted in blue) and the Sigma model (plotted in red) show similar trends and deviation from the expected trend (plotted in yellow~\cite{schultz2013reynolds}).}
\label{fig:C_fplot}
\end{figure}

In order to provide first-order analysis on the reasonableness of the channel flow scaling results, the implied skin friction coefficient, $C_f$ is plotted, as well as the expected inverse $log(\Rey)$ scaling that is supported by~\cite{schultz2013reynolds}. These results are shown in figure~\ref{fig:C_fplot}. As seen, the implied $C_f$ is decreasing for the first 4 Reynolds numbers, and is still reasonably following the overall trend. However, the last Reynolds number suggests an increase in $C_f$, suggesting that the results are less trustworthy at the highest Reynolds number. However, it is very likely this is more of a wall modelling error as compared to a subgrid stress model error, as in these really coarse simulations there are 8 points across the channel and the wall model's accuracy is suspect at these extreme Reynolds number cases ($\Rey_{\tau}$ of 500,000,000). This argument is further enhanced when the Sigma model results are taken into account, as it exhibits similar trends as compared to the S4ND model. Therefore, the S4ND model likely extrapolates well to these extremely out of training set Reynolds numbers, and the error that is seen can be attributed to the non-data driven wall model. Even if this largest Reynolds number data point is discarded, the S4ND model still obtains 500x friction Reynolds number scaling when the model is only trained on one friction Reynolds number. The largest Reynolds number data point is mainly kept to show that even if the results are less trustworthy, the subgrid stress model still remains stable even in this sort of highly extrapolatory situation. 

While another way to ensure that the neural networks are stable during Reynolds number scaling is to use clipping, where values of the SGS tensor are set to zero if the local dissipation is negative. While this is an effective method, it is ultimately unphysical. In all these cases, no clipping is utilized and the S4ND model is stable without clipping.

\subsection{Computational Cost}
\begin{figure}
  \centerline{\includegraphics[scale=0.7]{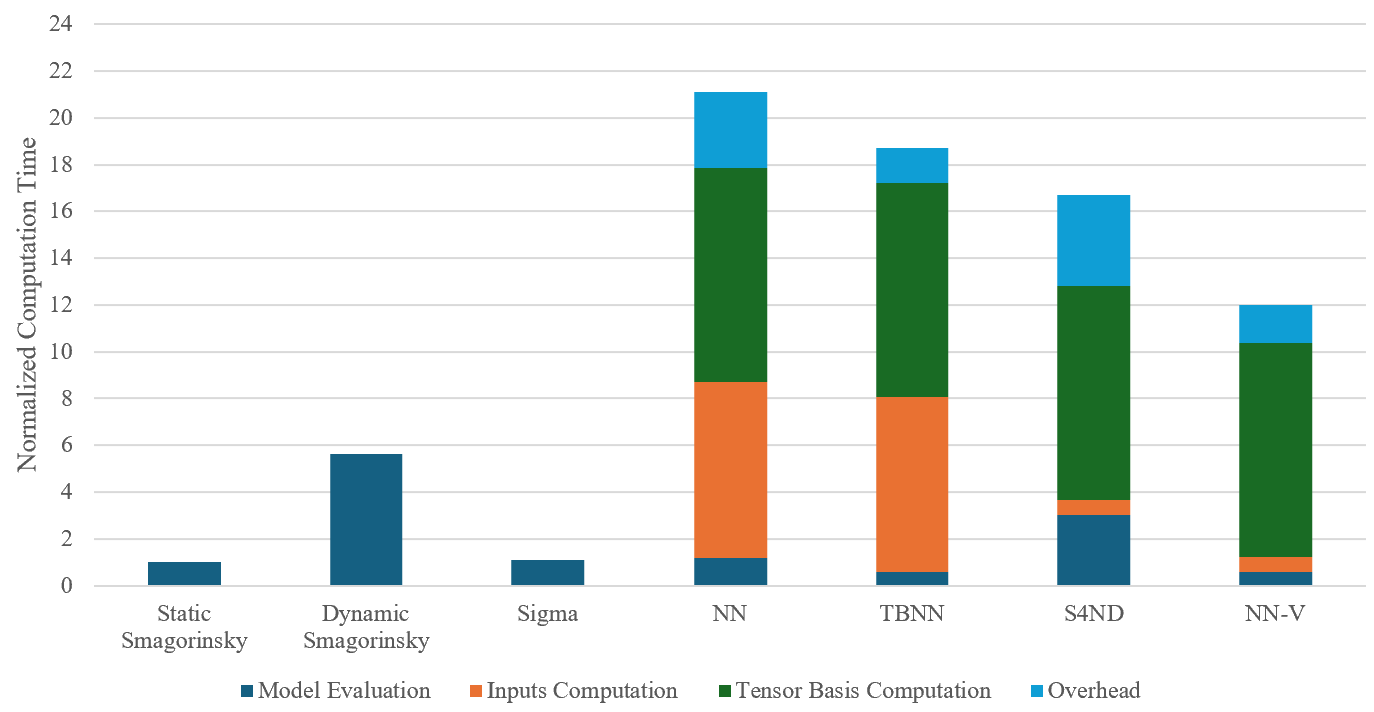}}
  \caption{Computational cost of various subgrid stress models. Overhead refers to the time it takes to load the neural network from disk and send the inputs from the LES to the neural network for computation, and then returning the neural network subgrid stress output back to the LES.}
\label{fig:Computationcost}
\end{figure}

The computational cost of various models are listed, normalized by the Static Smagorinsky computation time. In general, using the tensor basis neural network backend is more expensive due to the computation of the invariants and the tensor basis, which are the dominant computational costs of the model, as seen in figure~\ref{fig:Computationcost}.
Overall, the actual cost to evaluate the neural network models is cheaper than a Dynamic Smagorinsky evaluation, but the computation of the invariants and associated tensor basis significantly dominate the time. Overall, the neural network models are around 3-4 times more expensive than a Dynamic Smagorinsky evaluation. Due to the use of the velocity gradient tensor instead of the invariants, the S4ND model is less costly than the TBNN and the NN model. However, even with the increase in computation cost, the simulations are still fairly computationally inexpensive, requiring less than a day to complete even when evaluating the most expensive neural network models. 

Due to the architectural sophistication of the S4ND model, the model is approximately 2.5 times more expensive than the NN model when evaluating the neural network. However, the S4ND model is able to gain extrapolatory capability by being grid-spacing aware as opposed to the discrete convolutions that the NN model uses. In addition, the tradeoff between directly enforcing physical constraints and additional runtime is clear. If directly enforcing physical constraints (for example through a tensor basis neural network backend), the cost to compute the tensor basis can be as much as two Dynamic Smagorinsky evaluations. If the invariants are used as inputs, then the cost to compute the invariants would cost around another Dynamic Smagorinsky evaluation. Therefore, removing the invariants allows the computational time to be allocated to the more sophisticated S4ND model. Overall, this results in the overall compute time of the S4ND model to be less than the NN and TBNN models. If one did not care about directly enforcing physical constraints through a TBNN, then the overall compute time could potentially be around the same as 1 to 1.5 times that of a Dynamic Smagorinsky evaluation. While the S4ND model provides a step in the right direction over convolutional neural networks for scaling to more complex flows with anisotropic grids and different grid resolutions (such as channel flow), it is unable to handle stretched, curvilinear, or unstructured grids natively. The S4ND architecture would be a good choice for many canonical flows in a wall-modeled context.

\section{Conclusions}
A new neural network subgrid stress model based on state space sequence models (S4ND) was introduced, and bandlimiting was used to increase the generalization capabilities of the model to untrained grid spacings. \textit{A priori}, the use of bandlimiting does not restrict the expressivity of the learned representation as compared to other traditional and neural network models, while learning a continuous convolution kernel and using bandlimiting allows the neural network to extrapolate to more aggressive, untrained grid spacings. \textit{A posteriori}, when testing on untrained grid widths, the S4ND model is more accurate than traditional models and is able to generalize where other neural network models fail on both forced HIT and channel flow LES. In addition, even when extrapolating in grid width and Reynolds number at the same time, the S4ND model is stable for both forced HIT and channel flow even when evaluated on Reynolds numbers that are over 500000 times larger than the training set Reynolds numbers, achieving one-shot Reynolds number extrapolation capability. Overall, this suggests that the S4ND model produces more filter-consistent behavior over a range of grid resolutions and is able to respect the scale-dependent structure of turbulence to create a more robust subgrid stress model.

\backsection[Acknowledgements]{This work used CPU and GPU resources via Darwin at the University of Delaware and via Bridges2 at the Pittsburgh Supercomputing Center through allocation PHY230025 from the Advanced Cyberinfrastructure Coordination
Ecosystem: Services \& Support (ACCESS) program, which is supported by National Science Foundation grants
\#2138259, \#2138286, \#2138307, \#2137603, and \#2138296. DNS data is provided by the Johns Hopkins Turbulence Database. We appreciate the comments of the anonymous referees whose comments have made the manuscript better.}

\backsection[Funding]{Andy Wu is partially supported by NASA Cooperative Agreement number 80NSSC22M0108 and Northrop Grumman, as well as the NDSEG Fellowship.}

\backsection[Declaration of Interests]{\bf Declaration of Interests. The authors report no conflict of interest.} 

\backsection[Data Availability Statement]{All DNS data that was filtered to produce the training data for the neural network is provided publicly on JHTDB. A public repository of the S4ND model used can be found in the references section, linking to the author's implementation.} 

\appendix

\section{The Bandlimit Parameter for S4ND}\label{appA}
\begin{figure}
\centering
     \begin{subfigure}{0.45\textwidth}
         \centering
         \includegraphics[width=\textwidth]{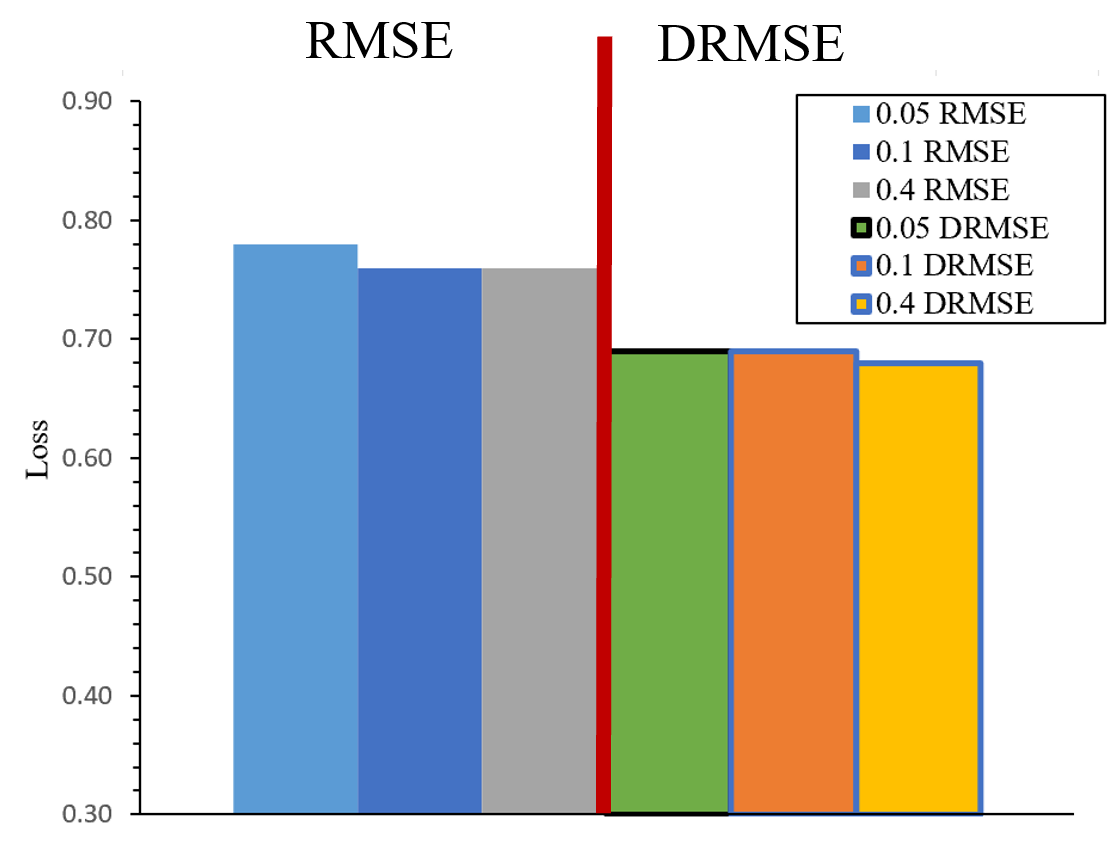}
         \caption{Losses on the test set for HIT when varying $\alpha$.}
         \label{fig:HITablation}
     \end{subfigure}
     \hspace{1em}
     \begin{subfigure}{0.45\textwidth}
         \centering
         \includegraphics[width=\textwidth]{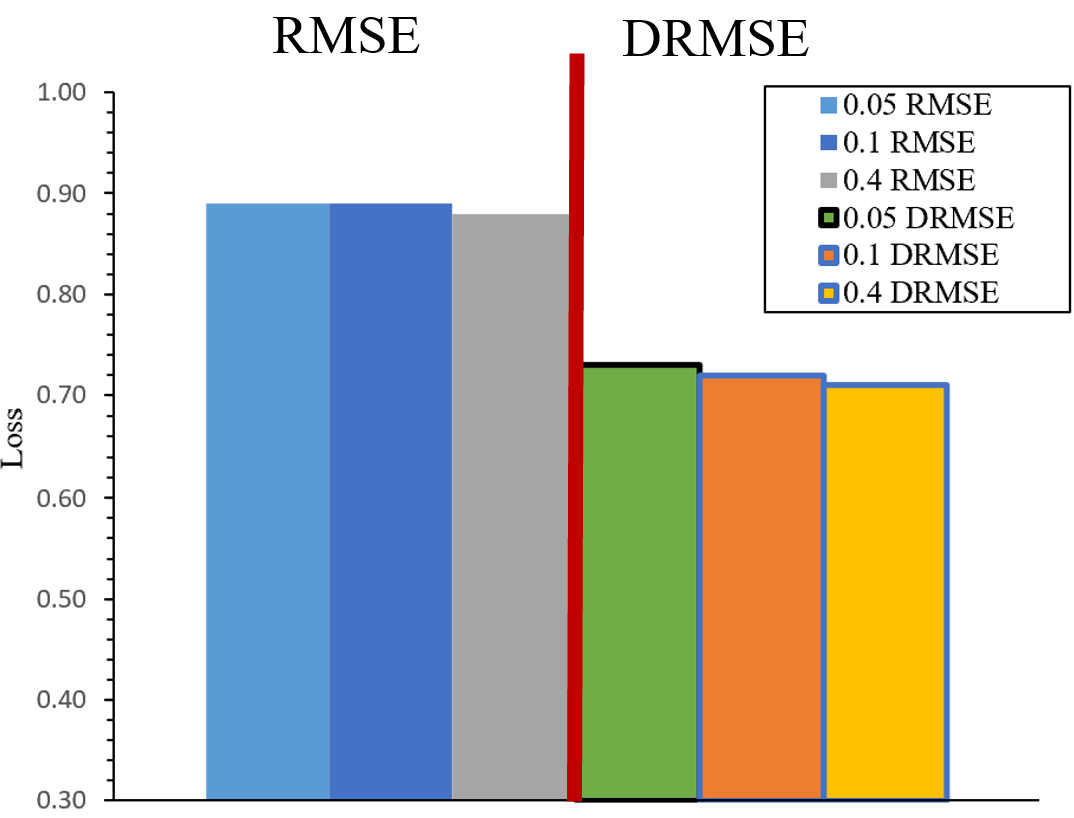}
         \caption{Losses on the test set for channel flow when varying $\alpha$.}
         \label{fig:Channelablation}
     \end{subfigure}
     \caption{Bandlimit parameter sensitivity sweeps for HIT and channel flow on the hold out test set. The bandlimit value of $\alpha = 0.05$ has a slight performance degradation as compared to the other two larger values.}
\end{figure}

As seen from the main text, the interpretation of the bandlimit parameter is that it masks out frequencies in the convolution kernel that are said to be spurious. Theoretically, a bandlimit parameter of 1 corresponds to masking out all frequencies above the Nyquist wavenumber. However, due to the finite state representation and also the decay associated with the real part of $a_n$, the bandlimit parameter $\alpha$ is often set lower numerically as seen in~\cite{nguyen2022s4nd}.~\cite{nguyen2022s4nd} study the effect of the bandlimit parameter and show that the accuracy of the extrapolation to different resolutions decreases quite significantly when $\alpha$ is set to be larger than $\alpha = 0.5$. For LES, one of the issues with \textit{a priori} versus \textit{a posteriori} evaluation is that the training data can be different than the neural network inputs that are seen in actual LES, partially due to the numerical method. Numerical discretization errors disproportionately contaminate the highest resolved wavenumbers near the grid cutoff~\cite{sagaut}, which suggests that for robustness, one would like to use as small as a $\alpha$ parameter as possible that still preserves accuracy. This can be evaluated with \textit{a priori} testing, as really low bandlimit values should decrease the expressiveness of the continuous convolution kernel, reducing its accuracy on the test set. Therefore, a parameter sweep of three $\alpha \in [0.05, 0.1, 0.4]$ is shown, with the \textit{a priori} results on the test set seen in~\crefrange{fig:HITablation}{fig:Channelablation} for both the in training set HIT and channel flow results.

As seen, the results are not very sensitive to the bandlimit parameter, although there is a slight performance improvement when comparing the $\alpha = 0.05$ bandlimit cases and the $\alpha = 0.1$ bandlimit cases, where the $\alpha = 0.1$ case either does just as well or slightly better than the $\alpha = 0.05$ bandlimit cases. Meanwhile, the $\alpha = 0.4$ cases are comparable to $\alpha = 0.1$ cases. Therefore, this suggests that the bandlimit parameter of $\alpha=0.05$ is too small as it produces too much regularization to the convolution kernel. This is seen as there is a slight performance decrease (higher errors) for this value of the bandlimit parameter. As such, $\alpha = 0.1$ is chosen as the bandlimit parameter as it is the smallest value (when compared with $\alpha = 0.4$) where there is no performance degradation, since $\alpha = 0.05$ has a slight performance degradation.

\section{Sigma Model Coefficient Sweep for HIT}\label{appB}
\begin{figure}
\centering
     \begin{subfigure}{0.45\textwidth}
         \centering
         \includegraphics[width=\textwidth]{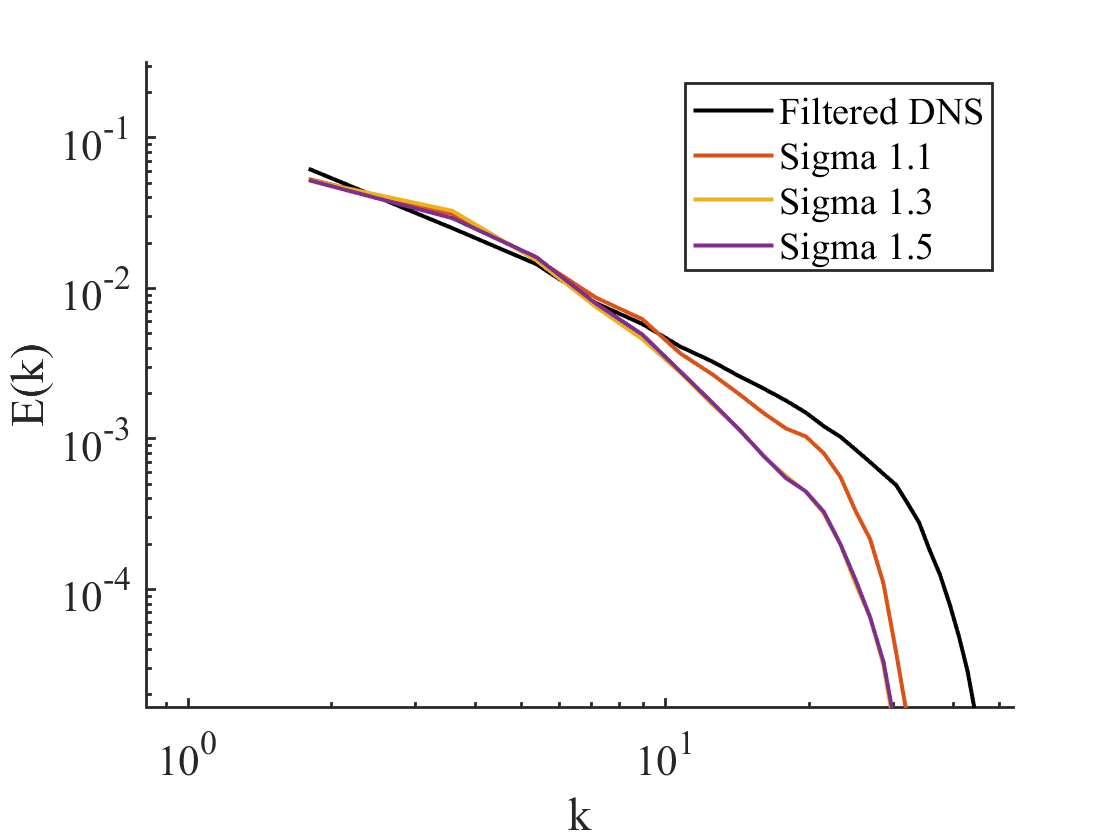}
         \caption{Spectra of forced HIT for various Sigma model coefficient values.}
         \label{fig:SigmaSpectra}
     \end{subfigure}
     \hspace{1em}
     \begin{subfigure}{0.45\textwidth}
         \centering
         \includegraphics[width=\textwidth]{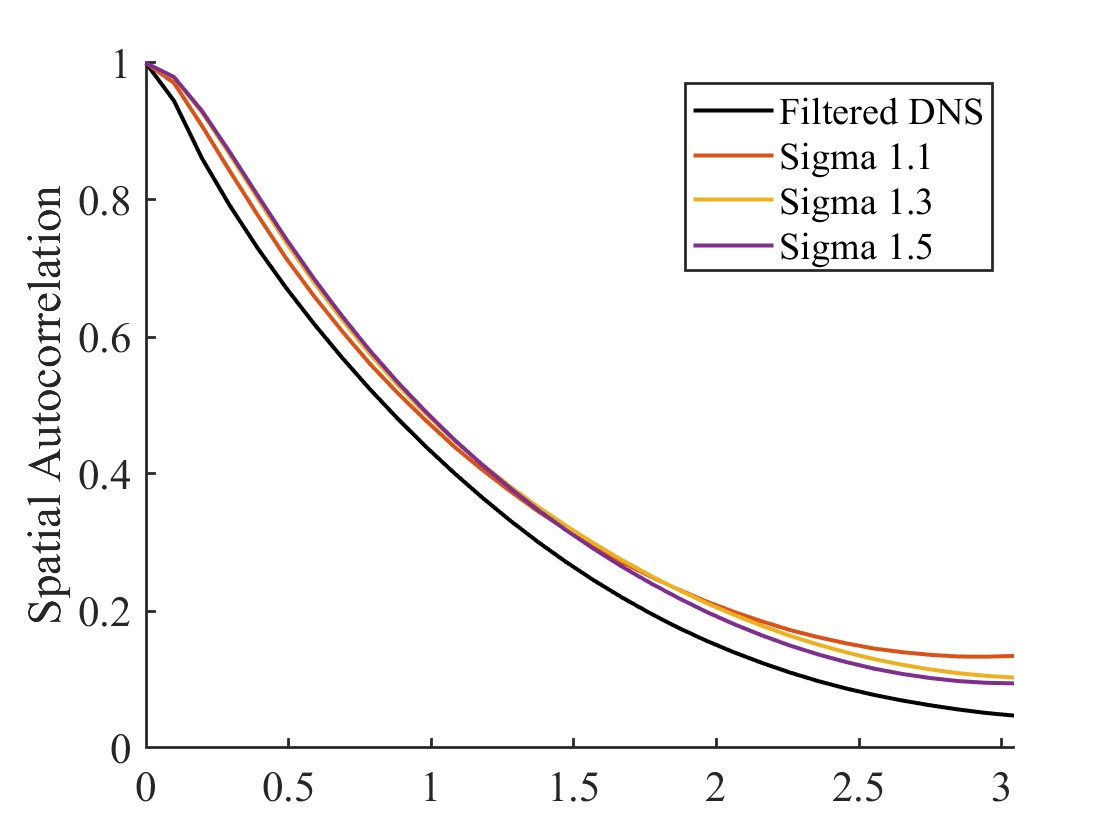}
         \caption{Forced HIT spatial correlations for various Sigma model values}
         \label{fig:SigmaCorr}
     \end{subfigure}
     \caption{Sensitivity study of the sigma model to coefficients in the recommended range and also non-standard coefficients. The values in the legend correspond to the coefficient used.}
     \label{fig:SigmaAblation}
\end{figure}

Here, the sensitivity of the Sigma Model to the coefficient is investigated for coarse grid forced HIT. The Sigma model is run for forced HIT using three different model coefficients. This case would correspond to the same simulation parameters as Section 4.1.1. The smallest model coefficient is a non-standard model coefficient, and was used since the Sigma model was found to be over-dissipative. The other two are within the standard range. In total, the three coefficients are 1.1, 1.3, and 1.5 respectively. Fig.~\ref{fig:SigmaAblation} shows the spectra and spatial correlation results for forced HIT at these three coefficient values.

As seen, when in the recommended range, the spectra for the Sigma model with coefficients of 1.3 and 1.5 are very similar, while the non-standard coefficient of 1.1 is less dissipative. This is expected as it is outside the recommended range, and reduces the magnitude of the SGS tensor, resulting in less dissipation. While this means that it is closer to filtered DNS, there is a corresponding trade-off in the spatial correlations, as the Sigma model with the non-standard coefficient performs the worst at larger spatial spacings. The spatial correlations indicate that at smaller spacings, all the model coefficients are relatively similar, but at larger grid spacings, the smaller coefficients perform worse than the larger coefficients when compared to filtered DNS. Therefore, even for coarser grids, adopting the ``middle" coefficient of 1.3 is reasonable for subgrid stress model comparisons.

\bibliographystyle{jfm}
\bibliography{jfm}

@article{Choi_2024, 
title={A recursive neural-network-based subgrid-scale model for large eddy simulation: application to homogeneous isotropic turbulence}, 
volume={1000}, 
DOI={10.1017/jfm.2024.992}, 
journal={Journal of Fluid Mechanics}, 
author={Cho, Chonghyuk and Park, Jonghwan and Choi, Haecheon}, year={2024}, pages={A76}}

@article{Wu_PrF,
  title={Two neural network Unet architecture for subfilter stress modeling},
  author={Wu, Andy and Lele, Sanjiva K},
  journal={Physical Review Fluids},
  volume={10},
  number={1},
  pages={014601},
  year={2025},
  publisher={APS}
}

@article{smagorinsky1963general,
  title={General circulation experiments with the primitive equations: I. The basic experiment},
  author={Smagorinsky, Joseph},
  journal={Monthly weather review},
  volume={91},
  number={3},
  pages={99--164},
  year={1963},
  publisher={American Meteorological Society}
}

@article{nicoud2011using,
  title={Using singular values to build a subgrid-scale model for large eddy simulations},
  author={Nicoud, Franck and Toda, Hubert Baya and Cabrit, Olivier and Bose, Sanjeeb and Lee, Jungil},
  journal={Physics of fluids},
  volume={23},
  number={8},
  year={2011},
  publisher={AIP Publishing}
}

@article{clark1979evaluation,
  title={Evaluation of subgrid-scale models using an accurately simulated turbulent flow},
  author={Clark, Robert A and Ferziger, Joel H and Reynolds, William Craig},
  journal={Journal of fluid mechanics},
  volume={91},
  number={1},
  pages={1--16},
  year={1979},
  publisher={Cambridge University Press}
}

@inbook{pope,
  title={Turbulent {F}lows},
  edition={1},
  author={Pope, S B},
  year={2000},
  publisher={Cambridge University},
}

@inbook{sagaut,
  title={Large {E}ddy {S}imulation for {I}ncompressible {F}lows: {A}n {I}ntroduction},
  edition={3},
  author={Sagaut, P},
  year={2006},
  publisher={Springer Science \& Business Media},
}

@article{evans,
  title={Invariant data-driven subgrid stress modeling in the strain-rate eigenframe for large eddy simulation},
  author={Prakash, P and Jansen, K E and Evans, J A},
  journal={Computer Methods in Applied Mechanics and Engineering},
  volume={399},
  number={115457},
  year={2022},
  publisher={Elsevier},
  doi = {10.1016/j.cma.2022.115457}
}

@article{GenSmag,
  title={Generalized {Smagorinsky} model for anisotropic grids },
  author={Scotti, A and Meneveau, C and Lilly, DK},
  journal={Physics of Fluids},
  volume={5},
  year={1993},
  publisher={AIP},
  doi = {10.1063/1.858537}
}

@article{vreman,
  title={An eddy-viscosity subgrid-scale model for turbulent shear flow: {A}lgebraic theory and applications},
  author={Vreman, A W},
  journal={Physics of Fluids},
  volume={16},
  number={10},
  year={2004}
}

@article{ghate,
  title={Subfilter-scale {E}nrichment of {P}lanetary {B}oundary {L}ayer {L}arge {E}ddy {S}imulation {U}sing {D}iscrete {F}ourier–{G}abor modes.},
  author={Ghate, A S and Lele, S K},
  journal={Journal of Fluid Mechanics},
  volume={819},
  year={2017}
}

@inbook{bardina,
  title={Mathematics of {L}arge {E}ddy {S}imulation of {T}urbulent {F}lows},
  edition={1},
  author={Berselli, LC and Iliescu, T and Layton, W J},
  year={2006},
  publisher={Springer},
}

@book{bardina1983improved,
  title={Improved turbulence models based on large eddy simulation of homogeneous, incompressible, turbulent flows},
  author={Bardina, Jorge},
  year={1983},
  publisher={Stanford University}
}

@article{beckreview,
  title={A {Perspective on Machine Learning Methods in Turbulence Modelling}},
  author={Beck, A and Kurz, M},
  journal={Surveys for Applied Mathematics and Mechanics},
  volume={44},
  number={1},
  year={2021}
}

@article{sarg,
  title={Neural networks based subgrid scale modeling in large eddy simulations},
  author={Sarghini, F and De Felice, G and Santini, S},
  journal={Computers \& Fluids},
  volume={32},
  number={1},
  pages={97-108},
  year={2003}
}

@article{beck,
  title={Deep neural networks for data-driven {LES} closure models},
  author={Beck, A and Flad, D and Munz, C-D},
  journal={Journal of Computational Physics},
  volume={398},
  number={108910},
  year={2019}
}

@article{xie,
  title={Modeling subgrid-scale forces by spatial artificial neural networks in large eddy simulation of turbulence},
  author={Xie, C and Wang, J and Weinan, E},
  journal={Physical Review Fluids},
  volume={5},
  number={5},
  year={2020}
}

@article{stoffer,
  title={Development of a large-eddy simulation subgrid model based on artificial neural networks: a case study of turbulent channel flow},
  author={Stoffer, R and Leeuwen, C M and Podareanu, D and Codreanu, V and Veerman, M A and Janssens, M and Hartogensis, O K and Heerwaarden, C C},
  journal={Geoscientific Model Development},
  volume={14},
  number={6},
  year={2021}
}

@article{ling,
  title={Reynolds averaged turbulence modelling using deep neural networks with embedded invariance},
  author={Ling, J and Kurzawski, A and Templeton, J},
  journal={Journal of Fluid Mechanics},
  volume={807},
  pages={155-166},
  year={2016}
}

@article{park,
  title={Toward neural-network-based large eddy simulation: application to turbulent channel flow},
  author={Park, J and Choi, H},
  journal={Journal of Fluid Mechanics},
  volume={914},
  year={2021}
}

@inproceedings{stallcup,
  title={{Adaptive Scale-Similar Closure for Large Eddy Simulations. Part 1: Subgrid Stress Closure}},
  author={Stallcup, E W and Kshitĳ, A and Dahm, W J},
  booktitle={AIAA SciTech},
  year={2022},
  publisher={AIAA},
  address={San Diego, CA},
  pages = {AIAA 2022-0595},
  doi={10.2514/6.2022-0595}
}

@article{JHTB1,
  title={{A public turbulence database cluster and applications to study Lagrangian evolution of velocity increments in turbulence}},
  author={Li, Y and Perlman, E and Wan, M and Yang, Y and Meneveau, C and Burns, R and Chen, S and Szalay A and Eyink, G},
  journal={Journal of Turbulence},
  volume={9},
  number={31},
  year={2008}
}

@inproceedings{JHTB2,
  title={{Data Exploration of Turbulence Simulations using a Database Cluster}},
  author={Perlman, E and Burns, R and Li, Y and Meneveau, C},
  booktitle={Supercomputing SC07, ACM},
  year={2007},
  publisher={IEEE},
}

@article{xie2,
  title={Artificial neural network-based nonlinear algebraic models for large eddy simulation of turbulence},
  author={Xie, C and Yuan, Z and Wang, J},
  journal={Physics of Fluids},
  volume={32},
  number={11},
  year={2020}
}

@article{Kang,
  title={Neural-network-based mixed subgrid-scale model for turbulent flow},
  author={Kang, M and Jeon, Y and You, D},
  journal={Journal of Fluid Mechanics},
  volume={962},
  year={2023}
}

@article{yuan2021dynamic,
  title={Dynamic iterative approximate deconvolution models for large-eddy simulation of turbulence},
  author={Yuan, Zelong and Wang, Yunpeng and Xie, Chenyue and Wang, Jianchun},
  journal={Physics of Fluids},
  volume={33},
  number={8},
  year={2021},
  publisher={AIP Publishing}
}

@article{yuan2020deconvolutional,
  title={Deconvolutional artificial neural network models for large eddy simulation of turbulence},
  author={Yuan, Zelong and Xie, Chenyue and Wang, Jianchun},
  journal={Physics of Fluids},
  volume={32},
  number={11},
  year={2020},
  publisher={AIP Publishing}
}

@article{stolz2001approximate,
  title={An approximate deconvolution model for large-eddy simulation with application to incompressible wall-bounded flows},
  author={Stolz, Steffen and Adams, Nikolaus A and Kleiser, Leonhard},
  journal={Physics of fluids},
  volume={13},
  number={4},
  pages={997--1015},
  year={2001},
  publisher={American Institute of Physics}
}

@article{bou2005scale,
  title={{A scale-dependent Lagrangian dynamic model for large eddy simulation of complex turbulent flows}},
  author={Bou-Zeid, Elie and Meneveau, Charles and Parlange, Marc},
  journal={Physics of fluids},
  volume={17},
  number={2},
  year={2005},
  publisher={AIP Publishing}
}

@article{sirignano2023deep,
  title={Deep learning closure models for large-eddy simulation of flows around bluff bodies},
  author={Sirignano, Justin and MacArt, Jonathan F},
  journal={Journal of Fluid Mechanics},
  volume={966},
  pages={A26},
  year={2023},
  publisher={Cambridge University Press}
}

@article{Guan2022,
title = {Stable a posteriori {LES} of 2D turbulence using convolutional neural networks: Backscattering analysis and generalization to higher Re via transfer learning},
journal = {Journal of Computational Physics},
volume = {458},
pages = {111090},
year = {2022},
issn = {0021-9991},
doi = {https://doi.org/10.1016/j.jcp.2022.111090},
url = {https://www.sciencedirect.com/science/article/pii/S0021999122001528},
author = {Guan, Yifei and Chattopadhyay, Ashesh and Subel, Adam and Hassanzadeh, Pedram}
}

@article{cheng2022deep,
  title={Deep learning for subgrid-scale turbulence modeling in large-eddy simulations of the convective atmospheric boundary layer},
  author={Cheng, Yu and Giometto, Marco G and Kauffmann, Pit and Lin, Ling and Cao, Chen and Zupnick, Cody and Li, Harold and Li, Qi and Huang, Yu and Abernathey, Ryan and others},
  journal={Journal of Advances in Modeling Earth Systems},
  volume={14},
  number={5},
  pages={e2021MS002847},
  year={2022},
  publisher={Wiley Online Library}
}

@article{wang2018investigations,
  title={Investigations of data-driven closure for subgrid-scale stress in large-eddy simulation},
  author={Wang, Zhuo and Luo, Kun and Li, Dong and Tan, Junhua and Fan, Jianren},
  journal={Physics of Fluids},
  volume={30},
  number={12},
  year={2018},
  publisher={AIP Publishing}
}

@article{Pawar_2020,
   title={A priori analysis on deep learning of subgrid-scale parameterizations for {K}raichnan turbulence},
   volume={34},
   ISSN={1432-2250},
   url={http://dx.doi.org/10.1007/s00162-019-00512-z},
   DOI={10.1007/s00162-019-00512-z},
   number={4},
   journal={Theoretical and Computational Fluid Dynamics},
   publisher={Springer Science and Business Media LLC},
   author={Pawar, Suraj and San, Omer and Rasheed, Adil and Vedula, Prakash},
   year={2020},
   month=jan, pages={429–455} }

@article{Kurz_2023,
   title={Deep reinforcement learning for turbulence modeling in large eddy simulations},
   volume={99},
   ISSN={0142-727X},
   url={http://dx.doi.org/10.1016/j.ijheatfluidflow.2022.109094},
   DOI={10.1016/j.ijheatfluidflow.2022.109094},
   journal={International Journal of Heat and Fluid Flow},
   publisher={Elsevier BV},
   author={Kurz, Marius and Offenhäuser, Philipp and Beck, Andrea},
   year={2023},
   month=feb, pages={109094} }

@article{bae2022scientific,
  title={Scientific multi-agent reinforcement learning for wall-models of turbulent flows},
  author={Bae, H Jane and Koumoutsakos, Petros},
  journal={Nature Communications},
  volume={13},
  number={1},
  pages={1443},
  year={2022},
  publisher={Nature Publishing Group UK London}
}

@misc{li2021fourierneuraloperatorparametric,
      title={Fourier Neural Operator for Parametric Partial Differential Equations}, 
      author={Zongyi Li and Nikola Kovachki and Kamyar Azizzadenesheli and Burigede Liu and Kaushik Bhattacharya and Andrew Stuart and Anima Anandkumar},
      year={2021},
      eprint={2010.08895},
      archivePrefix={arXiv},
      primaryClass={cs.LG},
      url={https://arxiv.org/abs/2010.08895}, 
}

@inproceedings{kawai,
  title={Coarse-Grid Large-Eddy Simulation by Unsupervised-Learning-Based Sub-Grid Scale Modeling},
  author={Maejimam, Soju and Kawai, Soshi},
  booktitle={AIAA SciTech 2024 Forum},
  year={2024},
  pages={AIAA 2024-1361},
}

@article{gu2021efficiently,
  title={Efficiently modeling long sequences with structured state spaces},
  author={Gu, Albert and Goel, Karan and R{\'e}, Christopher},
  journal={arXiv preprint arXiv:2111.00396},
  year={2021}
}

@article{nguyen2022s4nd,
  title={S4nd: Modeling images and videos as multidimensional signals with state spaces},
  author={Nguyen, Eric and Goel, Karan and Gu, Albert and Downs, Gordon and Shah, Preey and Dao, Tri and Baccus, Stephen and R{\'e}, Christopher},
  journal={Advances in neural information processing systems},
  volume={35},
  pages={2846--2861},
  year={2022}
}

@article{germano1991dynamic,
  title={A dynamic subgrid-scale eddy viscosity model},
  author={Germano, Massimo and Piomelli, Ugo and Moin, Parviz and Cabot, William H},
  journal={Physics of Fluids A: Fluid Dynamics},
  volume={3},
  number={7},
  pages={1760--1765},
  year={1991},
  publisher={American Institute of Physics}
}

@article{schultz2013reynolds,
  title={Reynolds-number scaling of turbulent channel flow},
  author={Schultz, Michael P and Flack, Karen A},
  journal={Physics of Fluids},
  volume={25},
  number={2},
  year={2013},
  publisher={AIP Publishing}
}

@article{Bogacki,
title = {An efficient {R}unge-{K}utta (4,5) pair},
journal = {Computers \& Mathematics with Applications},
volume = {32},
number = {6},
pages = {15-28},
year = {1996},
issn = {0898-1221},
doi = {https://doi.org/10.1016/0898-1221(96)00141-1},
url = {https://www.sciencedirect.com/science/article/pii/0898122196001411},
author = {Bogacki, P. and Shampine, L.F.},
}

@book{roberts1987digital,
title={Digital signal processing},
author={Roberts, Richard A and Mullis, Clifford T},
year={1987},
publisher={Addison-Wesley Longman Publishing Co., Inc.}
}

@article{goc2021large,
  title={Large eddy simulation of aircraft at affordable cost: a milestone in computational fluid dynamics},
  author={Goc, Konrad A and Lehmkuhl, Oriol and Park, George Ilhwan and Bose, Sanjeeb T and Moin, Parviz},
  journal={Flow},
  volume={1},
  pages={E14},
  year={2021},
  publisher={Cambridge University Press}
}

@misc{Wu2025_Git,
  author       = {Wu, Andy},
  title        = {{S4ND\_SGS\_Model}},
  year         = {2025},
  publisher    = {GitHub},
  note         = {\url{https://github.com/awu1018/S4ND_SGS_Model/tree/main}}
}

@article{graham2016web,
  title={A web services accessible database of turbulent channel flow and its use for testing a new integral wall model for LES},
  author={Graham, Jason and Kanov, K and Yang, XIA and Lee, M and Malaya, N and Lalescu, CC and Burns, R and Eyink, G and Szalay, A and Moser, RD and others},
  journal={Journal of Turbulence},
  volume={17},
  number={2},
  pages={181--215},
  year={2016},
  publisher={Taylor \& Francis}
}

@article{meyers2006model,
  title={On the model coefficients for the standard and the variational multi-scale Smagorinsky model},
  author={Meyers, Johan and Sagaut, Pierre},
  journal={Journal of Fluid Mechanics},
  volume={569},
  pages={287--319},
  year={2006},
  publisher={Cambridge University Press}
}

@article{meneveau1996lagrangian,
  title={A Lagrangian dynamic subgrid-scale model of turbulence},
  author={Meneveau, Charles and Lund, Thomas S and Cabot, William H},
  journal={Journal of fluid mechanics},
  volume={319},
  pages={353--385},
  year={1996},
  publisher={Cambridge University Press}
}

\end{document}